\RequirePackage[2020-02-02]{latexrelease}
\documentclass{aastex631}
\shorttitle{OJ 287 and Other $\gamma$-ray Blazar Light Curves from K2}
\shortauthors{Wehrle et al.}
\begin{document}

\title{K2 Optical Emission from OJ~287 and Other {$\gamma$}-Ray Blazars on Hours-to-Weeks Timescales from 2014--2018}
\author[0000-0003-4737-1477]{Ann E.~Wehrle}
\affiliation{Space Science Institute, 
 4765 Walnut Street, Suite B,
Boulder, CO 80301, USA}

\author[0000-0001-8961-2465]{Michael Carini}
\affiliation{Western Kentucky University,
Department of Physics and Astronomy,
1906 College Heights Blvd.,
Bowling Green, KY 42101, USA}

 \author[0000-0002-1029-3746]{Paul J.~Wiita}
 \affiliation{The College of New Jersey,  
 Department of Physics, 
 2000 Pennington Rd., 
 Ewing, NJ 08628-0718, USA}

\author[0000-0002-3827-8417]{Joshua Pepper}
\affiliation{Lehigh University,
Department of Physics,
413 Lewis Laboratory, 16 Memorial Drive East, 
Bethlehem, PA 18015, USA}

\author[0000-0003-0395-9869]{B.~Scott Gaudi}
\affiliation{The Ohio State University,
Department of Astronomy,
140 W. 18th Ave.,
Columbus, OH 43210, USA}

\author[0000-0003-1435-3053]{Richard W.~Pogge}
\affiliation{The Ohio State University,
Department of Astronomy,
140 W. 18th Ave.,
Columbus, OH 43210, USA}
 \affiliation{The Ohio State University,
Center for Cosmology \& AstroParticle Physics,
191 West Woodruff Ave.,
Columbus, OH 43210, USA}

\author[0000-0002-3481-9052]{Keivan G.~Stassun}
\affiliation{Vanderbilt University,
Department of Physics \& Astronomy,
6301 Stevenson Center Ln.,
Nashville, TN 37235, USA}

\author[0000-0001-6213-8804]{Steven Villanueva, Jr.}
\altaffiliation{NASA Postdoctoral Program Fellow}
\affiliation{The Ohio State University,
Department of Astronomy,
140 W. 18th Ave.,
Columbus, OH 43210, USA}
\affiliation{NASA Goddard Space Flight Center, Exoplanets and Stellar Astrophysics Laboratory (Code 667), Greenbelt, MD 20771, USA}

\date{Version of 29 March 2023. Accepted for publication in the Astrophysical Journal.}
\begin{abstract}

We present second observations  by K2 of OJ~287  and 7 other $\gamma$-ray AGNs obtained  in 2017--2018,  second and third observations of the lobe-dominated, steep spectrum quasar 3C~207, and  observations of  9 additional blazars not previously observed with K2.  The AGN were observed simultaneously with K2 and the Fermi Large Area Telescope for 51--81 days.   Our full sample, observed in 2014--2018, contained 16 BL Lac objects (BL Lacs),  9 Flat Spectrum Radio Quasars (FSRQs), and 4 other $\gamma$-ray AGNs.  Twelve BL Lacs and  7 FSRQs  exhibited fast, jagged light curves while 4 BL Lacs and  2 FSRQs had slow, smooth light curves.  Some objects changed their red-noise character significantly between repeated K2 observations.  The optical characteristics of OJ~287 derived from the short-cadence K2 light curves changed  between observations made before and after the predicted passage of the suspected secondary supermassive black hole through the accretion disk of the primary supermassive black hole. The average slopes of the periodogram power spectral densities  of the BL Lacs' and FSRQs' light curves differed significantly,   by $\approx 12$\%, with the BL Lac slopes being steeper, and a KS test with a $p$-value of  0.039 indicates that these samples probably come from different populations; however, this result is not as strongly supported by PSRESP analyses. Differences in the origin of the jets from the ergosphere or accretion disk in these two classes could produce such a disparity, as could  different sizes or locations of emission regions within the jets.   
\end{abstract}
\newpage
\setcounter{section}{0}

\section{Introduction}

Strong, rapid variability is one of the prime characteristics of the blazar class of active galactic nuclei.  Both the historic sub-categories of blazars, the BL Lacertae objects (BL Lacs) and the flat spectrum radio quasars (FSRQs), show these fluctuations in every observable band of the electromagnetic spectrum. This paper is the last of three to focus on the optical variations of those blazars measured by the Kepler satellite in its extended  K2 mission that were also bright enough in $\gamma$-rays to have been detected by the Large Area Telescope on the Fermi Gamma-Ray Space Telescope (Fermi-LAT) in the first four years of its mission  as listed in the Fermi-LAT Second AGN Catalog or 2LAC (Ackermann et al.\ 2011).  Our earlier observations were reported in Wehrle et al.\ (2019), hereafter Paper 1, and Carini et al.\ (2020), hereafter Paper 2.  These  K2  measurements provide unique  data, as they are made nearly continuously over nominally  $\sim 80$ day long campaigns at a cadence of 29.4 minutes.

All blazars have long been understood to possess plasma jets with bulk relativistic flows pointing close to our line of sight (Blandford \& Rees 1978;  Urry \& Padovani 1995). The strong Doppler boosting of the flux from the approaching jet means that the its emission dominates over other sources of radiation in most, if not all,  bands.  The first main difference between FSRQs and BL Lacs are that the former are quasars with clear emission lines visible in their optical and UV spectra while the latter have undetectable or extremely weak emission lines.   Both BL Lacs and FSRQs have non-thermal double-humped spectral energy distributions (SEDs)  (usually plotted as  log($\nu F_{\nu}$) vs.\ log($\nu$)), but the frequencies at which the intrinsically weaker BL Lac SEDs peak are normally higher than the corresponding ones for the more powerful FSRQs (e.g., Fossati et al.\ 1998).  The lower-frequency SED humps  are  dominated by synchrotron emission from the ultrarelativistic electrons in the jet, though quasi-thermal radiation from accretion disks is detectable in many FSRQs but generally not in BL Lacs (e.g., Ghisellini et al.\ 2017).   The high-energy humps can usually be nicely fitted by leptonic models, where the same ultrarelativistic electrons upscatter photons to produce $\gamma$-rays (and sometimes hard X-rays); the seed photons can be the synchrotron emission itself (synchrotron-self-Compton process) (e.g., Ghisellini et al. 1998) but in many cases they apparently come from the accretion disk, broad line clouds or the dust torus (external Compton process) (e.g., Dermer et al.\ 2009; Arsioli \& Chang 2018).  Alternative, hadronic, models for the high-energy hump involve contributions from proton-synchrotron radiation and $\gamma \gamma$ pair production and can sometimes provide better fits to the SEDs (e.g., B{\"o}ttcher et al. 2013; Paliya et al.\ 2018).

In Paper 1 we described and discussed simultaneous Fermi-LAT observations of nine $\gamma$-ray blazars, including the famous BL Lac, OJ 287, that were made in 2014--2015.   OJ 287 is the best candidate for a blazar with a  supermassive black hole (SMBH) binary at its core (e.g., Sillanp{\"a}{\"a} et al.\ 1988; Valtonen et al.\ 2016).  Paper 2 similarly treated an additional ten such blazars, including the first identified quasar, 3C 273, that were observed in 2015--2017.  In Paper 2 we also presented a reanalysis of some of the results from Paper 1, so that specific comparisons could be made uniformly.  In this paper we consider the final observations of $\gamma$-ray blazars made by K2  in 2017--2018.  Included in the total of 18 sources for which new observations are presented in this paper are seven sources, including OJ 287, that we had previously observed (Paper 1) and one source observed three times.  These repeated K2 observations allow us to investigate changes in the characteristics of multi-month-long light curves  and their  power spectral densities (PSDs) over spans of a few years. 

We describe the selection of $\gamma$-ray blazars in Section 2.  In Section 3, we describe the K2 observations in Campaigns 14--19, the data reduction, and the light curve analysis.   We present the results of the K2 optical observations, including the  PSDs, in Section 4. Section 5 describes contemporaneous ground-based observations obtained for  three objects which provided independent confirmation of the K2 brightness corrections.  We describe the contemporaneous  Fermi-LAT data and our analysis of them in Section 6. We discuss our results in Section 7 and we summarize our conclusions in Section 8.

\begin{deluxetable}{lrrrclccc}
\tablecaption{Targets in 2017--2018\tablenotemark{a}}
\tablewidth{0pt}
\singlespace
\tablecolumns{10}
\tabletypesize{\small}
\tablehead{
             \colhead{Name}
             &\colhead{EPIC ID}
             & \colhead{Fermi-LAT}
             & \colhead{$K_p$\tablenotemark{b}}
             & \colhead{$z$}
             & \colhead{Class}
             & \colhead{Campaign}
             & \colhead{Notes}
             \\
             \colhead{}
             & \colhead{}
             & \colhead{Name}
             & \colhead{}
             & \colhead{}
             & \colhead{}
             & \colhead{Fields}
             & \colhead{}
            \\
             }
\startdata
NVSS J110735+022225	 & 	201621388	 & 	4FGL~J1107.6+0222	 & 	18.584	 & 	$>$1.0735	 & 	BL Lac	 & 	14	 & 	1	\\
(NVSS J105151+010312	 & 	248438564	 & 	 4FGL~J1051.9+0103	 & 	18.831	 & 	\nodata	 & 	BL Lac	 & 	14	 & 	2, 3)	\\
(PMN J1018+0530	 & 	248586210	 & 	4FGL~J1018.4+0528	 & 	17.457	 & 	1.94525	 & 	FSRQ	 & 	14	 & 	4)	\\
4C +06.41	 & 	248611911	 & 	see note	 & 	16.888	 & 	1.27	 & 	FSRQ	 & 	14	 & 	5	\\
TXS 1013+054	 & 	251457104	 & 	4FGL~J1016.0+0512	 & 	18.75	 & 	1.71272	 & 	FSRQ	 & 	14	 & 	\nodata	\\
(PMN J1059+0225	 & 	251457105	 & 	see note	 & 	19.0	 & 	\nodata	 & 	unknown	 & 	14	 & 	6)	\\
3C 207	 & 	211504760	 & 	4FGL~J0840.8+1317	 & 	18.264	 & 	0.6808	 & 	LDRQ	 & 	5, 16, 18	 & 	\nodata	\\
TXS 0836+182	 & 	211852059	 & 	4FGL~J0839.4+1803	 & 	17.046	 & 	\nodata	 & 	BL Lac	 & 	16	 & 	2, 7	\\
NVSS J090226+205045	 & 	212035517	 & 	4FGL~J0902.4+2051	 & 	15.741	 & 	\nodata	 & 	unknown	 & 	16	 & 	2	\\
TXS 0853+211	 & 	212042111	 & 	4FGL~J0856.8+2056	 & 	18.345	 & 	\nodata	 & 	BL Lac	 & 	16	 & 	2	\\
NVSS J090900+231112	 & 	251376444	 & 	4FGL~J0908.9+2311	 & 	17.939	 & 	$>$ 0.432	 & 	BL Lac	 & 	16	 & 	8	\\
PKS 1335-127	 & 	212489625	 & 	4FGL~J1337.6$-$1257	 & 	17.851	 & 	0.539	 & 	FSRQ	 & 	6, 17	 & 	\nodata	\\
PMN J1318-1235	 & 	212507036	 & 	 4FGL~J1318.7$-$1234	 & 	18.027	 & 	\nodata	 & 	unknown	 & 	17	 & 	\nodata	\\
PKS 1352-104	 & 	212595811	 & 	4FGL~J1354.8$-$1041	 & 	17.399	 & 	0.332	 & 	FSRQ	 & 	6, 17	 & 	\nodata	\\
RBS 1273	 & 	212800574	 & 	 4FGL~J1329.4$-$0530	 & 	15.201	 & 	0.57587	 & 	X-ray QSO	 & 	6, 17	 & 	\nodata	\\
PKS B1329-049	 & 	229227170	 & 	4FGL~J1332.0$-$0509	 & 	18.2	 & 	2.15	 & 	FSRQ	 & 	6, 17	 & 	9	\\
PKS B1310-041	 & 	251502828	 & 	 4FGL~J1312.8$-$0425	 & 	18.3	 & 	0.8249	 & 	FSRQ	 & 	17	 & 	\nodata	\\
RGB J0847+115	 & 	211394951	 & 	 4FGL~J0847.2+1134	 & 	17.331	 & 	0.1982	 & 	BL Lac	 & 	5, 18	 & 	\nodata	\\
WB J0905+1358	 & 	211559047	 & 	 4FGL~J0905.6+1358	 & 	17.213	 & 	0.2239	 & 	BL Lac	 & 	5, 18	 & 	10	\\
OJ 287	 & 	211991001	 & 	4FGL~J0854.8+2006	 & 	14.819	 & 	0.3056	 & 	BL Lac	 & 	5, 18	 & 	\nodata	\\
BZB J0816+2051	 & 	212035840	 & 	4FGL~J0816.9+2050	 & 	17.486	 & 	\nodata	 & 	BL Lac	 & 	5, 18	 & 	\nodata	\\
(PKS 2335-027	 & 	246327456	 & 	4FGL~J2338.0$-$0230	 & 	18.318	 & 	1.072	 & 	FSRQ	 & 	19	 & 	11)	\\
(PKS 2244-002	 & 	251721269	 & 	4FGL~J2247.4$-$0001	 & 	18.359	 & 	0.094	 & 	BL Lac	 & 	19	 & 	11)	\\
\enddata
\tablecomments{
1. Two faint objects are 8\arcsec ~and 12\arcsec ~from the blazar. The blazar light curve was separated with a custom aperture.
2. The SDSS automated photometric redshift cited in NED is not reliable because quasar spectral template does not apply to blazars.
3. The blazar has a stellar object of similar brightness within 4\arcsec ~(1 pixel). Light curves could not be separated.
4. Target was too faint during K2 observations to have sufficient signal-to-noise ratio for analysis.
5. See text for discussion of association.
6. See text for explanation of why K2 target was not associated with a 4FGL source.
7.  A redshift of 0.28 for TXS 0836+182 is an estimate from galaxy morphological fitting (Abraham et al. 1991). 
8. The blazar has a fainter non-AGN companion galaxy 3.8\arcsec ($\sim 1$ pixel) away (see Rosa-Gonz{\'a}lez et al.\ 2017). Both have the same absorption line redshift. In our custom aperture, the variability comes from the blazar, not the inactive companion.
9. Target has alternate EPIC ID 229228144.
10. Target has alternate EPIC ID 211559044 which does not appear in all MAST searches.
11. Target omitted from final sample due to poor  K2 data quality in Campaign 19 during the last days of the mission.}
\tablenotetext{a}{Entries for the five objects omitted from the final sample are indicated with parentheses.} 
\tablenotetext{b}{Kepler magnitude as tabulated in the EPIC  catalog.} 

\end{deluxetable}

\section {Sample Selection}

 We initially chose 23 targets known to emit $\gamma$-rays (Table 1) by searching the Fermi-LAT Second AGN Catalog (2LAC, Ackermann et al.\ 2011) for the optically-brightest AGN that would be observed in the K2 Campaigns 14--19 fields (programs GO14027, GO16027, GO17010, GO18010, and GO19010, led by  principal investigator A.\ Wehrle), following the criteria we used in Papers 1 and 2.  Although 21 of these targets are indeed blazars, we determined that 2 are not and and we do not have enough information to classify 3 targets as either FSRQs or BL Lacs. Thirteen blazars overlapped with the infrared and optically bright samples  selected by programs GO14040, GO16040, GO17020, GO18020, and GO19020 that were led by principal investigator M.\ Carini.  We give updated $\gamma$-ray names from the Fermi-LAT 8-year Source Catalog (4FGL, Ajello et al.\ 2020) and Fermi-LAT 10-year Source Catalog (4FGL-DR2,  Ballet et al.\ 2020) in Table 1.  
The $\gamma$-ray blazars in the five fields had estimated K2 optical magnitudes $K_p$ in the 420--900 nm band from 14.818 to 18.831  in the K2 Ecliptic Plane Input Catalog (EPIC; Huber et al.\ 2016; Huber \& Bryson 2018 and references therein) which was generally drawn from the USNO-B and SDSS catalogs. OJ~287  is by far the brightest, with $K_p = 14.818$.    The  K2 mission ended a few days into Campaign 19 when the spacecraft could no longer be accurately pointed.  Our two Campaign 19 targets were dropped from the sample because the data quality was poor during the last few days of the mission. One target, NVSS J105151+010312	 = EPIC 	248438564, was dropped because its light curve could not be separated from that of a nearby star. Another target, PMN J1018+0530 = EPIC 	248586210, was dropped because it was too faint during the mission. A target, PMN J1059+0225 = EPIC 251457105, that turned out to be not associated with a 4FGL source was also dropped from the sample.
In our final sample of 18 unique targets observed in 2017--2018, eight are classed as BL Lacs, six as FSRQs, one as an X-ray QSO, one as a lobe-dominated radio quasar (LDRQ), and two as blazars of unknown class.

\begin{deluxetable}{cccr}
\tablecaption{Dates of K2 and Fermi-LAT Observations}
\tablewidth{0pt}
\singlespace
\tablecolumns{4}
\tabletypesize{\small}
\tablehead{
             \colhead{K2 Campaign}
             & \colhead{Dates}
             & \colhead{MJD}
             &\colhead{Duration} 
             \\
             }
\startdata
14 & 2017 Jun 1 -- 2017 Aug 19 & 57905-57985 & 81 days \\
16\tablenotemark{a} & 2017 Dec 7 -- 2018 Feb 25 & 58095-58174 & 80 days \\
17\tablenotemark{a} & 2018 Mar 2 -- 2018 May 8 & 58179-58246 & 67 days \\
18 & 2018 May 13 -- 2018 Jul 2 & 58251-58301 & 51 days \\
\enddata
\tablenotetext{a}{Fields 16 and 17 were forward-facing campaigns which enabled simultaneous observations from the ground.}
\end{deluxetable}
 
 \newpage
\section{ K2 Observations, Data Reduction and Analysis}

A journal of the K2 and Fermi-LAT observations is given in Table 2.  Campaigns 14, 16, 17 and 18 had no anomalies as described in the K2 Data Release Notes \url {https://keplergo.github.io/KeplerScienceWebsite/k2-data-release-notes.html}. Campaigns 16 and 17 were ``forward-facing'' campaigns to enable simultaneous ground-based data to be obtained.  The Campaign 5 and 18 fields,  the  Campaign 6 and 17 fields, and the Campaign 16 and 18 fields overlapped, hence, the targets with more than one K2 observation are all from Campaigns 5, 6, 16, 17, and 18.  Campaign 18 was ended early to download its data when it appeared that the cryogen was running low. Campaign 19 had poor data quality while the cryogen was running out, and  ended  after 10 days.
 
As in our earlier observations, the two longest campaigns' observations contained  $\sim$3,500 29.4-minute (``30-minute''  or ``long cadence'') samples, with the 67 day Campaign 17 containing $\sim$3000 data points, and the 51-day Campaign 18 containing $\sim$ 2200 data points.  We used data corrected by the EVEREST algorithm Version 2, hereafter ``EVEREST'' developed by Luger et al.\ (2016, 2018) to compensate for the drift of the spacecraft caused by photon pressure and the motions introduced by the firings of the spacecraft thruster approximately every 6 or 12 hours  (Van Cleve et al.\ 2016); see Paper 2 for details.   Anomalous data caused by thermal drift during the first couple of days at the beginnings of the campaigns were discarded.    As in Papers 1 and 2, we were also able to obtain ``short cadence'' ($\sim 1$ minute intervals) for the brightest blazar in our sample, here OJ~287.
 
As we found with the standard {\it Kepler} pipeline products for AGN light curves (Wehrle et al.\ 2013; Revalski et al.\ 2014),  standard K2 pipeline output for AGN often removed true astrophysical brightness variations in the process of removing instrumental effects. As in Papers 1 and 2, we examined the light curves produced by the standard pipeline processing (SAP, PDCSAP), the K2SFF processing (Vandenburg \& Johnson 2014, and updates online at MAST), and the EVEREST processing (Luger et al.\ 2016, 2018). The EVEREST light curves had the fewest residual instrumental errors  coming from thermal drifts in the first two days of campaigns, isolated low points, and ``sawtooth'' amplitude variations at thruster firing (6 and 12 hour) intervals.

We compared the EVEREST light curves of targets that fell on the same modules to see if they varied simultaneously. We found no evidence of significant correlated variability.  We also checked for evidence of nonlinear amplifier effects (Kolodziejczak et al.\ 2010; tabulated in Table 13 of the {\it Kepler} Instrument Handbook, Van Cleve \& Caldwell 2016). Two and six targets, respectively, fell on modules classed as having ``strong" and ``medium'' (``Moir\'{e}") effects, however,  detailed examination of the amplitude variations  showed no visible evidence of ``Moir\'{e}" distortions.  We  found no evidence of (unrelated)``rolling band" effects. 

Our long cadence observations of the brightest target, OJ 287, reached a noise level (standard deviation) after EVEREST processing of  13 ct s$^{-1}$  ($\sim$ 0.09\%).  The other targets' long cadence observations had noise levels of  3--13 ct s$^{-1}$  while the average brightnesses were 
244  to 11,728 ct s$^{-1}$.  The noise was measured  during $\sim$ 0.5 day intervals when the light curves  exhibited variations below 1\%. The OJ 287 short cadence data were custom processed by us through the EVEREST software in the same fashion as described in Paper 1.
  
We examined the  K2 ``postage stamp'' apertures used in processing for each target to make sure the targets were isolated. EPIC 211394951 was not present in the EVEREST pipeline processing data products available at MAST; we applied the EVEREST processing ourselves utilizing a custom aperture.  EPIC 251376444 has a close companion only 3.8\arcsec away; we used a custom aperture where the variable signal comes from the AGN, not the companion.   The original EVEREST aperture of EPIC 212035840 had moved the object out of the aperture at the end of the campaign; hence, it needed custom processing with a carefully chosen aperture.   EPIC 201621388 had two other faint sources in the EVEREST pipeline aperture, so a custom aperture was designed to exclude contribution from these sources. EPIC 251502828 has another similar-brightness object about  10\arcsec  ~away, so we reprocessed it with a custom aperture that excluded the other object.    The final sample in Campaigns 14 --18  contained 18 unique targets with good quality data, including 3C~207 observed in both Campaigns 16 and 18.  In addition, we found that for 3 of the objects from our previous papers (2 from Paper 1 (EPIC 211559047, 211394951) and 1 from Paper 2 (EPIC 217154395)) the default EVEREST aperture was contaminated with other objects.  We re-reduced these sources with custom apertures to eliminate the contribution  from these nearby sources to the blazar's light curve.  The field of EPIC 217154395 was found to be too crowded to produce a reliable light curve containing only the AGN flux.

\section{K2 Optical Results}

The K2 EVEREST light curves for our new and reprocessed targets are shown in the top  sub-panels of  Figures 1--4.   Key features of these light curves are given in Table 3. 
During the K2 observations, the greatest variation (maximum/minimum) detected was the factor of 12.24 flaring we observed in the FSRQ PKS 1352-104 (EPIC 212595811) during Campaign 6. The smallest  variations, a factor of 1.01 (long cadence) or 1.02 (short cadence), were seen in 3C~273 (EPIC 229151988) and  in the X-ray QSO, RBS 1273 (EPIC 212800574), with factors of 1.02 and 1.01 in Campaigns 6 and 17, respectively. Substantial variations by factors of  1.07--8.50 were detected in the other objects. All of these variations are much larger than the respective noise levels (Table 3).

\subsection {Overall Appearance}

We examined the light curves for all the sources in our K2 samples as listed in Table 3 and shown  in Figs.\ 1 -- 4 in this paper and in Figs.\ 3 -- 13 in Paper 1 and Figs.\ 2 -- 5 in Paper 2. We  qualitatively characterized the observed fluctuations into two broad categories: fast, jagged changes for light curves that contained multiple clear changes of flux trends during the campaigns; and slow, smooth variations that showed only a few flux modulations over that period.   These characterizations are given in Table 3, in the Max/Min column, with S denoting smooth light curves and J denoting jagged. Considering each individual campaign separately (so some sources are double counted and one is triple counted) we have a total of 39 light curves, of which  29 are jagged and 10 are smooth.  There is  no significant difference between the blazar types, with there being 15 jagged light curves for BL Lacs and  10 for FSRQs and 5 smooth  light curves for BL Lacs and  2 for FSRQs.   The sources in our sample that are  of uncertain blazar type or not true blazars were also  split, with 4 jagged and 3 smooth light curves.   If we do not count second observations of blazars independently, there are 12  BL Lacs and 7  FSRQs with jagged light curves while 4 BL Lacs and 2 FSRQs have smooth light curves.  We also checked for any difference with source redshift; however,  both the more common jagged and the rarer smooth categories included both  lower and higher redshift objects of both BL Lac and FSRQ  types, so there is no evidence of a trend with redshift. 

We   also considered the ranges of optical fluxes in terms of the ratio of the maximum to minimum count rates seen during these observations, as given in Table 3.  The mean value of Maximum/Minimum for the 12 long cadence FSRQ light curves is 2.96 if we include the very large flare seen in PKS 1532$-$104 during Campaign 6 but 2.08 if we exclude that flare from that light curve.  The corresponding mean value for the 20 BL Lac light curves is 1.49, so it appears that FSRQs are more active by this measure.  If we consider the median of the Maximum/Minimum values the difference is somewhat less: 1.52 for FSRQs (regardless of the flare's inclusion) and 1.32 for BL Lacs.  The fact that FSRQs have higher average Doppler boosting factors than BL Lacs (e.g., Liodakis et al.\ 2017) could explain this difference. 
 We also tabulate, in the ninth column of Table 3, another quantity characterizing the variations,  the ``Coefficient of Variation" (defined as (average count rate)/(standard deviation during full campaign)).  This variability measure is more commonly used at other wavebands and so may be useful for multi-wavelength studies.

\subsection {Slopes of Optical Power Spectral Densities}

The primary analysis technique for the K2 data in this paper, as in Papers 1 and 2, is the computation and examination of PSDs. A key difference in this paper is that the PSDs were constructed utilizing a periodogram-based method, rather than a direct application of the discrete Fourier transform (e.g., Edelson \& Krolik 1988), as in Papers 1 and 2. Our rationale for this choice is provided in detail in the Appendix. These PSDs are shown in the bottom sub-panels of Figures 1--4 and their slopes and errors are listed in Table 3. These PSDs are usually characterized by red noise, with $P(\nu) \propto \nu^{\alpha}$, with $\alpha < 0$, for $\nu < \nu_b$; above this ``break frequency'' it has a white noise character ($\alpha = 0$) which is expected when the signal is dominated by measurement errors. To minimize the red noise leak (transfer of variability power from low to high frequencies), we computed the PSDs after removing any linear trend and then convolved them with a Hamming window function (Harris 1978), as in Papers 1 and 2. To best estimate the PSD slopes, we then binned the logarithm of the power in intervals of 0.08 in log~$\nu$ and found a linear fit to the portion of the PSD displaying power law behavior (bottom  sub-panels of  Figs.\ 1--4) as defined in Paper 2.  This ``sweet spot'' in the PSDs, where the abundant, adequately-sampled and low-noise data were well fitted by a single power law,  was generally between log $\nu [Hz] = -5.0$ and $-6.4$, corresponding to timescales between 1.16 and 36.6 days. Error bars in the PSD plots represent the rms scatter in the data points in each bin and where there is only one point per bin, no error bar is shown.

It is useful to consider the effects of flares on the PSDs. We note that there is only one light curve in which a strong isolated flare was observed: PKS 1352$-$104 during Campaign 6 (Figure 2d). In this case we saw that a single PSD slope is not a good fit so we removed the last 16.5 days of data and recomputed the PSD, as shown in Figure 2e. The results with and without the big flare are both given in Table 3. All of the other source light curves are typical of blazars, with multiple flares of varying modest strengths observed during the length of these K2 campaigns. It would be unproductive to cut out any of these variations in this study of variability properties and we have not done so. Discontinuities with strong drops in power at certain frequencies are the dips associated with the weak peaks that originate in a few flares peaking in the light curve that have similar timescales of a few days. They are not due to individual strong flares because such flares cause flattening of the low frequency part of the PSD (compare Figures 2d and 2e). They are not due to windowing because windowing effects are minimized by using the Hamming window which produces dips in the light curve at exactly the same frequency for all light curves of a given length. The K2 campaigns are not all the same length: the PSDs obtained for each target during a campaign do not all show dips and peaks at the same frequencies.

In our power spectral response analysis (PSRESP; e.g., Uttley et al. 2002), the results of which are given in the last columns of Table 3, we test a model for the entire observed PSD; this model is a power law in the red noise portion of the PSD with a hard break to the white noise (slope = 0.0) portion.  The location of the hard break to white noise is identified by estimating the frequency at which the slope of the PSD transitions to 0.0. This ``turnover'' frequency varies over a decade or more in frequency from object to object and is seen to change for the same object between different observations on different CCD modules. Thus, the slopes from our PSRESP analysis represent the slope along the entire power law portion of the PSD, from the variable turnover to white noise down to the lowest sampled frequency. The confidence factors quoted in Table 3 represent the confidence of the adopted model (power law + white noise) fit. We note that only a small fraction (17\%) of these confidence factors are $>$ 90\%.   We discuss the effects of red noise leak and sampling window on the periodogram-derived PSDs in Appendix A. For consistency in comparing the PSDs of different sources or of the same source at different times, as we describe in Paper 2, our analysis was focused on the range of frequencies where we found the PSD to be well sampled and exhibiting the lowest noise (i.e., the ``sweet spot''; $-6.4 < $ log $\nu < -5.0$)  using the periodograms. 

For the short cadence data on OJ 287, we anticipated that the PSD, as we found in OJ~287 (Paper 1) and 3C 273 (Paper 2), would show a plateau induced by sampling effects between log~$\nu = -4.2$ and $-3.8$, corresponding to timescales between 4.4 and 1.8 hours.  A similar plateau in the Campaign 18 OJ 287 PSD did appear, and we corrected for it as we did in Papers 1 and 2, utilizing the  DFOURT and CLEAN methodology (Roberts et al.\ 1987; Hoegbom 1974).

The slopes of the optical PSDs  found by the periodogram method of the 14 new long cadence blazar light curves over the ``sweet spot''  red noise frequency ranges (generally for log ${\nu}$ ranging from $-5.0$ to  $-6.4$) vary  from $-1.18 \pm 0.38$ to $-2.95 \pm 0.35$ (Table 3).  The corresponding power-law fits over the entire range of red-noise found using the PSRESP method are between $-1.61\pm 0.10$ and $-2.58 \pm 0.32.$
We show histograms of both of these approaches to the PSD slopes in  Figure 5 for targets in  this paper combined with those in Paper 1 and Paper 2.  
The mean value of the 12 FSRQ PSD slopes determined via the periodogram method  is $-1.98 \pm 0.42$ and the mean value of the 20 BL Lac PSD slopes is $-2.22 \pm 0.38$.  The median values are $-1.93$ and $-2.27$, respectively. Objects with multiple observations have all measured slopes included in these histograms. A two-sided KS test yields a KS  statistic of 0.483 and a $p$ value of 0.039.  It is thus unlikely the two samples come from the same distribution and we conclude that there is a probable difference in the  typical PSD slopes between BL Lac objects and FSRQs measured on the timescales probed by K2 campaigns.

The mean value of the 12 FSRQs PSD slopes determined via the PSRESP method  is $-1.88 \pm 0.33$ and the mean value of the 20 BL Lacs is $-2.05 \pm 0.32$.  The median values are $-1.86$ and $-2.01$, respectively. Objects with multiple observations have all measured slopes included in these histograms. A two-sided KS test yields a KS  statistic of 0.35 and a $p$ value of 0.255.  We note that the majority of the slopes determined by the PSRESP method have low confidence factors (column 14, Table 3), due to the the fact that the PSRESP method fits the entire PSD, rather than the well defined portion of the PSD we define as our ``sweet spot''. Our PSRESP implementation includes an estimate of the high frequency turnover to white noise and the fit to the linear part of the PSD includes all PSD values down to the lowest frequency measured. Thus, while the probability returned by the two-sided KS test that the two distributions arise from different parent populations is not as high as that from the periodogram method, we can understand this because the PSRESP analysis includes data at low frequencies below the well-sampled ``sweet spot'' which tend to flatten the PSD slope.
   
\begin{figure*}
\figurenum{1}
\gridline{\fig{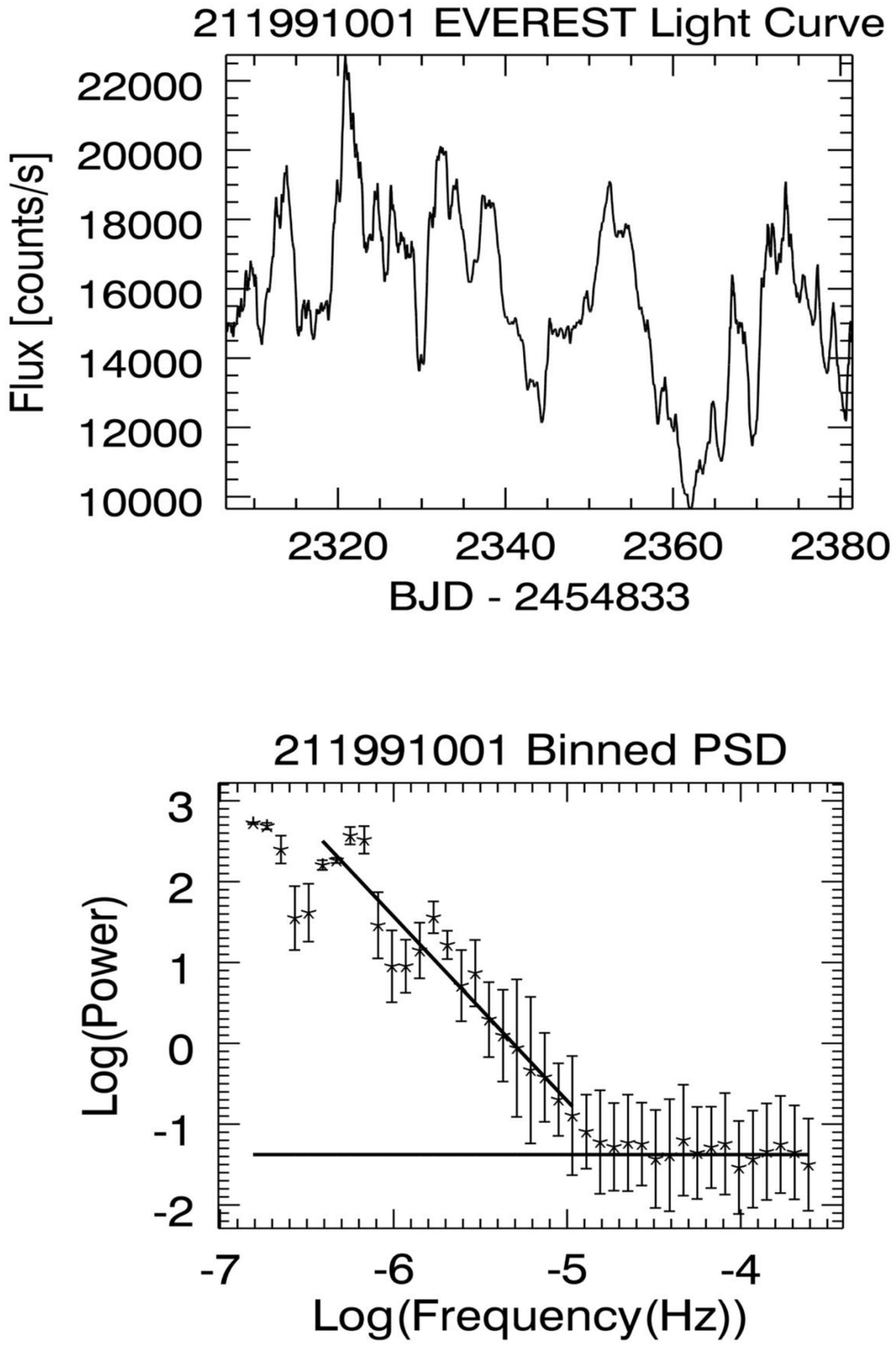}{0.44\textwidth}{(a)}
          \fig{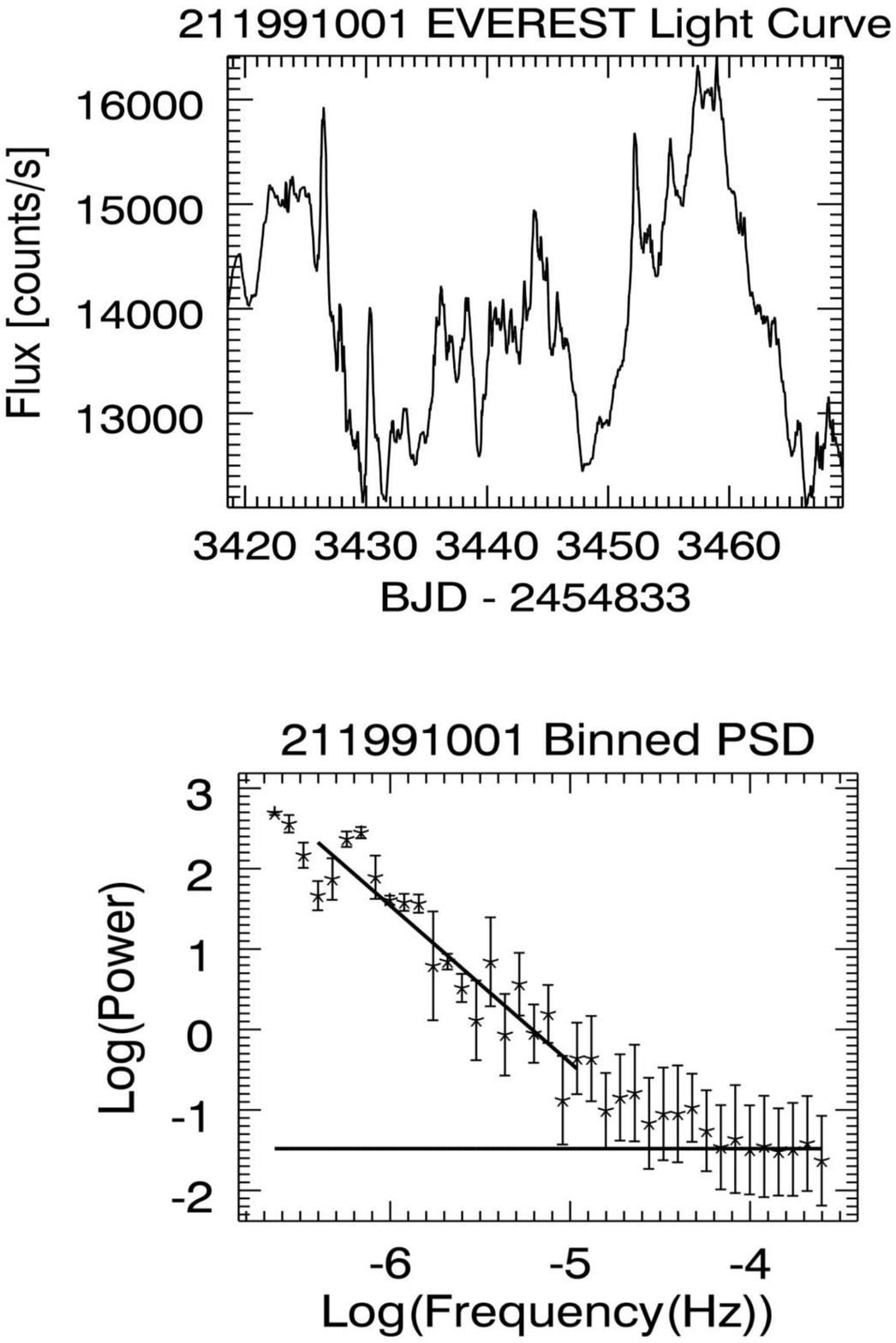}{0.44\textwidth}{(b)}
          }
\gridline{\fig{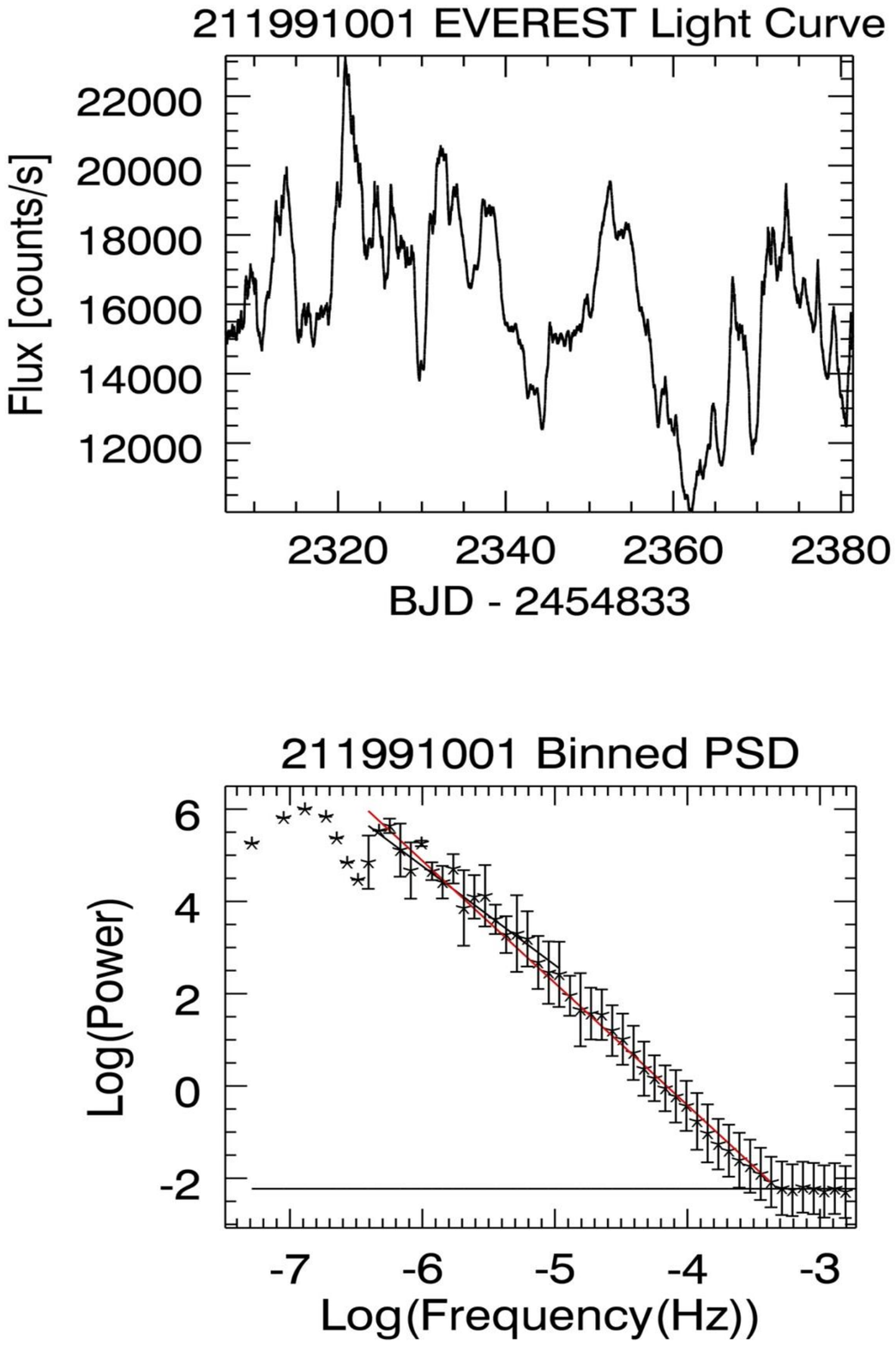}{0.44\textwidth}{(c)}          
          \fig{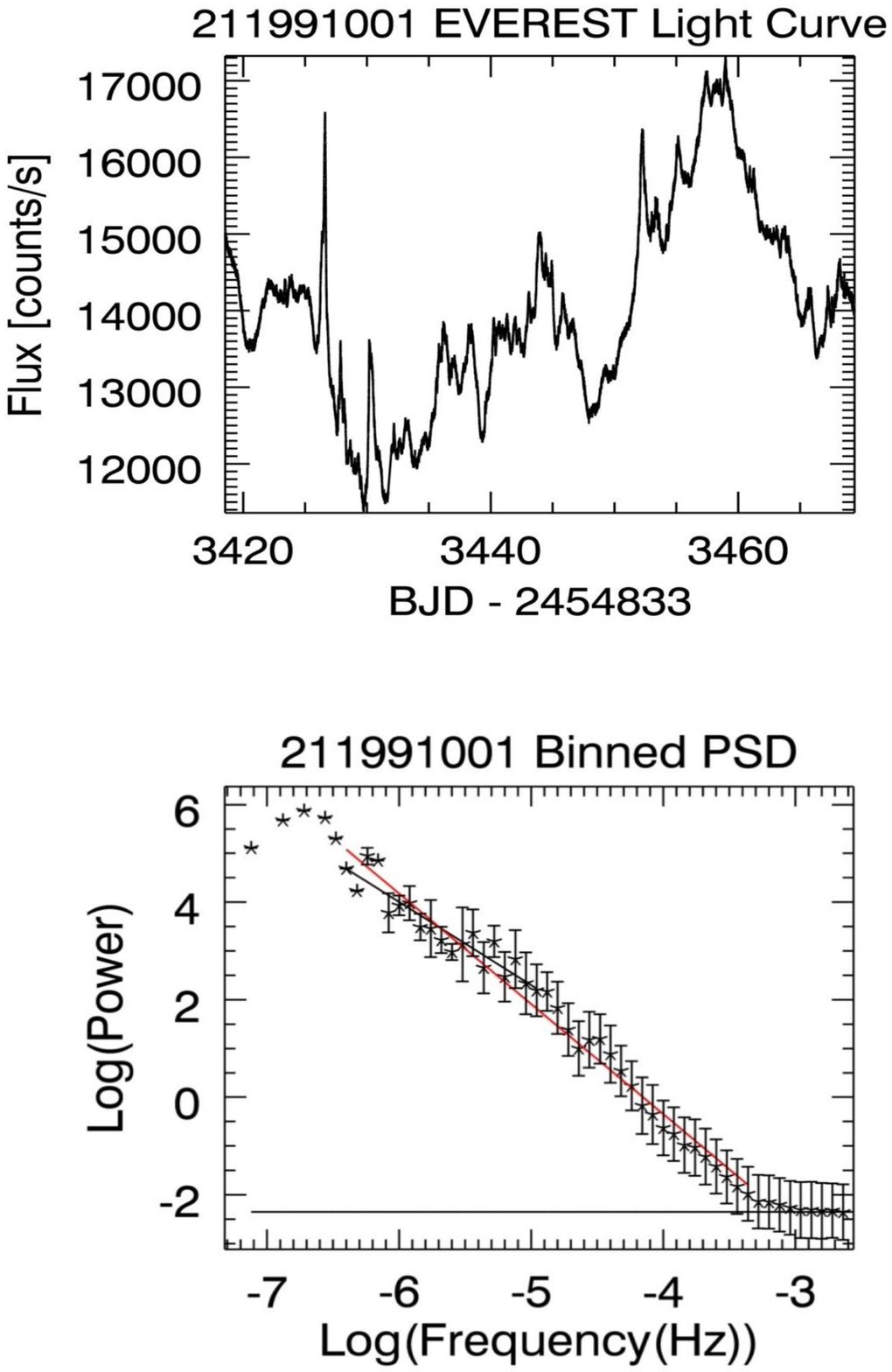}{0.44\textwidth}{(d)}
          }
\caption{OJ~287 ( EPIC 211991001) in Campaign 5  and 18. (a) Long cadence light curve and PSD in Campaign 5; (b) long cadence light curve and PSD in Campaign 18; (c) short cadence light curve and PSD in Campaign 5; (d) short cadence light curve and PSD in Campaign 18. The short cadence PSDs for both campaigns have been fitted twice: the black lines are fitted in the ``sweet spot'' between log $\nu = -5.0$ and $-6.4$, and the magenta lines are fitted between log $\nu = -3.4$ and $-6.4$; the latter PSDs have slopes $-2.65 \pm 0.05$ and $-2.26 \pm 0.06$, for Campaigns 5 and 18, respectively.}
\end{figure*}
\begin{figure*}
\figurenum{2}
\gridline{\fig{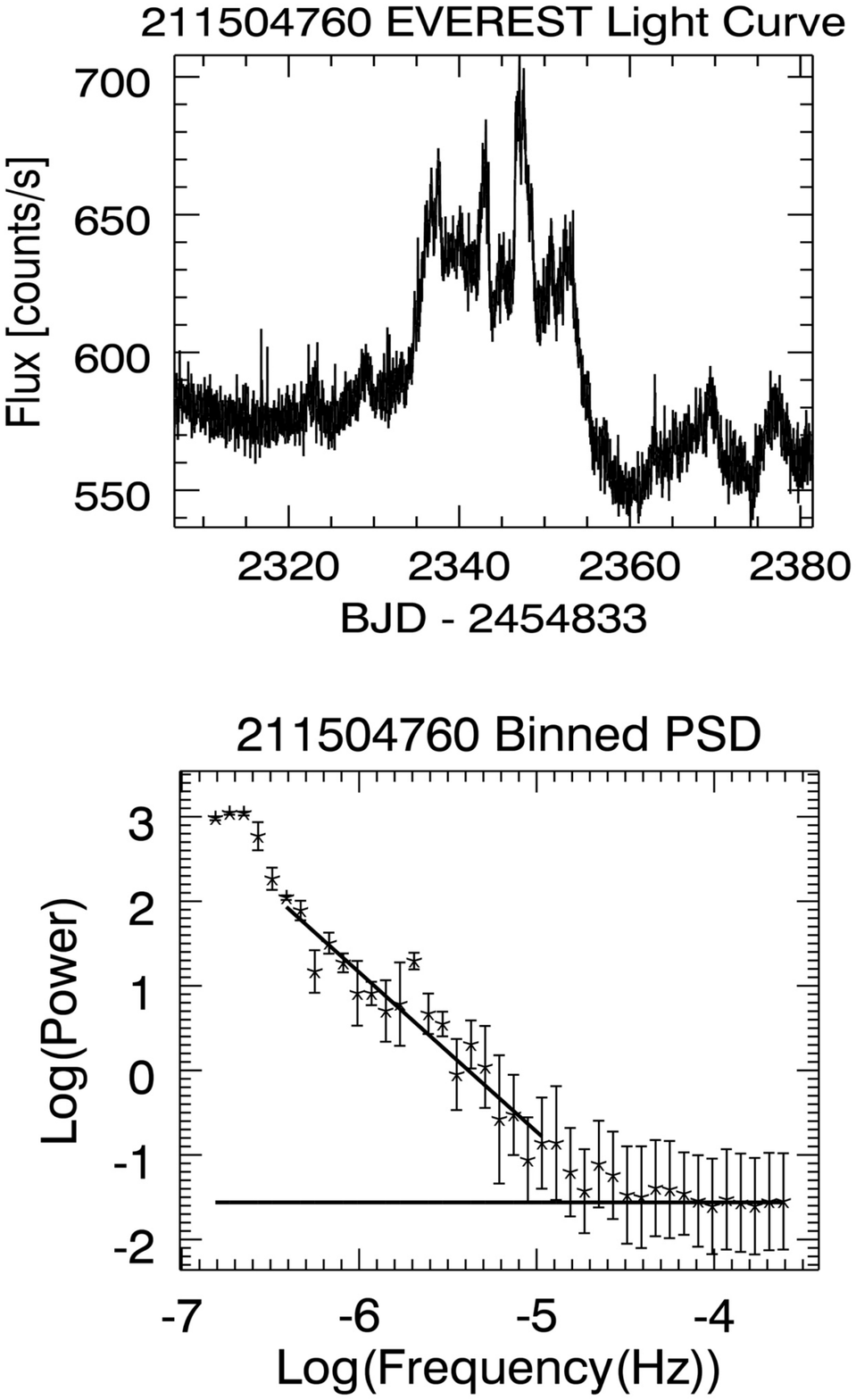}{0.35\textwidth}{(a)}
        \fig{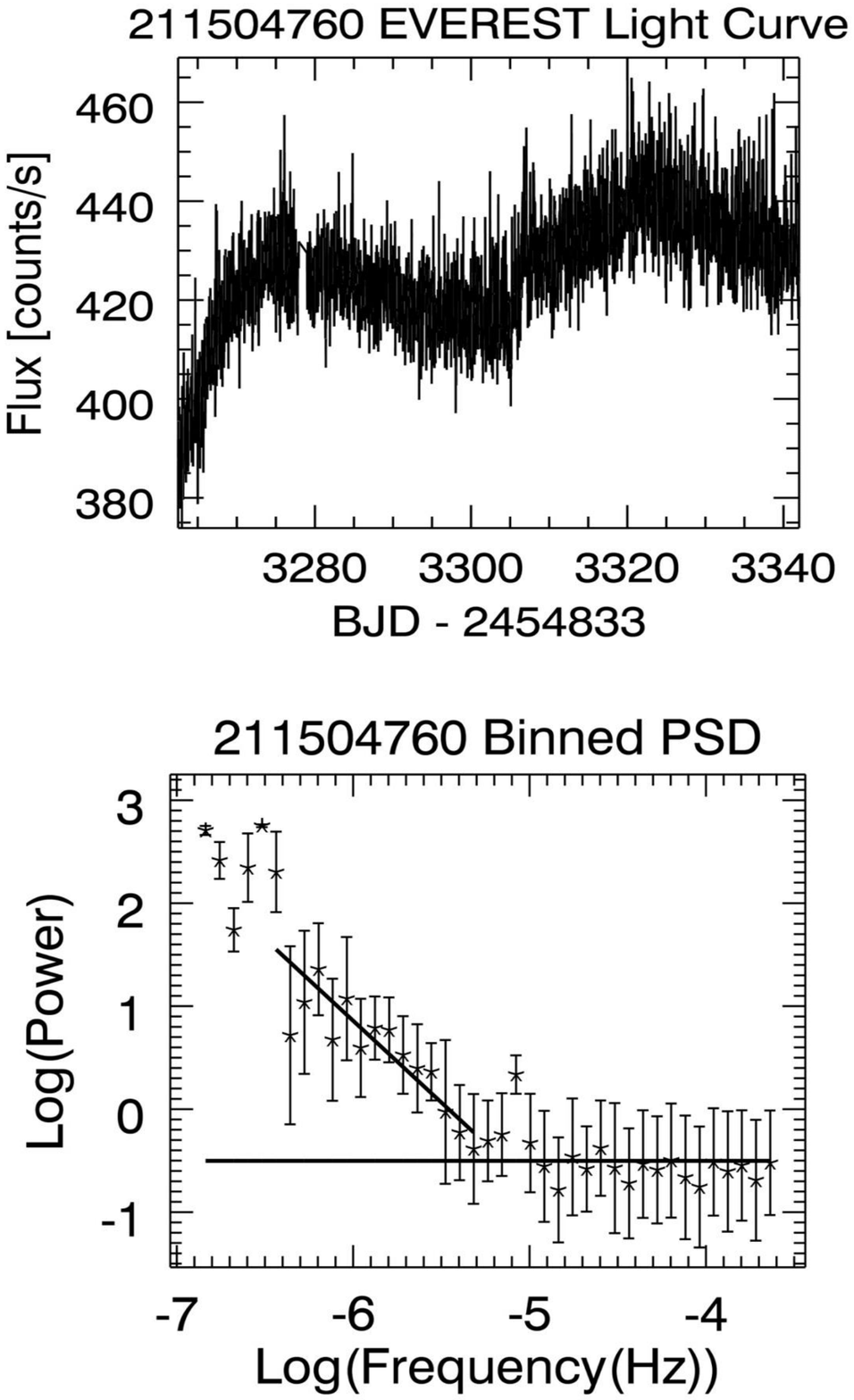}{0.35\textwidth}{(b)} 
          \fig{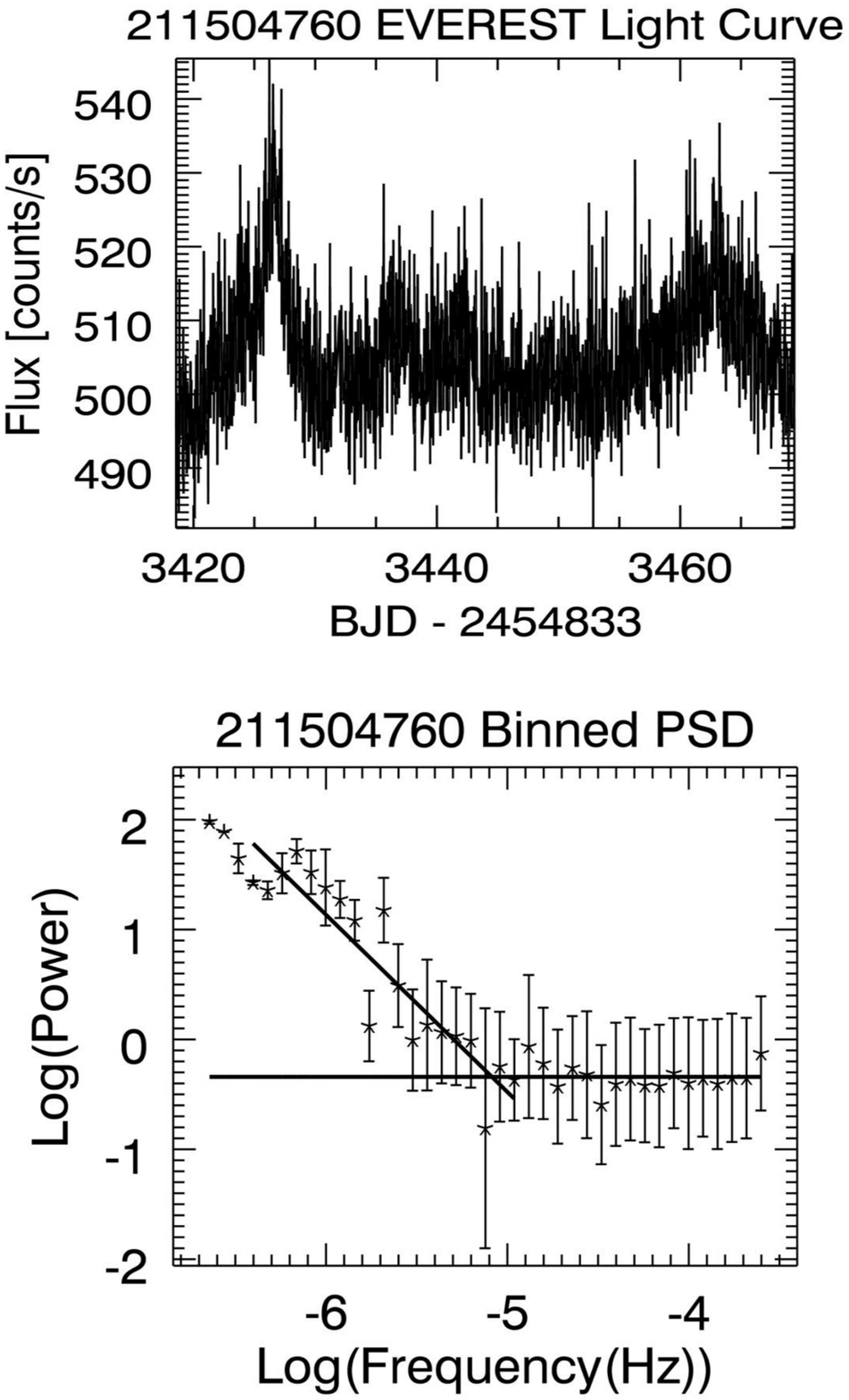}{0.35\textwidth}{(c)} 
          }        
\gridline{\fig{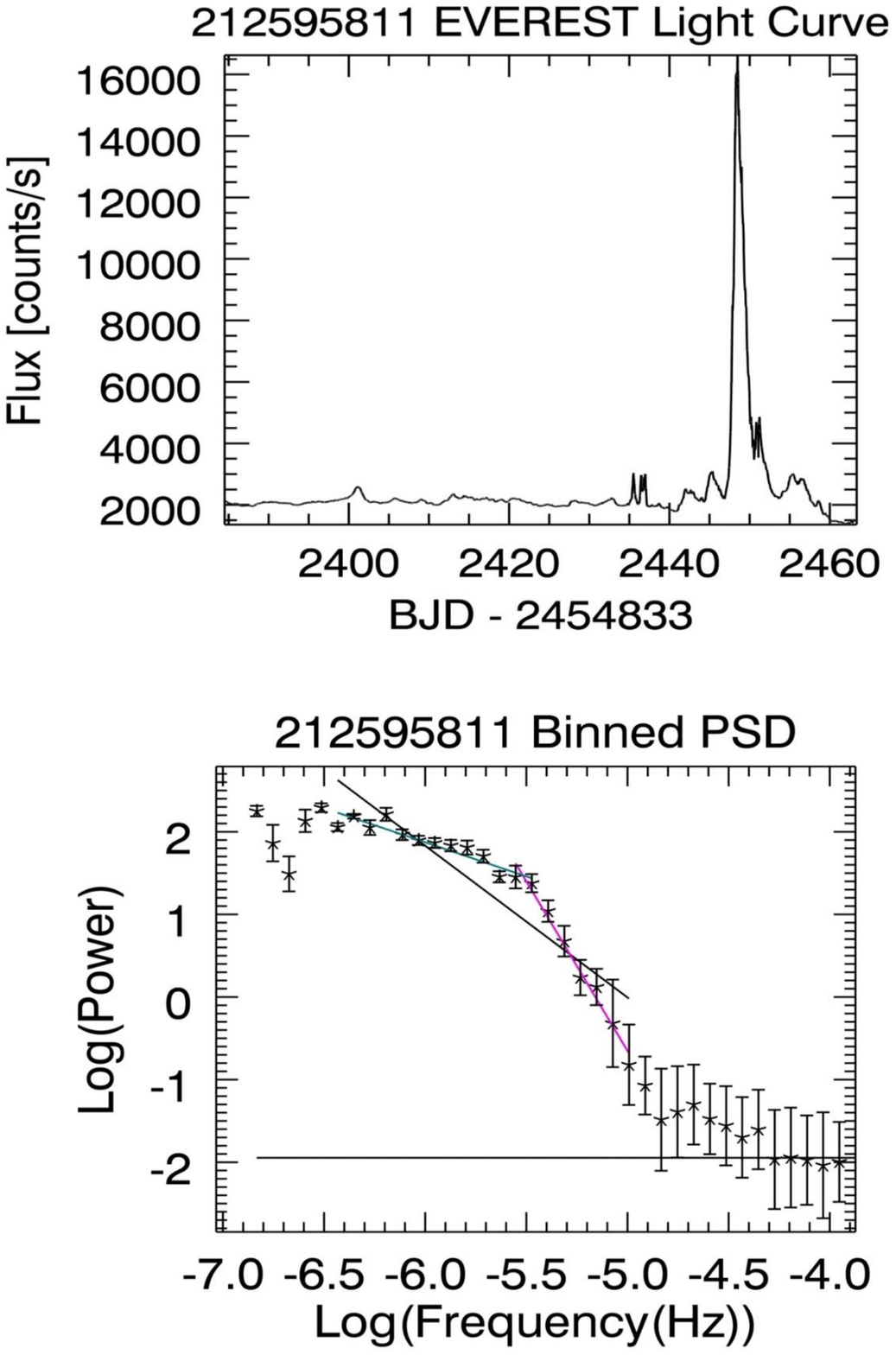}{0.35\textwidth}{(d)}
	\fig{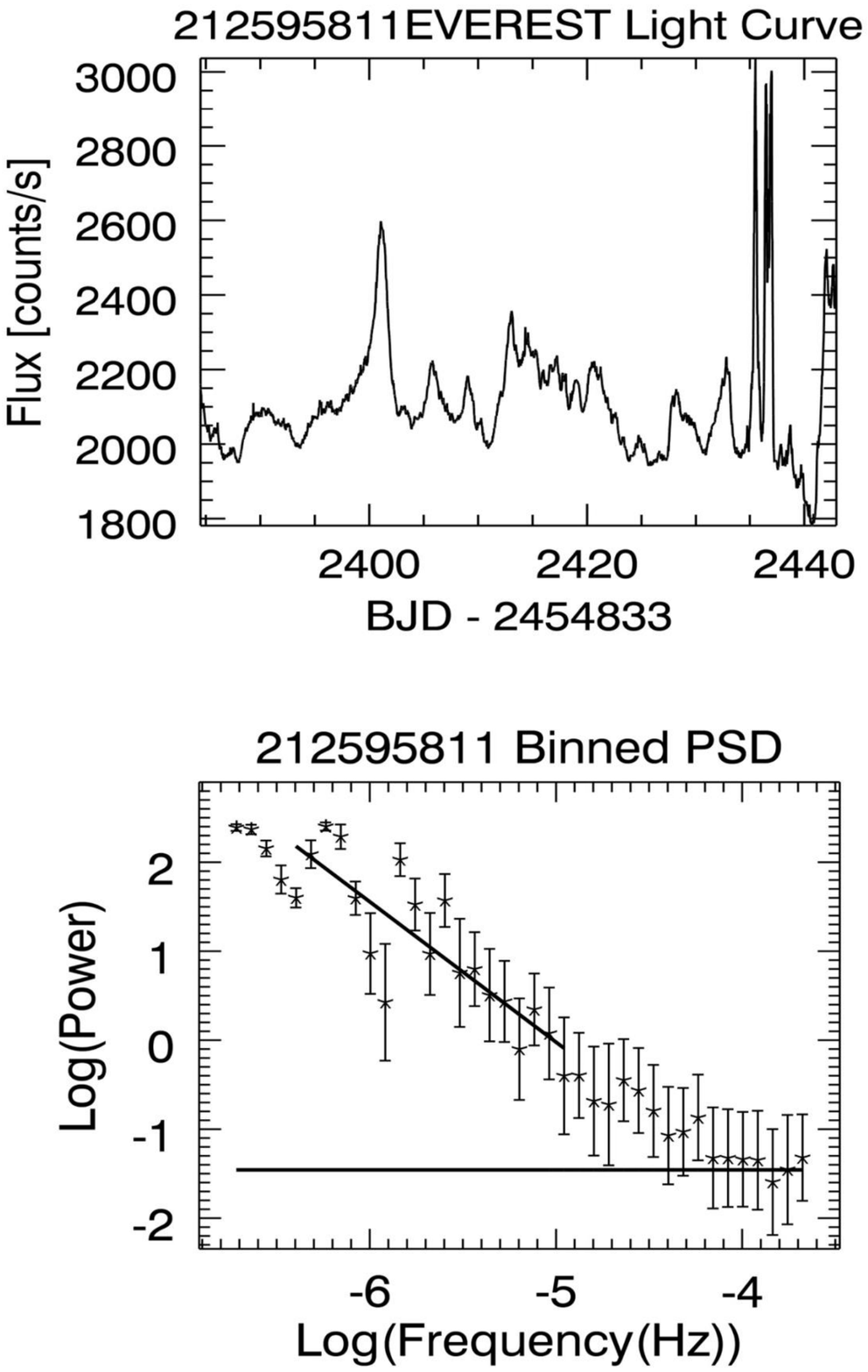}{0.35\textwidth}{(e)}
	\fig{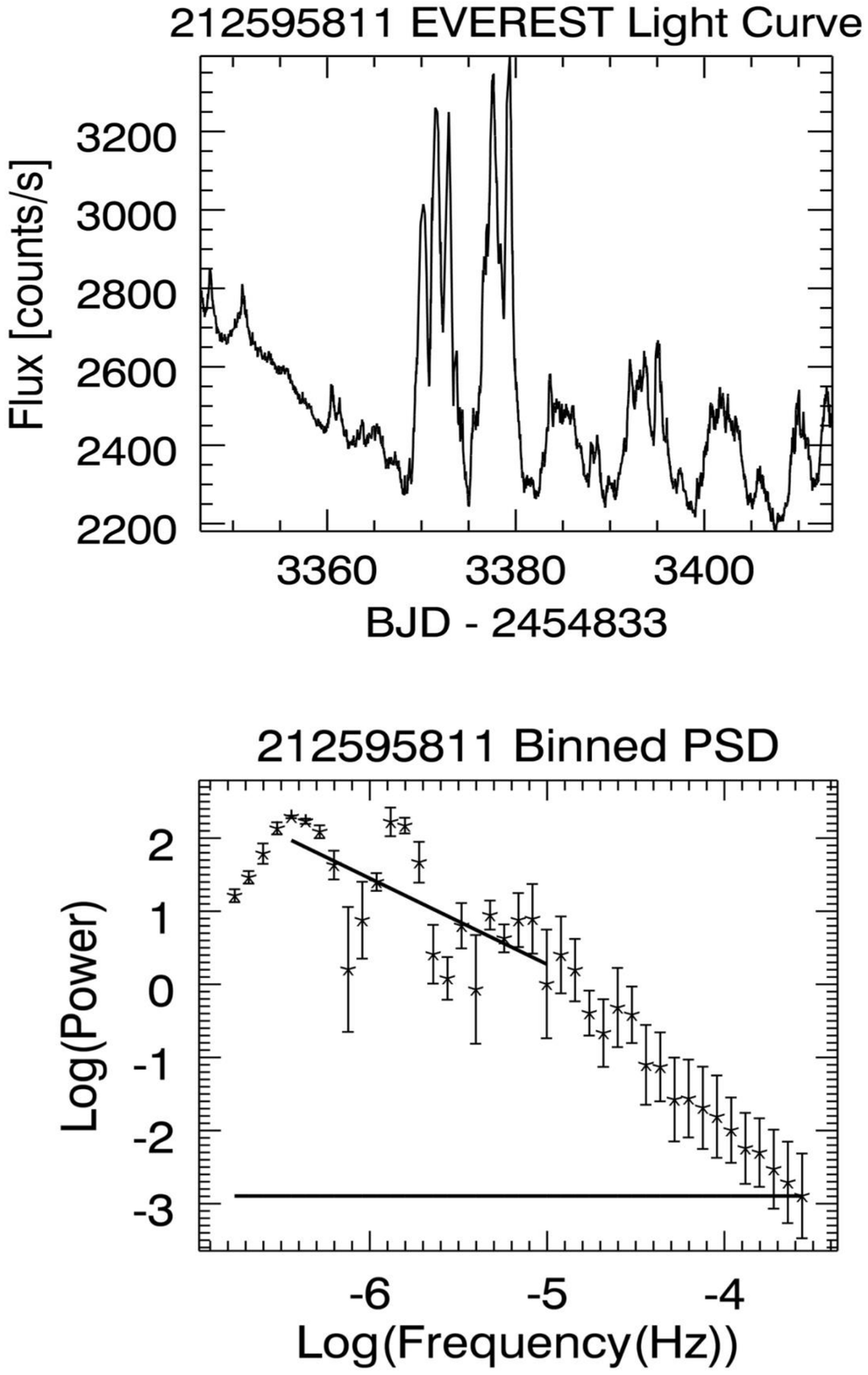}{0.35\textwidth}{(f)}
	}
\caption{Top: 3C~207 (EPIC 211504760), a lobe-dominated radio quasar, was observed in Campaigns 5, 16, and 18. Light curves and PSDs are shown for (a) Campaign 5,  (b)  Campaign 16, and (c) Campaign 18. Bottom: PKS 1352$-$104  (EPIC 212595811), the most variable object in our sample, was observed in Campaigns 6 and 17.   For the observations in Campaign 6, we show the light curves and PSDs: (d) with the giant flare and (e) without the giant flare; (f) Campaign 17 light curve and PSD. The Campaign 6 data with the giant flare (d) were fitted in the ``sweet spot'' between log $\nu = -5.0$ and $-6.4$ (shown in black), and then above and below the break in the power law at log $\nu =-5.5$ (shown in magenta and teal, respectively).}
\end{figure*}

\begin{longrotatetable}
\begin{deluxetable*}{lrrrrrrrrrrrrrl}
\tablecaption{ Optical Light Curve Characteristics For Combined Sample}
\tablewidth{0pt}
\singlespace
\tablecolumns{15}
\tabletypesize{\footnotesize}

\tablehead{
             \colhead{Name}
             &\colhead{EPIC}
             &\colhead{Class}
             &\colhead{C\tablenotemark{a}}
             & \colhead{Average}
             & \colhead{SD\tablenotemark{b}}
            & \colhead{Noise\tablenotemark{c}}
             & \colhead{Max/Min\tablenotemark{d}}
             & \colhead{CV\tablenotemark{e}}
             &\colhead{Slope\tablenotemark{f}} 
             & \colhead{Error}
              & \colhead{PSRESP}
             & \colhead{PSRESP}
             &\colhead{Con- } 
             &\colhead{Notes} 
             \\
              \colhead{}
             & \colhead{ID}
             & \colhead{}
             & \colhead{}
 	    & \colhead{Counts}
	     & \colhead{}
	     & \colhead{}
             & \colhead{Smooth or}
             &\colhead{} 
            & \colhead{}
            & \colhead{}
              & \colhead{Slope}
             & \colhead{Error}
             &\colhead{fidence\tablenotemark{g} } 
             &\colhead{} 
            \\
                         \colhead{}
               & \colhead{}
             & \colhead{}
             & \colhead{}
              & \colhead{(ct s$^{-1}$)}
              & \colhead{(ct s$^{-1}$)}
             & \colhead{(ct s$^{-1}$)}
              & \colhead{Jagged}
             & \colhead{}
              & \colhead{}
             &\colhead{} 
             &\colhead{} 
             & \colhead{}
              & \colhead{}{\%}
             & \colhead{}
            \\
                          \colhead{(1)}
               & \colhead{(2)}
             & \colhead{(3)}
             & \colhead{(4)}
              & \colhead{(5)}
              & \colhead{(6)}
             & \colhead{(7)}
              & \colhead{(8)}
             & \colhead{(9)}
              & \colhead{(10)}
             &\colhead{(11)} 
             &\colhead{(12)} 
             & \colhead{(13)}
              & \colhead{(14)}
             & \colhead{(15)}
            \\         
             }
\startdata
PKS B1130+008	&	201503438	&	BL Lac	&	1	&	875	&	17	&	4	&	1.10 J &	51.5	&	-2.47	&	0.14	&	-1.92	&	0.15	&	66	&	\nodata	\\
WB J0905+1358	&	211559047	&	BL Lac	&	5	&	3553	&	410	&	9	&	1.44	S &	8.7	&	-1.76	&	0.13	&	-2.07	&	0.14	&	49	&	1	\\
3C 207	&	211504760	&	LDRQ	&	5	&	589	&	31	&	4	&	1.32	J &	19.0	&	-1.89	&	0.15	&	-2.03	&	0.11	&	95	&	\nodata	\\
RGB J0847+115	&	211394951	&	BL Lac	&	5	&	1941	&	225	&	5	&	1.42	J &	8.6	&	-2.20	&	0.26	&	-2.26	&	0.16	&	55	&	 1	\\
OJ 287	&	211991001	&	BL Lac	&	5	&	15744	&	2417	&	58	&	2.36	J &	6.5	&	-2.28	&	0.17	&	-1.92	&	0.09	&	59	&	2	\\
OJ 287 	&	211991001	&	BL Lac	&	5	&	16038	&	2432	&	54	&	2.31	J &	6.6	&	-2.65	&	0.05	&	\nodata	&	\nodata	&	\nodata	&	3	\\
BZB J0816+2051	&	212035840	&	BL Lac	&	5	&	852	&	102	&	5	&	1.57	J &	8.4	&	-2.69	&	0.26	&	-2.69	&	0.16	&	24	&	1, 4	\\
PKS B1329-049	&	229227170	&	FSRQ	&	6	&	310	&	7	&	3	&	1.14	J &	44.3	&	-2.01	&	0.46	&	-1.86	&	0.36	&	99	&	5 	\\
PKS 1335-127	&	212489625	&	FSRQ	&	6	&	1233	&	130	&	2	&	1.64	J &	9.5	&	-2.35	&	0.21	&	-1.50 	&	0.10	&	68	&	\nodata	\\
PKS 1352-104	&	212595811	&	FSRQ	&	6	&	2465	&	1735	&	4	&	12.24 J &	1.4	&	-1.86	&	0.19	&	-1.85	&	0.08	&	5	&	6 	\\
PKS 1352-104	&	212595811	&	FSRQ	&	6	&	2465	&	1735	&	4	&	12.24 J &	1.4	&	-4.40	&	0.22	&	\nodata	&	\nodata	&	\nodata	&	7	\\
PKS 1352-104	&	212595811	&	FSRQ	&	6	&	2465	&	1735	&	4	&	12.24 J &	1.4	&	-0.83	&	0.09	&	\nodata	&	\nodata	&	\nodata	&	8	\\
PKS 1352-104	&	212595811	&	FSRQ	&	6	&	2106	&	148	&	7	&	1.71	J &	14.2	&	-1.58	&	0.24	&	-1.41	&	0.09	&	66	&	9	\\
RBS 1273	&	212800574	&	X-ray QSO	&	6	&	11738	&	98	&	5	&	1.02	S &	119.8	&	-1.38	&	0.14	&	-2.05	&	0.12	&	48	&	\nodata	\\
PKS B1908-201	&	217700467	&	BL Lac	&	7	&	965	&	10	&	4	&	1.08	J &	96.5	&	-1.63	&	0.22	&	-1.45	&	0.13	&	88	&	\nodata	\\
1H 1914-194	&	218129423	&	BL Lac	&	7	&	15426	&	332	&	20	&	1.11	J &	46.5	&	-2.20	&	0.22	&	-2.01	&	0.09	&	59	&	\nodata	\\
PKS B1921-293	&	229228355	&	BL Lac	&	7	&	2337	&	126	&	13	&	1.29	J &	18.5	&	-2.27	&	0.23	&	-1.81	&	0.10	&	55	&	10	\\
PKS 0047+023	&	220299433	&	BL Lac	&	8	&	661	&	125	&	6	&	3.15	J &	5.3	&	-1.45	&	0.26	&	-1.57	&	0.08	&	11	&	\nodata	\\
1RXS J120417.0-070959	&	201079736	&	BL Lac	&	10	&	5965	&	99	&	5	&	1.07	S &	60.3	&	-2.30	&	0.14	&	-2.13	&	0.14	&	62	&	11	\\
PKS 1216-010	&	201375481	&	BL Lac	&	10	&	5729	&	791	&	16	&	1.77 J	&	7.2	&	-2.50	&	0.29	&	-2.25	&	0.13	&	41	&	11	\\
3C 273	&	229151988	&	FSRQ	&	10	&	91737	&	353	&	10	&	1.01	J&	259.9	&	-2.19	&	0.19	&	-2.66	&	0.16	&	78	&	2, 11 	\\
3C 273	&	229151988	&	FSRQ	&	10	&	91751	&	403	&	52	&	1.02	S&	227.7	&	-2.31	&	0.15	&	\nodata	&	\nodata	&	\nodata	& 3	\\
1RXS J121946.0-031419	&	201247917	&	BL Lac	&	10	&	1554	&	87	&	4	&	1.29 J &	17.9	&	-2.43	&	0.23	&	-2.17	&	0.11	&	47	&	11 	\\
PKS B2320-035	&	246289180	&	FSRQ	&	12	&	2324	&	383	&	5	&	2.33	J &	6.1	&	-2.09	&	0.18	&	-1.93	&	0.10	&	23	&	\nodata	\\
PKS B2335-027	&	246327456	&	FSRQ	&	12	&	513	&	44	&	3	&	1.43	J &	11.7	&	-2.85	&	0.22	&	-2.18	&	0.12	&	49	&	4	\\
NVSS J110735+022225	&	201621388	&	BL Lac	&	14	&	1116	&	26	&	7	&	1.11	S &	42.9	&	-2.95	&	0.35	&	-2.58	&	0.32	&	88	&	1, 12 \\
4C +06.41	&	248611911	&	FSRQ	&	14	&	2910	&	109	&	6	&	1.18	J &	26.7	&	-2.17	&	0.13	&	-1.93	&	0.11	&	99	&	1	\\
TXS 1013+054	&	251457104	&	FSRQ	&	14	&	244	&	88	&	5	&	8.50	J &	2.8	&	-1.88	&	0.17	&	-2.02	&	0.11	&	90	&	\nodata	\\
3C 207	&	211504760	&	LDRQ	&	16	&	425	&	12	&	5	&	1.25	J &	35.4	&	-1.59	&	0.26	&	-2.06	&	0.12	&	67	&	13   	\\
TXS 0836+182	&	211852059	&	BL Lac	&	16	&	2066	&	102	&	3	&	1.26	S &	20.3	&	-2.38	&	0.17	&	-2.11	&	0.10	&	48	&	\nodata	\\
NVSS J090226+205045	&	212035517	&	Unknown	&	16	&	4288	&	973	&	13	&	2.48	J &	4.4	&	-2.67	&	0.14	&	-2.35	&	0.12	&	60	&	\nodata	\\
TXS 0853+211	&	212042111	&	BL Lac	&	16	&	642	&	49	&	4	&	1.43	J &	13.1	&	-1.76	&	0.18	&	-1.90 	&	0.11	&	83	&	\nodata	\\
NVSS J090900+231112	&	251376444	&	BL Lac	&	16	&	979	&	51	&	6	&	1.24	J &	19.2	&	-2.57	&	0.32	&	-2.41	&	0.11	&	34	&	  1, 4	\\
PKS 1335-127	&	212489625	&	FSRQ	&	17	&	581	&	60	&	4	&	1.61	J &	9.7	&	-1.74	&	0.17	&	-1.72	&	0.35	&	88	&	\nodata	\\
PMN J1318-1235	&	212507036	& unknown	&	17	&	468	&	62	&	4	&	1.71	J &	7.5	&	-2.55	&	0.21	&	-2.01	&	0.12	&	65	&	\nodata	\\
PKS 1352-104	&	212595811	&	FSRQ	&	17	&	2509	&	250	&	7	&	1.86	J &	10.0	&	-1.18	&	0.38	&	-1.62	&	0.07	&	17	&	14	\\
RBS 1273	&	212800574	&	X-ray QSO	&	17	&	11728	&	24	&	4	&	1.01	S &	488.7	&	-4.20	&	0.44	&	-3.12	&	0.47	&	92	&	 5	\\
PKS B1329-049	&	229227170	&	FSRQ	&	17	&	570	&	37	&	3	&	1.29	J &	15.4	&	-1.93	&	0.23	&	-1.92	&	0.12	&	86	&	\nodata	\\
PKS B1310-041	&	251502828	&	FSRQ	&	17	&	1686	&	84	&	6	&	1.29	S &	20.1	&	-1.73	&	0.10	&	-1.86	&	0.12	&	72	&	\nodata	\\
RGB J0847+115 &	211394951	&	BL Lac	&	18	&	1234	&	46	&	6	&	1.17	J &	26.8	&	-2.02	&	0.17	&	-1.82	&	0.12	&	98	&	1	\\
3C 207	&	211504760	&	LDRQ	&	18	&	506	&	8	&	7	&	1.13	S &	63.3	&	-1.57	&	0.22	&	-1.38	&	0.18	&	100	&	 4	\\
WB J0905+1358	&	211559047	&	BL Lac	&	18	&	4067	&	305	&	10	&	1.28	S &	13.3	&	-2.04	&	0.13	&	-1.94	&	0.16	&	82	&	1	\\
OJ 287	&	211991001	&	BL Lac	&	18	&	13919	&	1029	&	8	&	1.36	J &	13.5	&	-1.96	&	0.20	&	-1.61	&	0.10	&	60	&	2	\\
OJ 287	&	211991001	&	BL Lac	&	18	&	14025	&	1267	&	28	&	1.52	J &	11.1	&	-2.26	&	0.06	&	\nodata	&	\nodata	&	60	&	3	\\
BZB J0816+2051	&	212035840	&	BL Lac	&	18	&	1085	&	79	&	5	&	1.34	J &	13.7	&	-2.57	&	0.17	&	-2.31	&	0.13	&	89	&	1	\\
\enddata
\tablecomments{ 1. Custom aperture. 2. Long cadence. 3. Short cadence data fitted between log $\nu = -3.4$ and log $\nu = 6.4$. 4. Slope fitted between log $\nu = -5.2$ and $-6.4$. 5. Slope fitted between log $\nu = -5.6$ and $-6.4$. 6. Slope fit in the sweet spot of log $\nu =-5.0$ to $ -6.4$; however, there is a break in the power law around  log $\nu = -5.5$. 7. Fit above the break at log $\nu = -5.5$. 8. Fit below the break at log $\nu = -5.5$. 9. Without large flare. 10. PKS B1921-293 may be intermediate between BL Lac and FSRQ because its weak emission lines vary in equivalent width when the continuum emission changes (Wills and Wills 1981). 11. Slope fitted between log $\nu = -5.0$ and $-6.2$. 12. Slope fitted between log $\nu = -5.5$ and $-6.4$. 13. Slope fitted between log $\nu = -5.3$ and $-6.4$. 14. PSRESP used only one power law because no white noise was present.}
\tablenotetext{a}{Campaign}
\tablenotetext{b}{Standard deviation during full campaign.}
\tablenotetext{c}{Standard deviation during 0.2--0.5 day intervals when source variation was less than $\sim1\%$.}
\tablenotetext{d}{(Maximum count rate)/(Minimum count rate). Smooth or jagged character of variations is indicated with S or J.}
\tablenotetext{e}{Coefficient of Variation, defined as (average count rate)/(standard deviation during full campaign).}
\tablenotetext{f}{Slope fitted in the sweet spot between log $\nu = -5.0$ and $-6.4$ unless otherwise noted.}
\tablenotetext{g}{The confidence of the adopted model  (power law plus white noise) fit using PSRESP.} 
\end{deluxetable*}
\end{longrotatetable}

\begin{figure*}
\figurenum{3}
\gridline{\fig{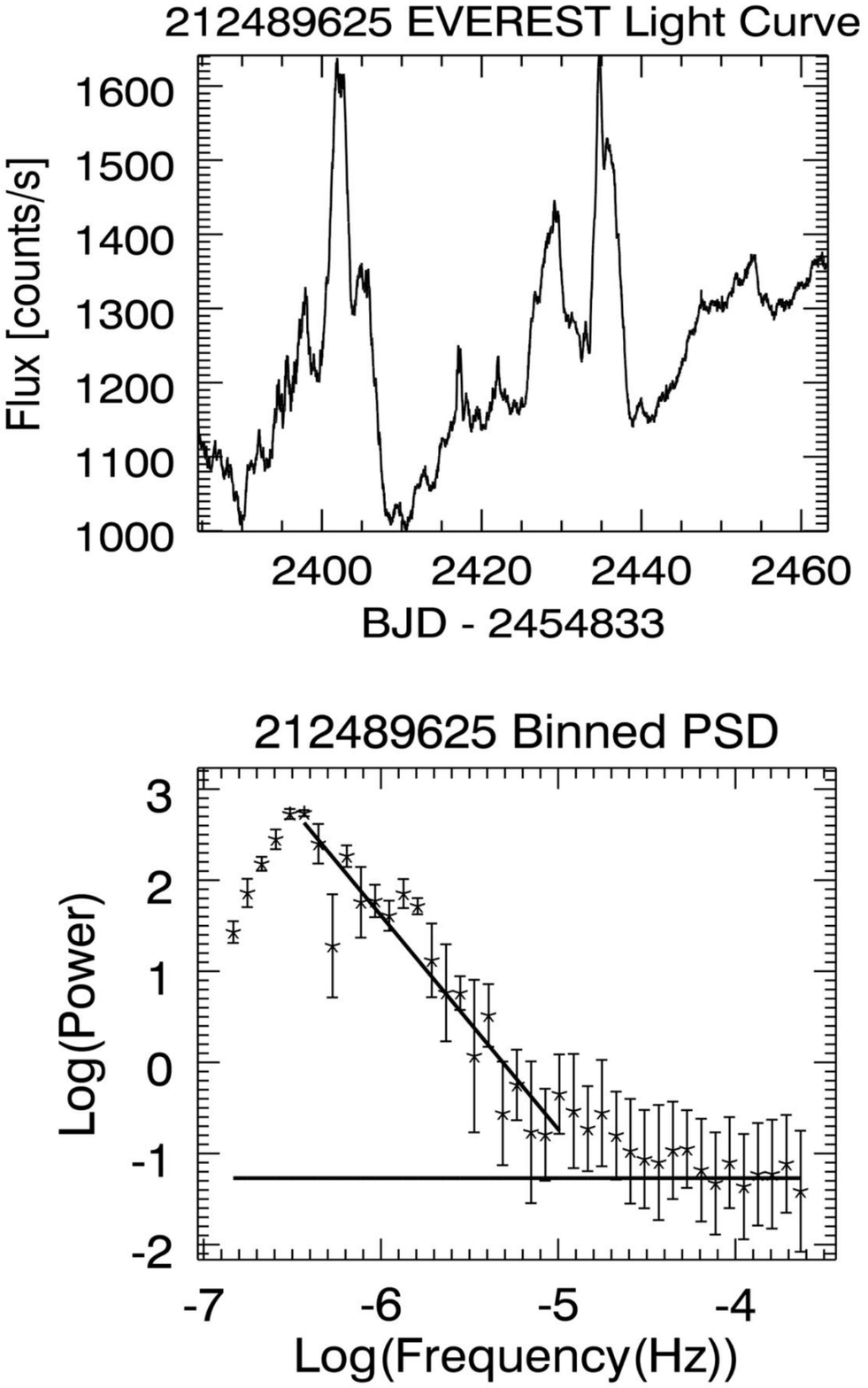}{0.47\textwidth}{(a)}
          \fig{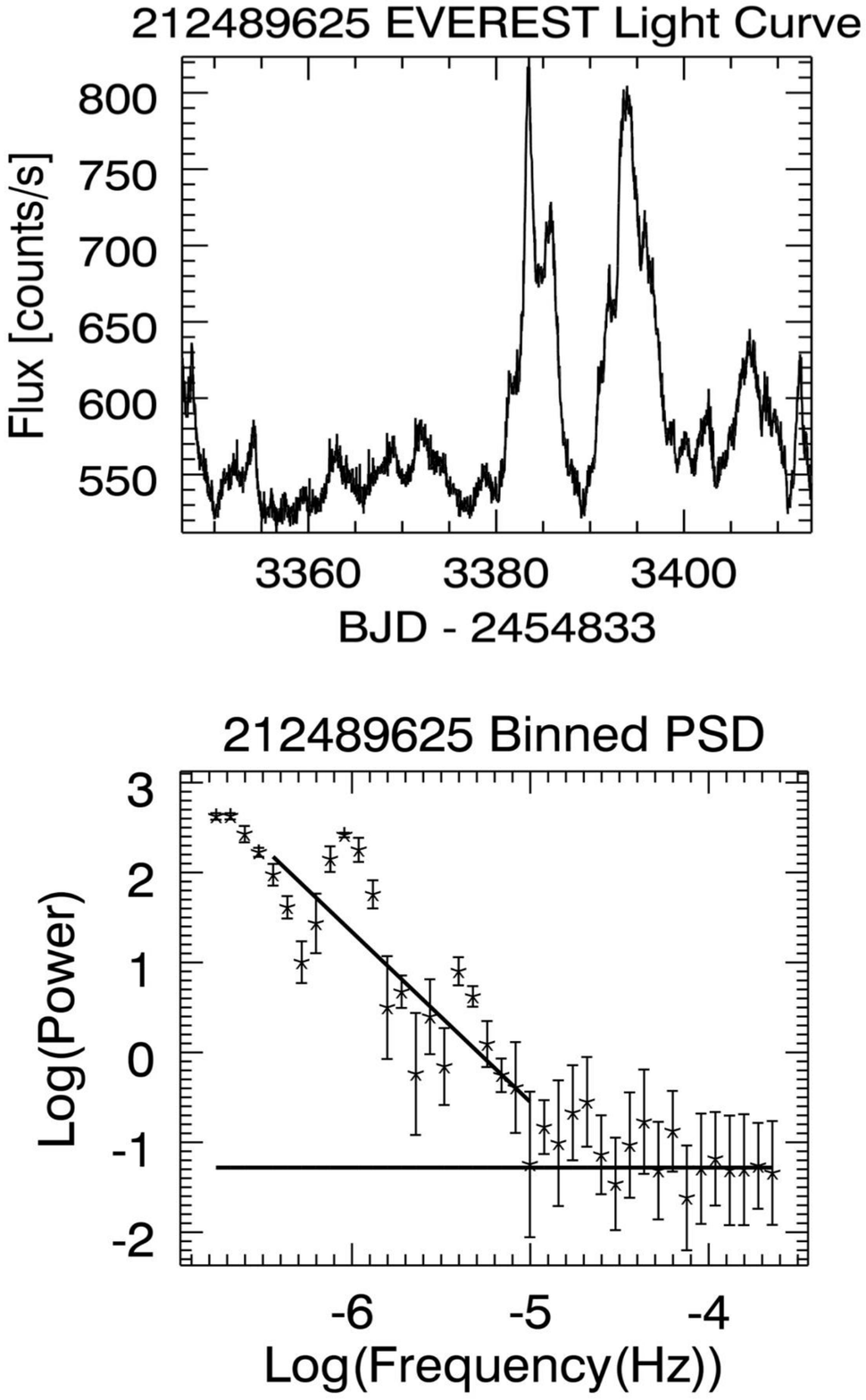}{0.47\textwidth}{(b)}
          }
\gridline{\fig{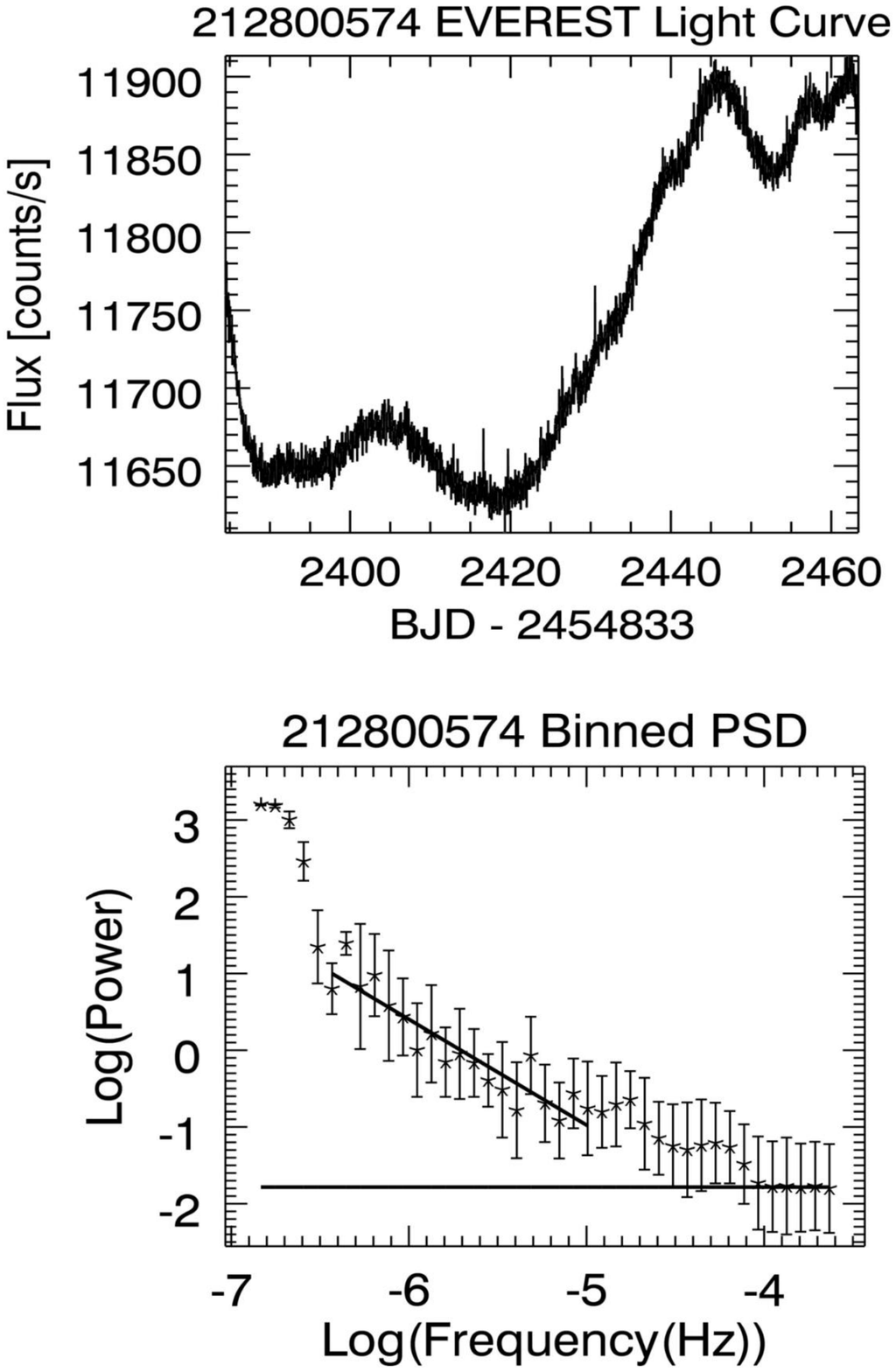}{0.47\textwidth}{(c)}
          \fig{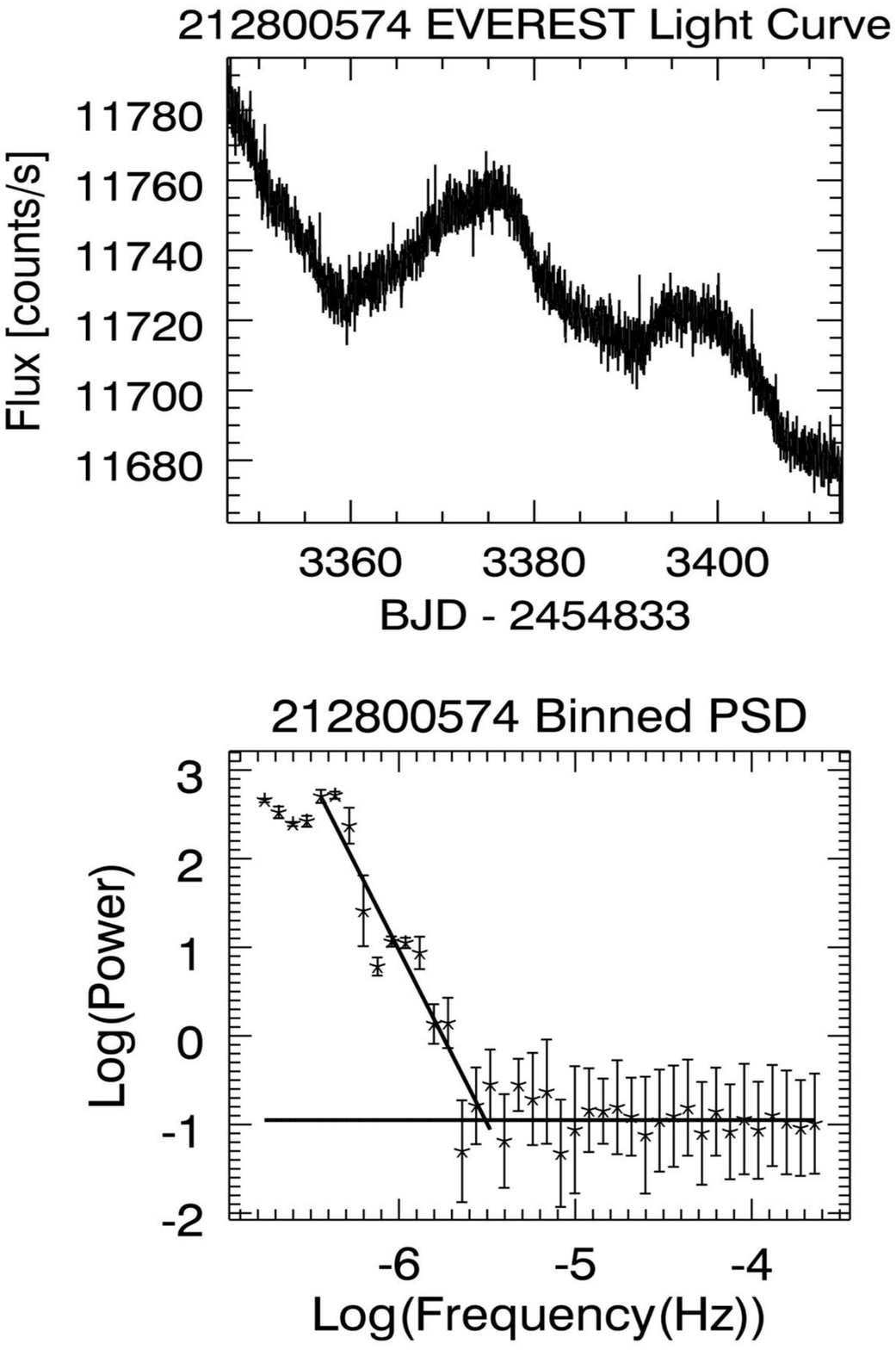}{0.47\textwidth}{(d)}
          }
\caption{Light curves and PSDs for EPIC 212489625 (PKS 1335$-$127)  in (a) Campaign 6 and  (b) Campaign 17; for EPIC 212800574 (RBS 1273) in (c) Campaign 6 and (d) Campaign 17.}.
\end{figure*}

\begin{figure*}
\figurenum{3, continued}
\gridline{\fig{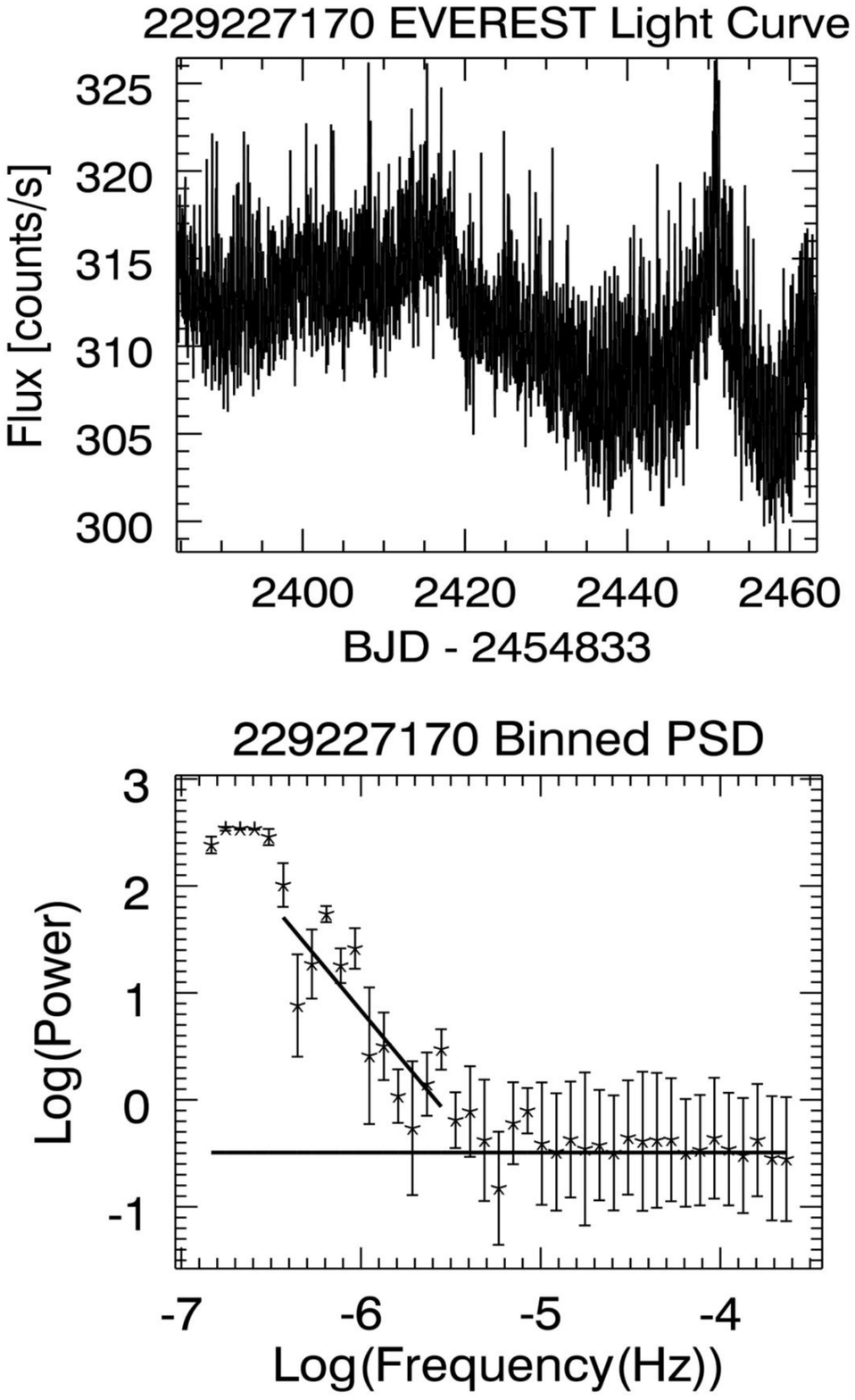}{0.47\textwidth}{(e)}
	\fig{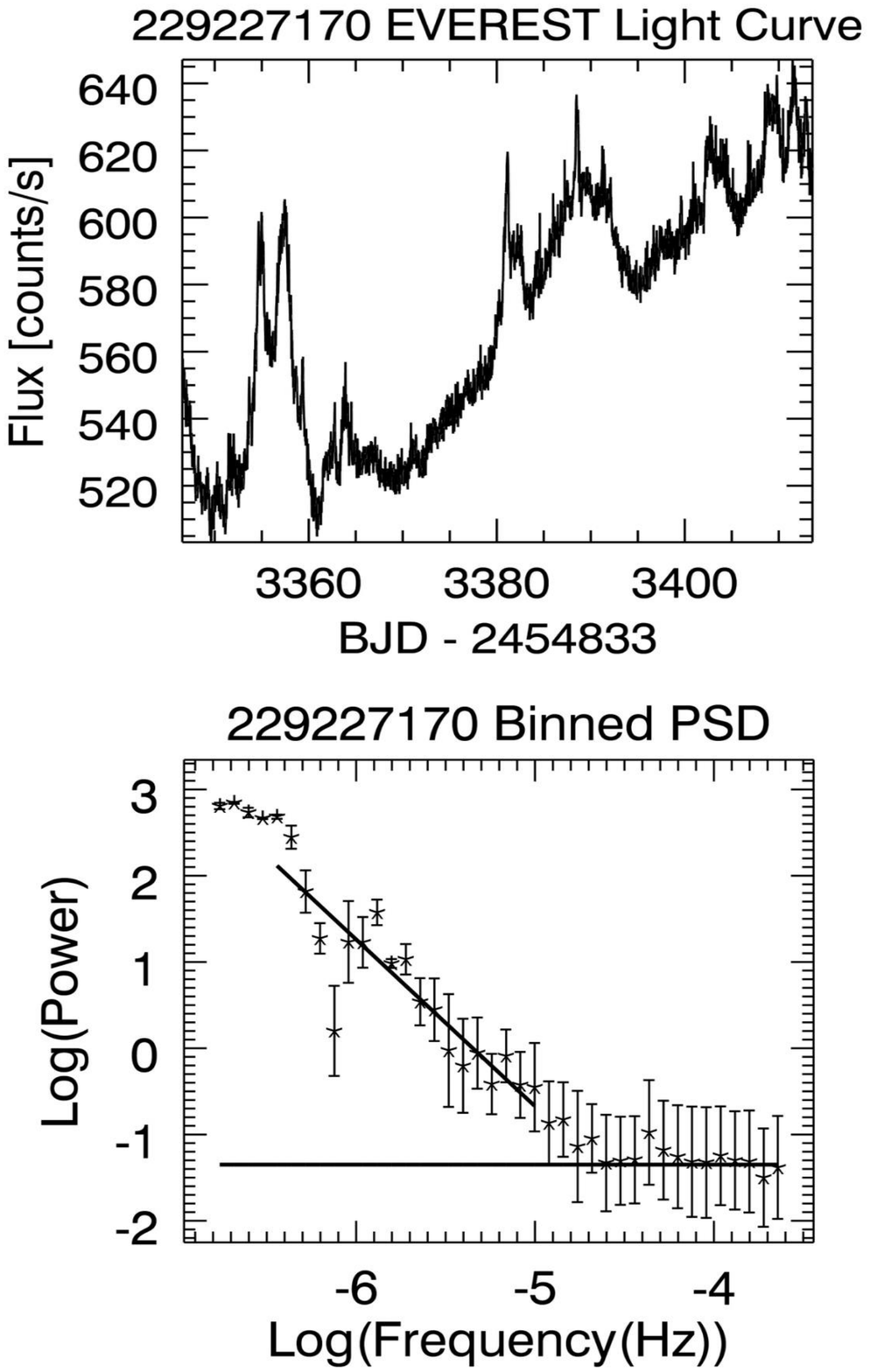}{0.47\textwidth}{(f)}
	}
	\gridline{\fig{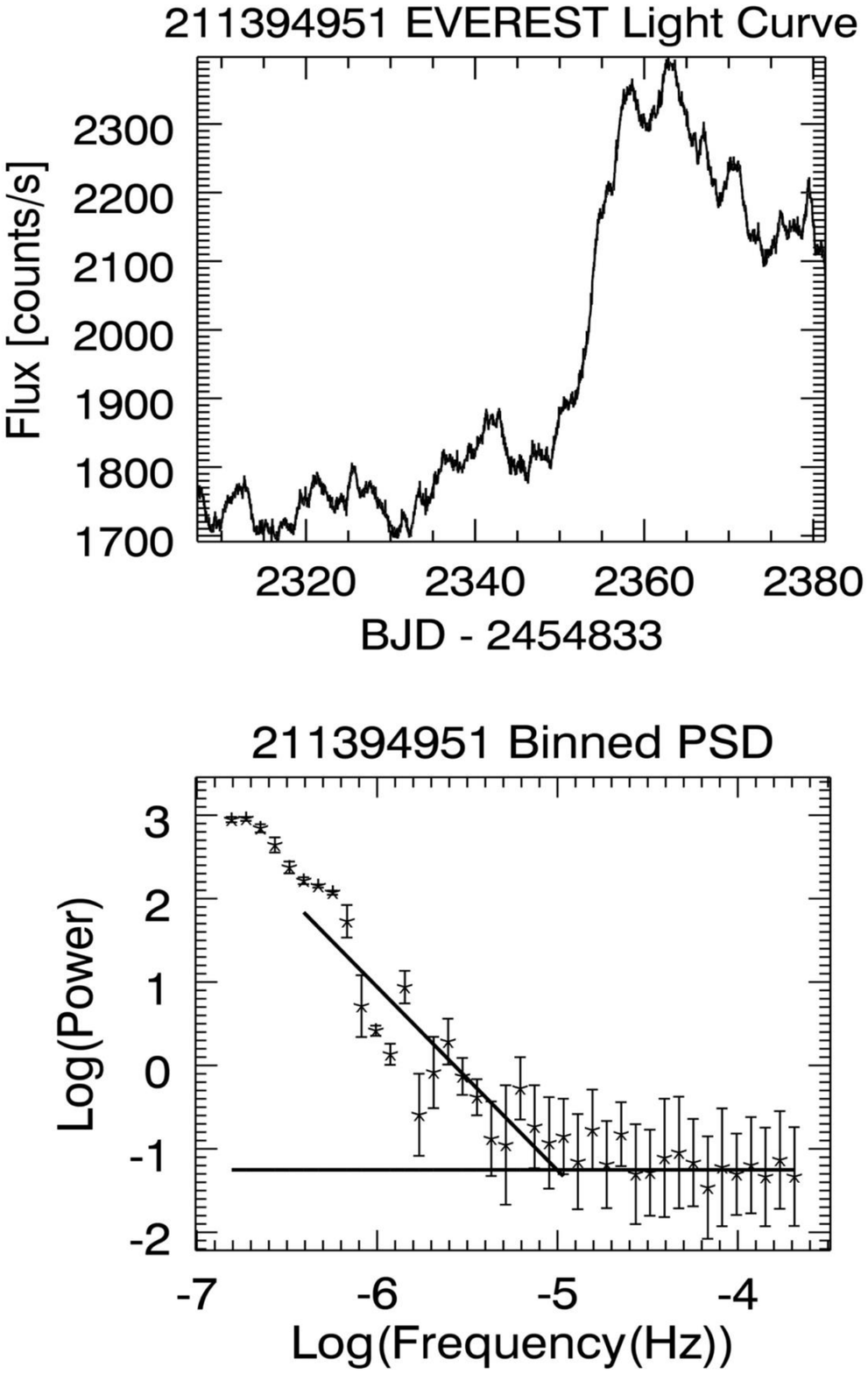}{0.47\textwidth}{(g)}
          \fig{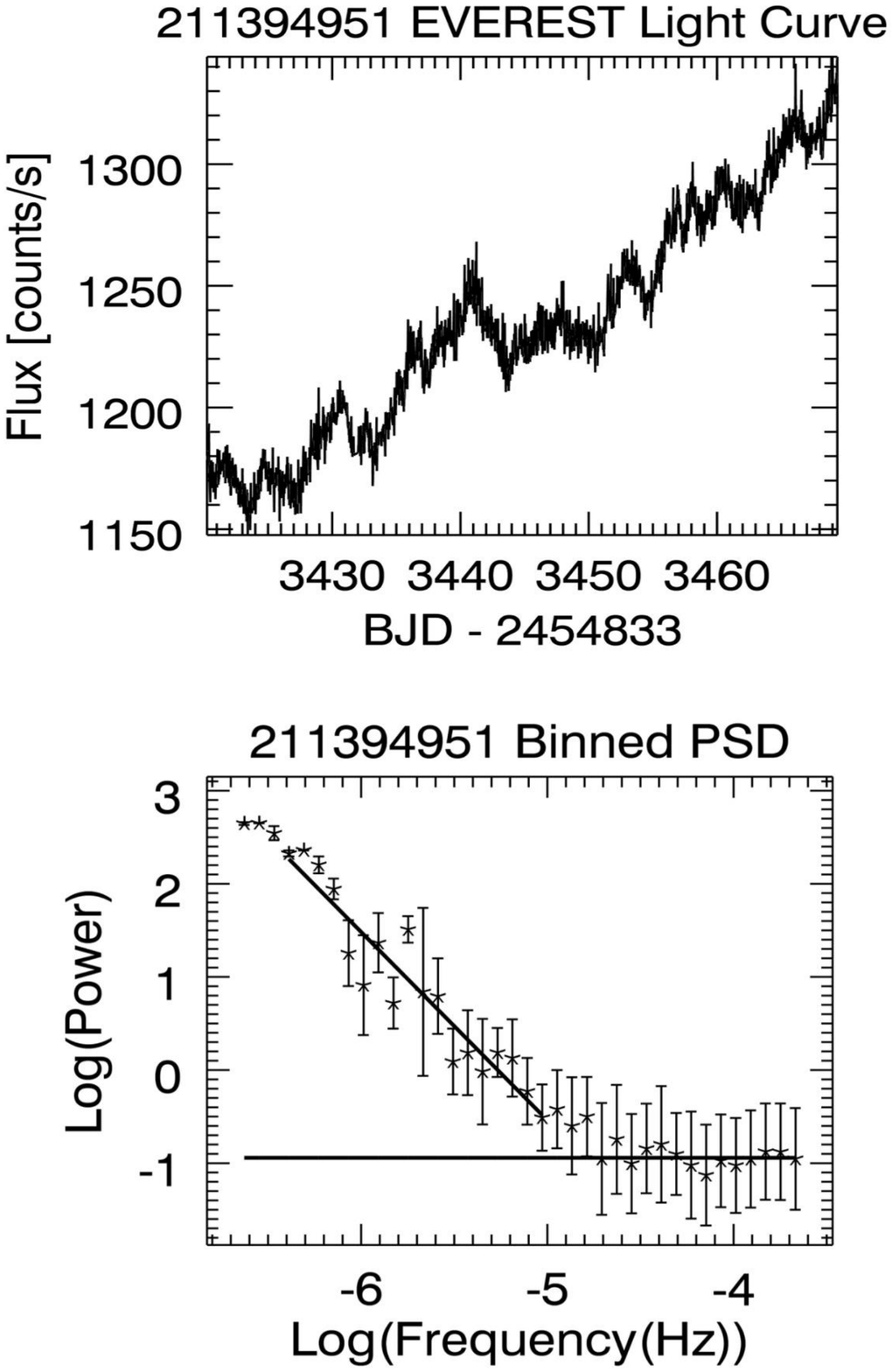}{0.47\textwidth}{(h)}
          }
\caption{EPIC 229227170 (PKS B1329$-$049) in (e) Campaign 6  and  (f) Campaign 17; EPIC 211394951 (RGB J0847+115) in  (g) Campaign 5 and (h) Campaign 18. }
\end{figure*}
\begin{figure*}
\figurenum{3, continued}
\gridline{\fig{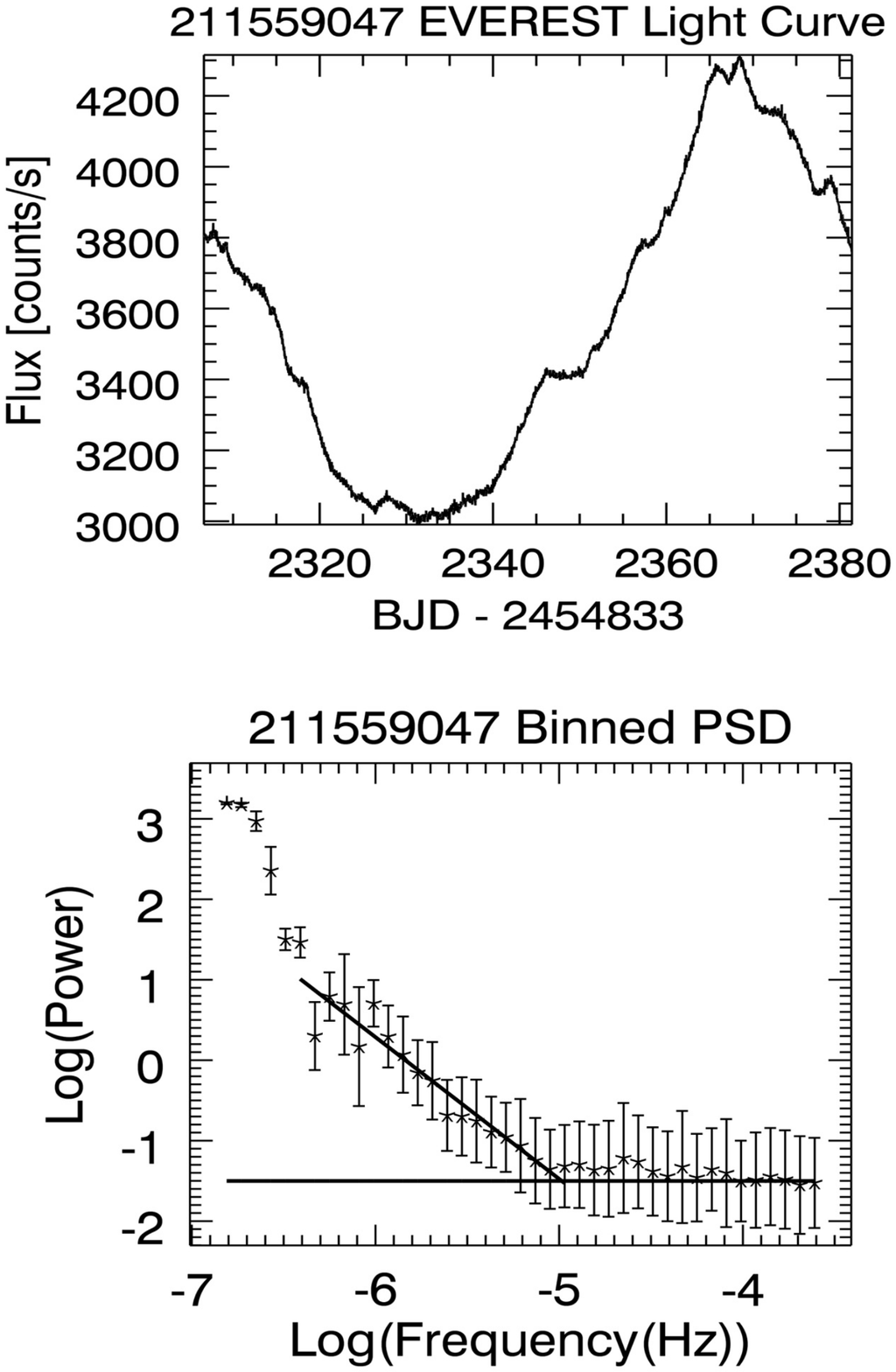}{0.47\textwidth}{(i)}
	\fig{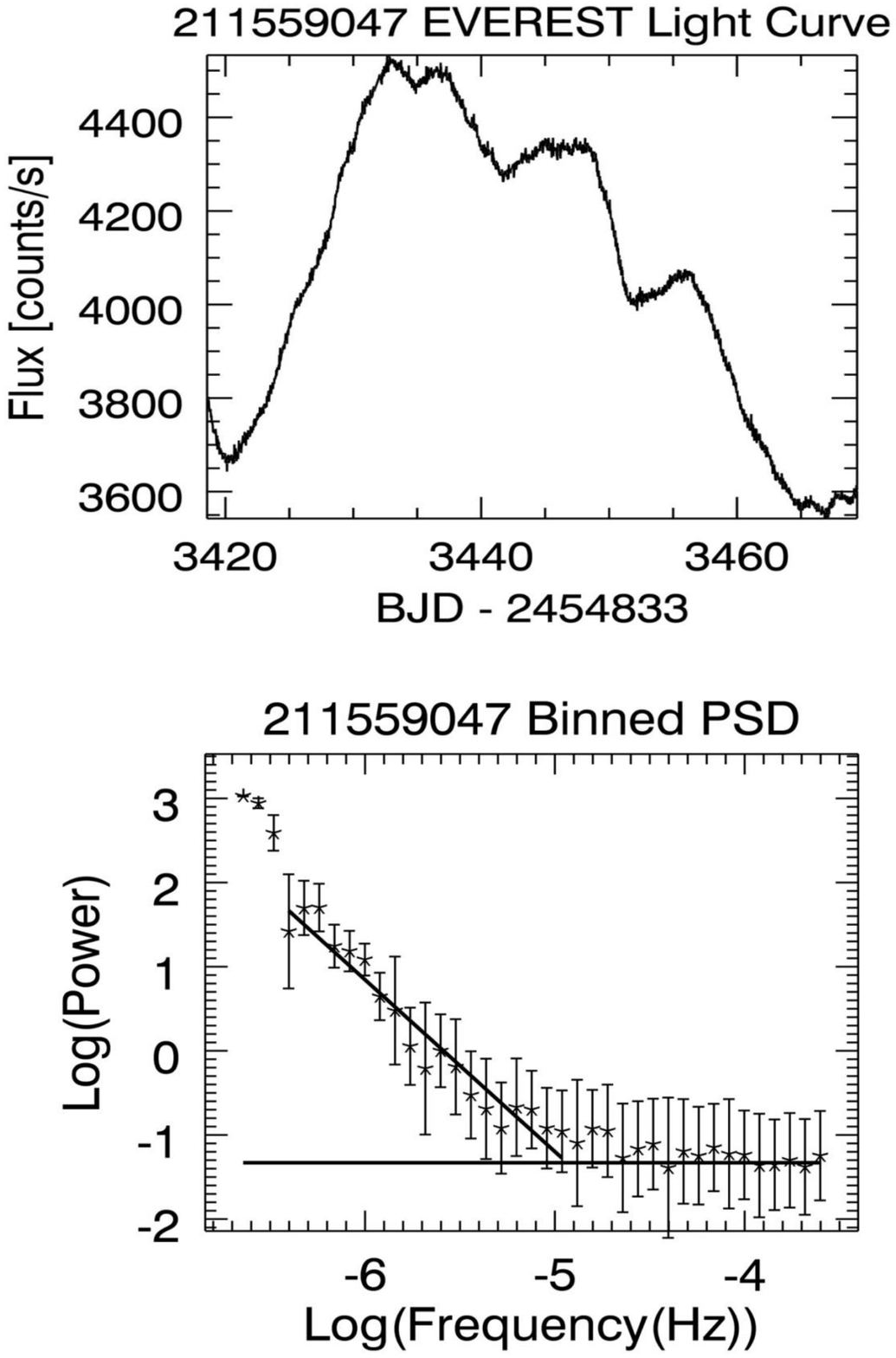}{0.47\textwidth}{(j)}
	}
	\gridline{\fig{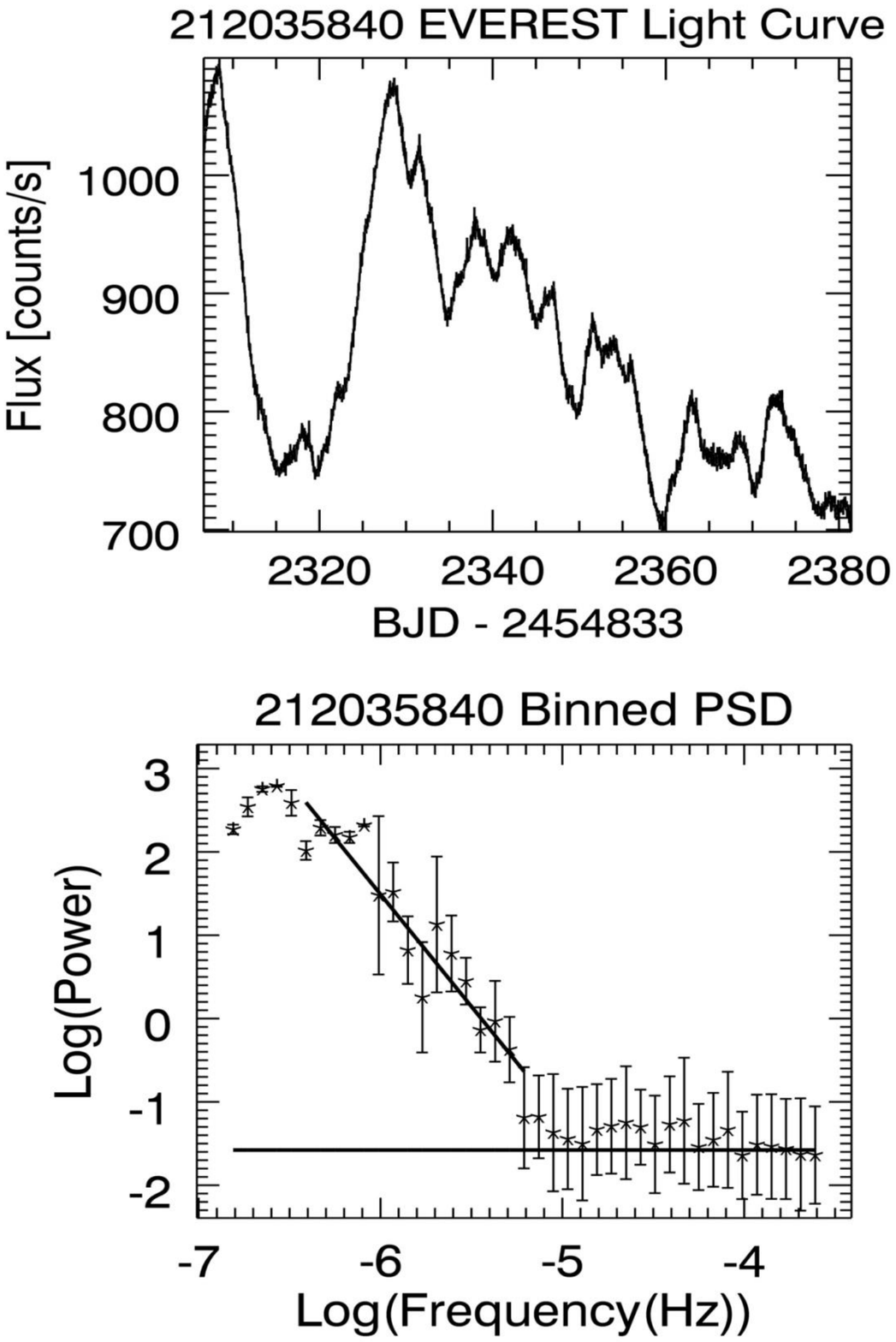}{0.47\textwidth}{(k)}
	\fig{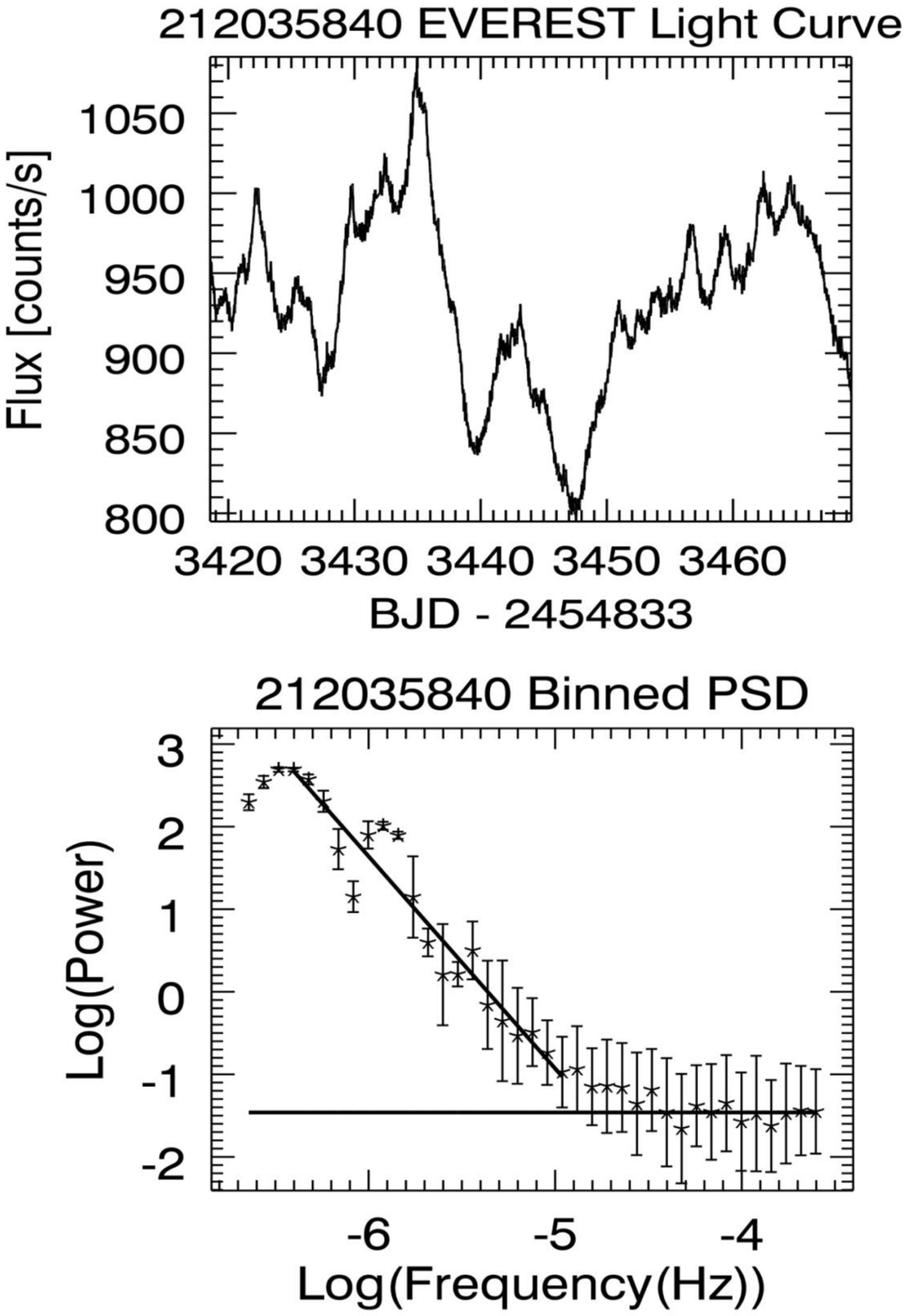}{0.47\textwidth}{(l)}
	}
\caption{EPIC 211559047 (WB J0905+1358) in (i) Campaign 5  and  (j) Campaign 18; EPIC 212035840 (BZB J0816+2051) in (k) Campaign 5  and  (l) Campaign 18. }
\end{figure*}

\begin{figure*}
\figurenum{4}
\gridline{\fig{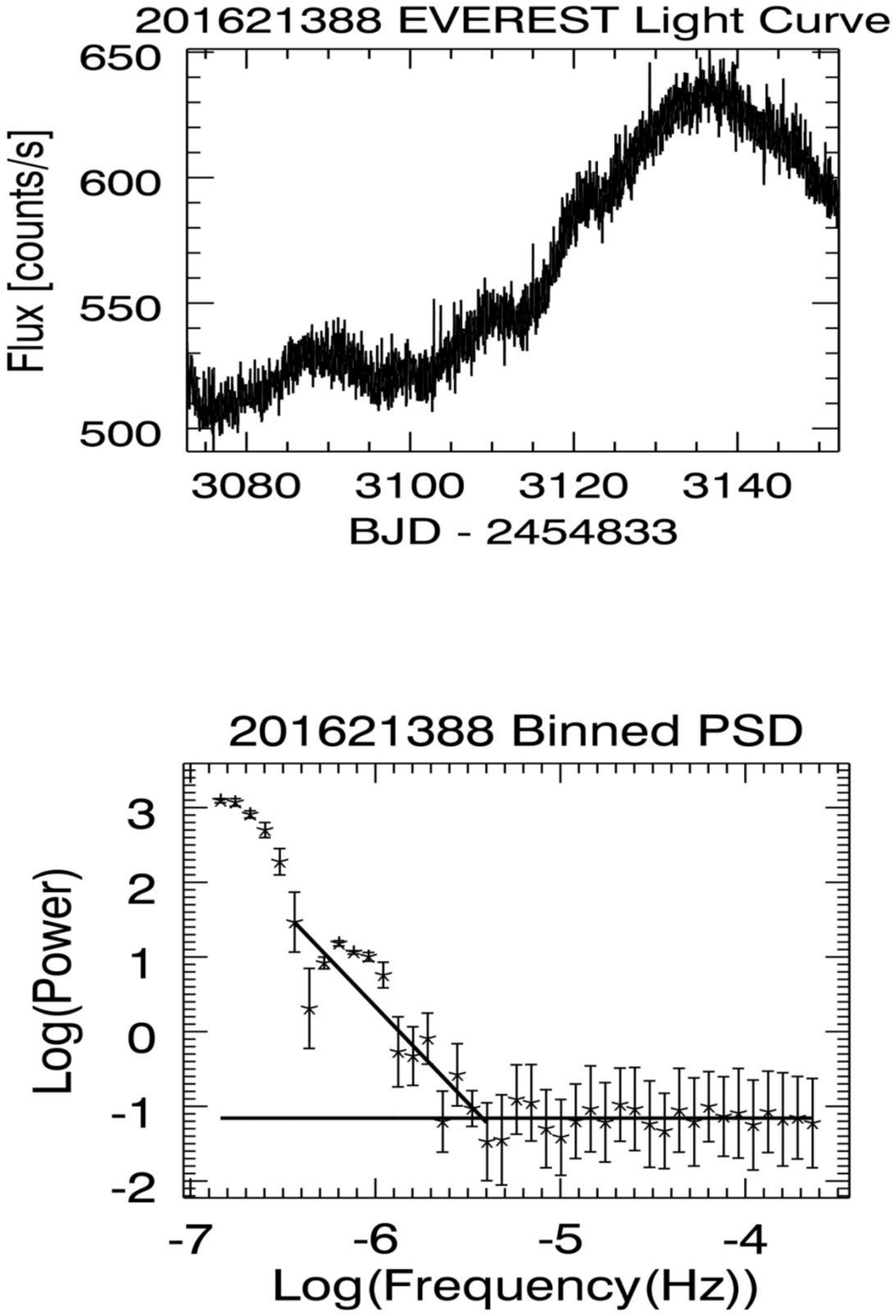}{0.47\textwidth}{(a)}
          \fig{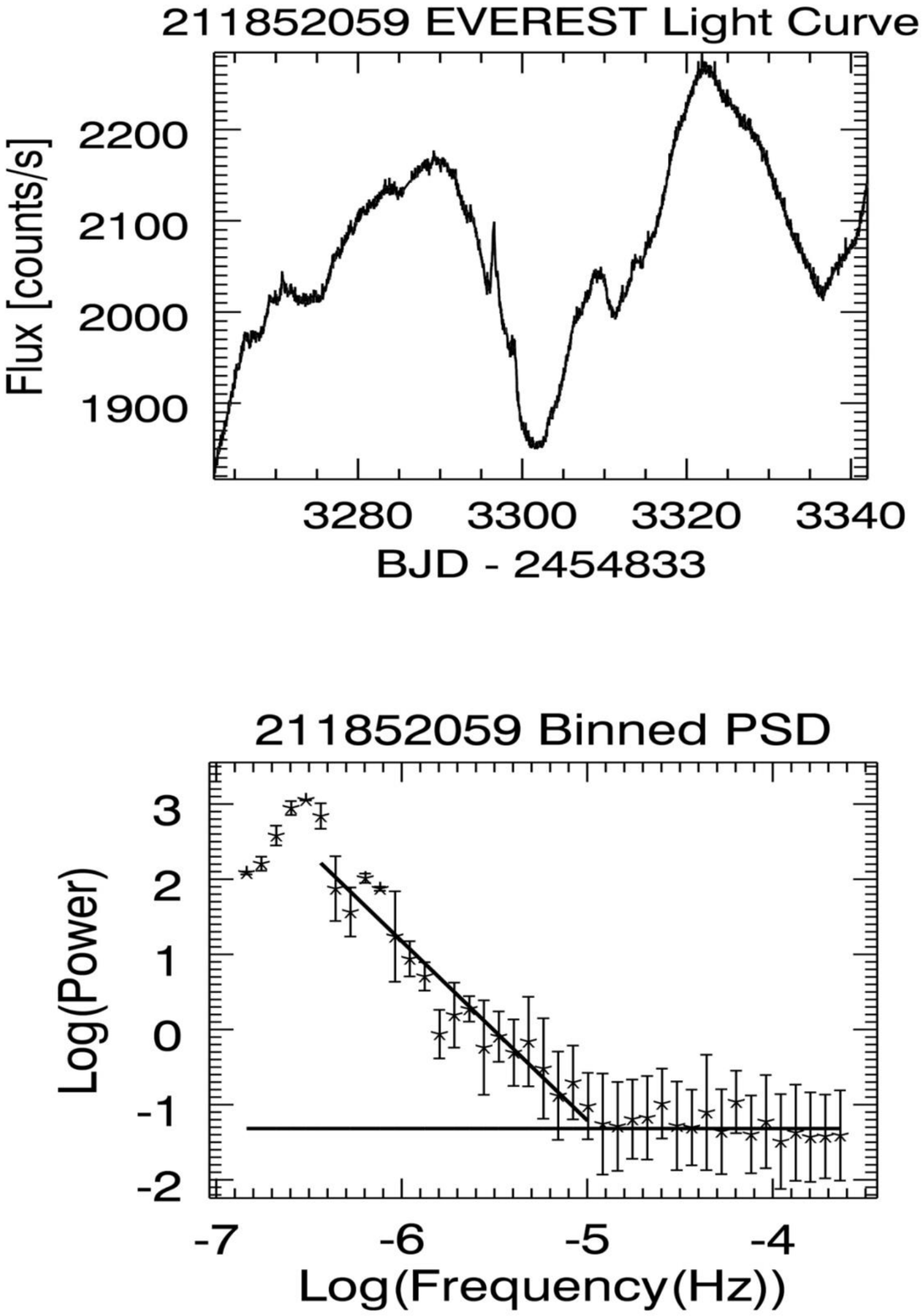}{0.46\textwidth}{(b)}        	
          }
\gridline{\fig{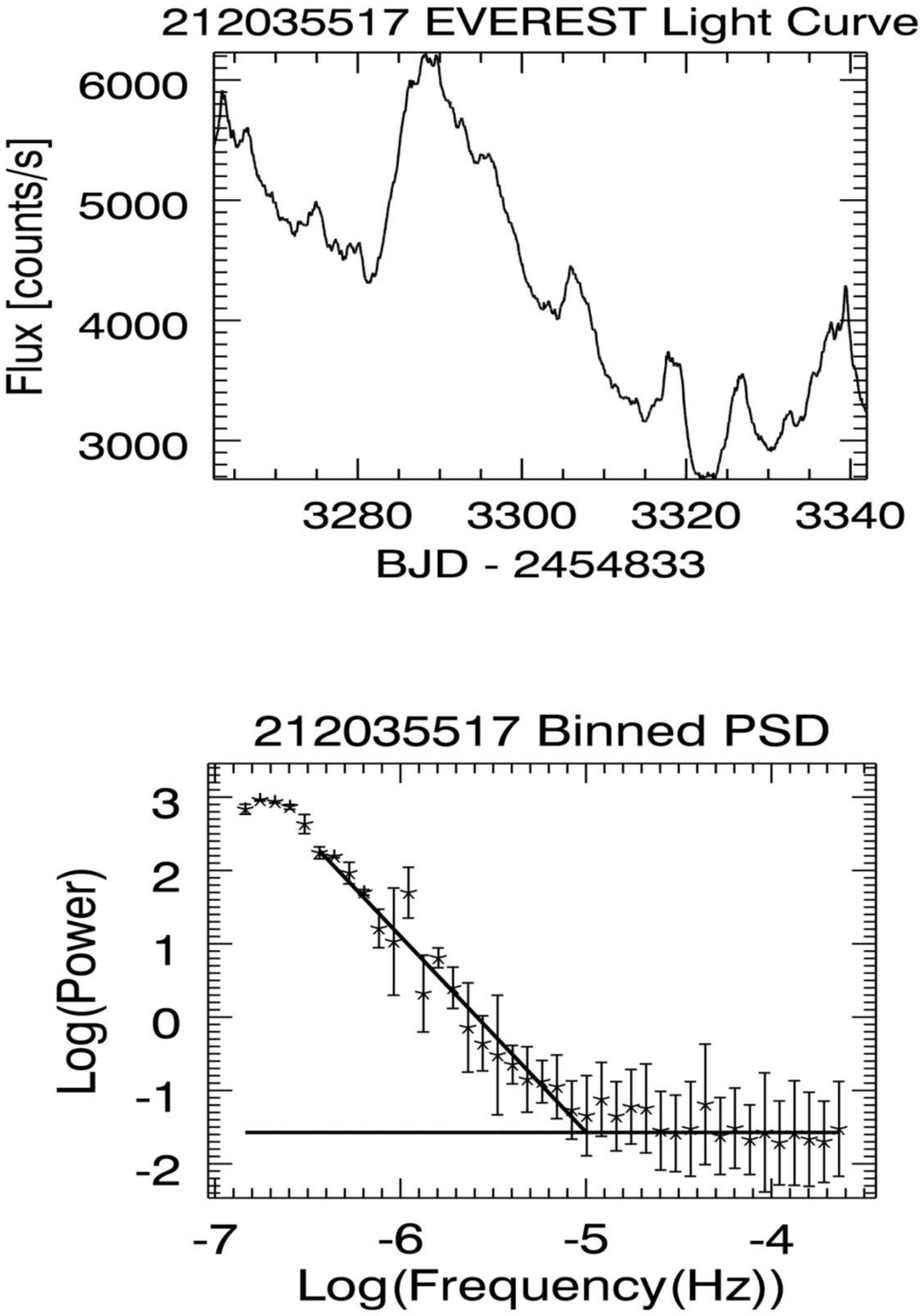}{0.465\textwidth}{(c)}
\fig{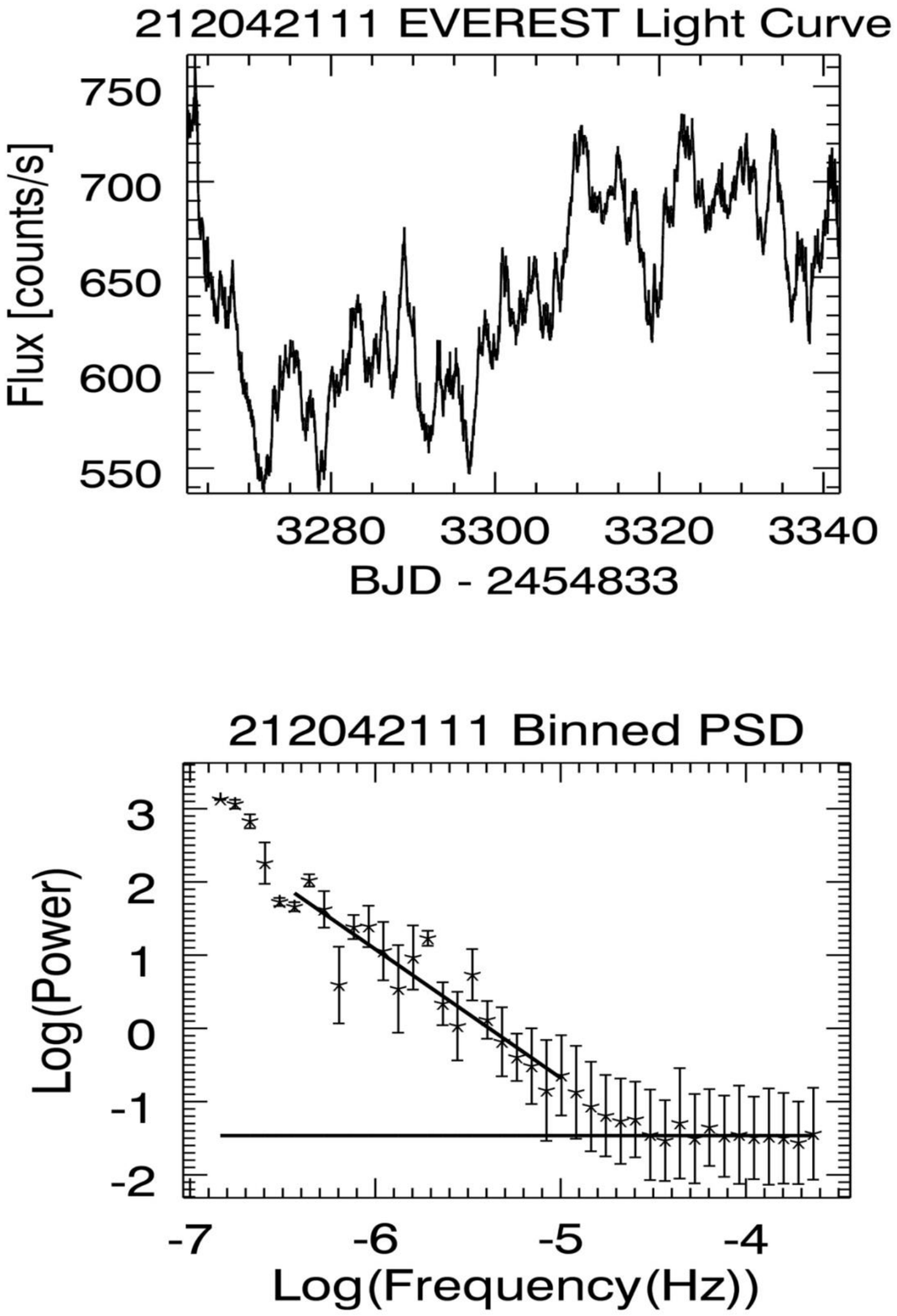}{0.46\textwidth}{(d)}
	}
\caption{Light curves and PSDs for single campaign observations of: (a) EPIC 201621388 (NVSS J110735+022225); (b) EPIC 211852059 (TXS 0836+182); (c) EPIC 212035517 (NVSS J090226+205045); (d) EPIC 212042111 (TXS 0853+211).}
\end{figure*}
\begin{figure*}
\figurenum{4, continued}
\gridline{\fig{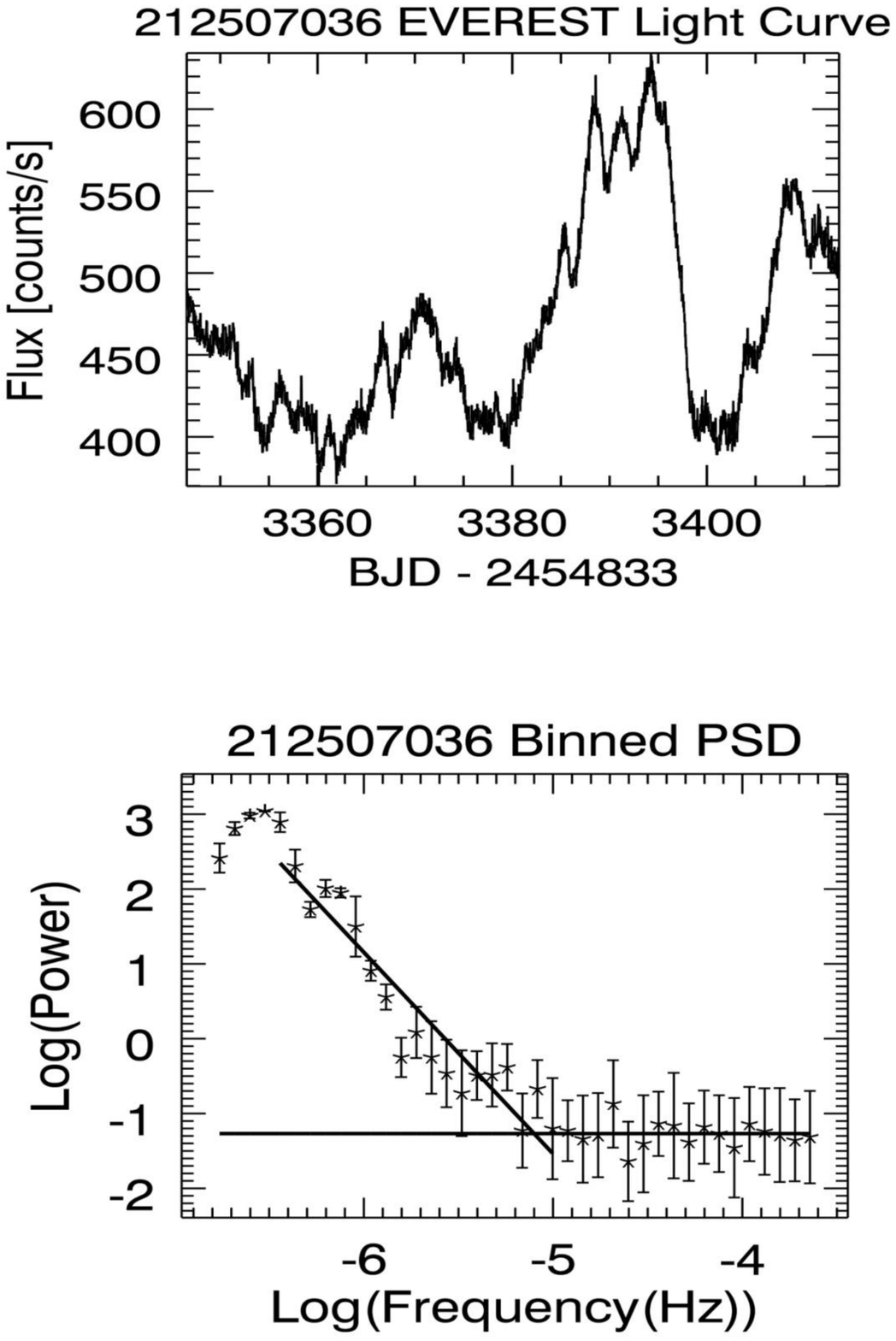}{0.47\textwidth}{(e)}
          \fig{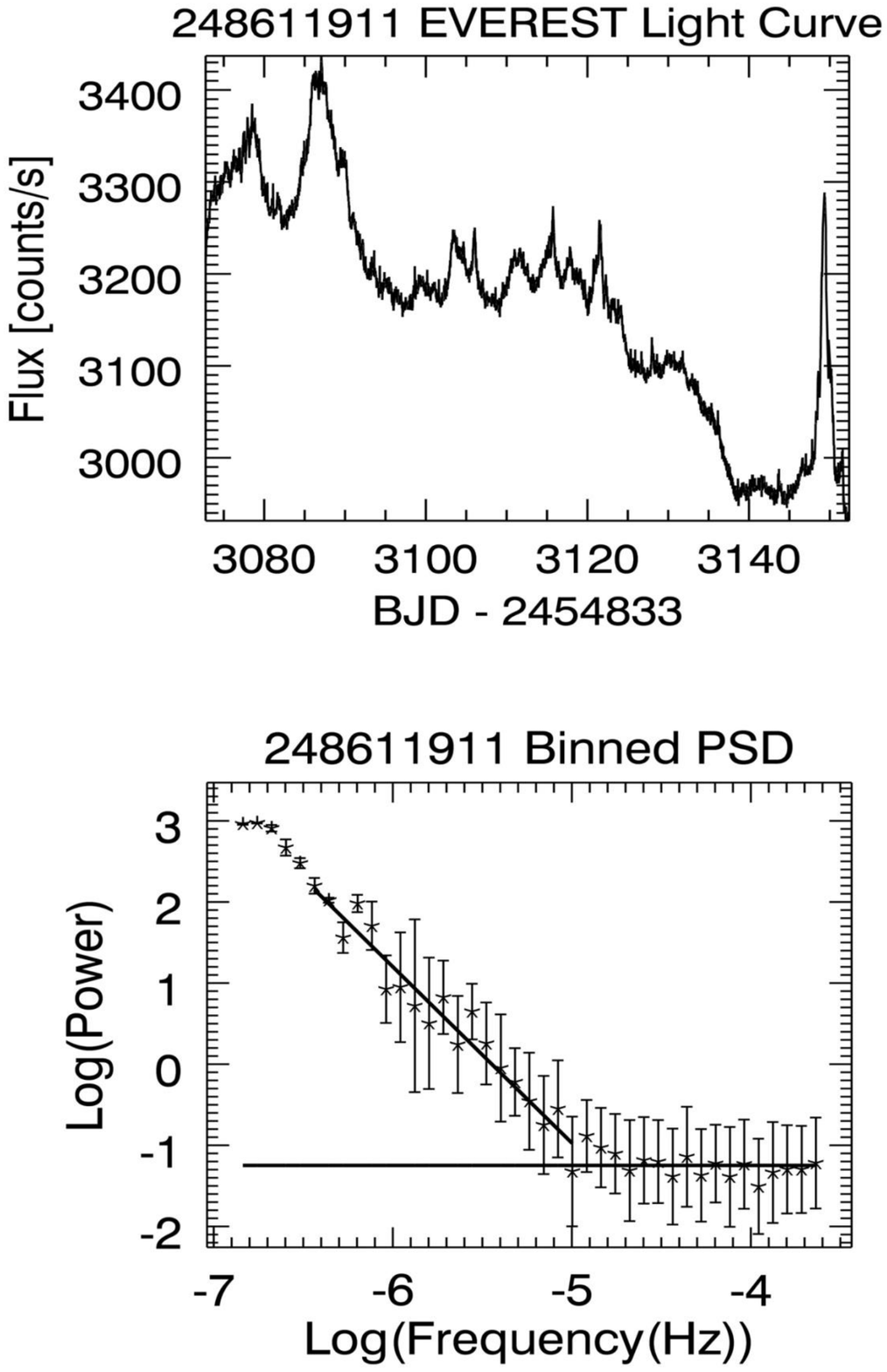}{0.45\textwidth}{(f)}
          }
\gridline{\fig{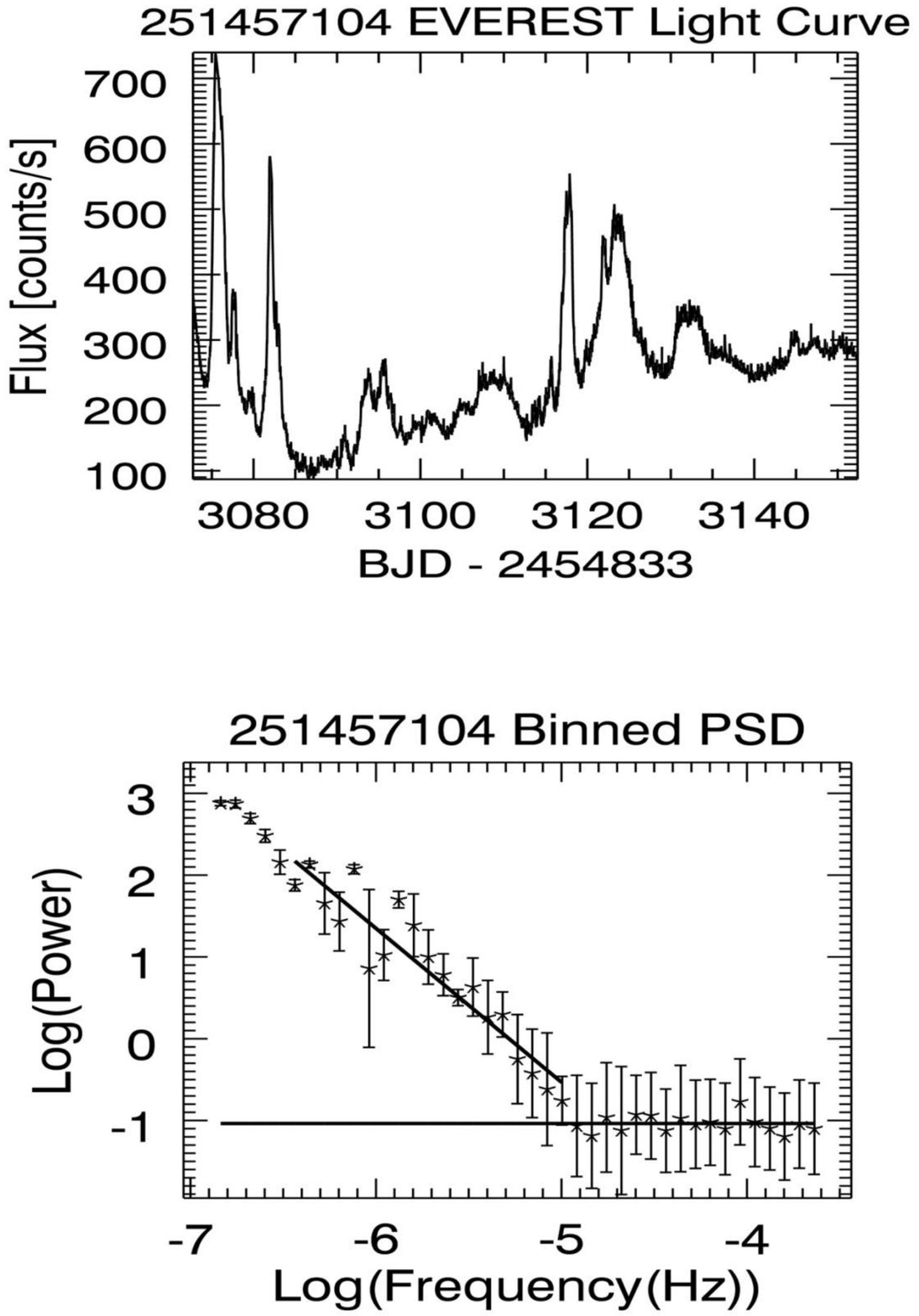}{0.47\textwidth}{(g)}
	\fig{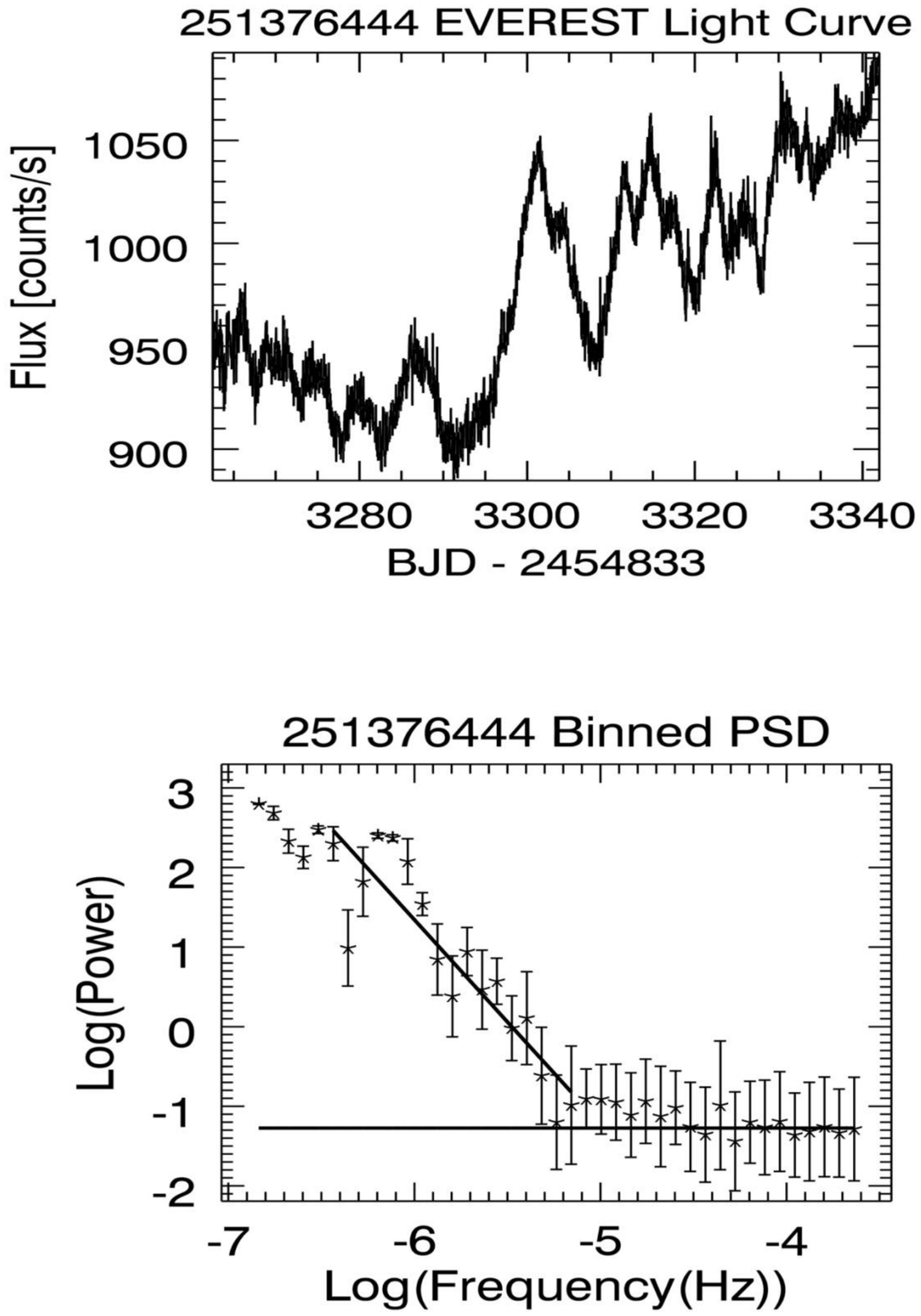}{0.46\textwidth}{(h)}
	}
\caption{ (e) EPIC 212507036 (PMN J1318$-$1235); (f) EPIC 248611911 (4C +06.41); (g) EPIC 251457104 (TXS 1013+054; (h) EPIC 251376444 (NVSS J090900+231112)}
\end{figure*}
\begin{figure*}
\figurenum{4, continued}
\gridline{\fig{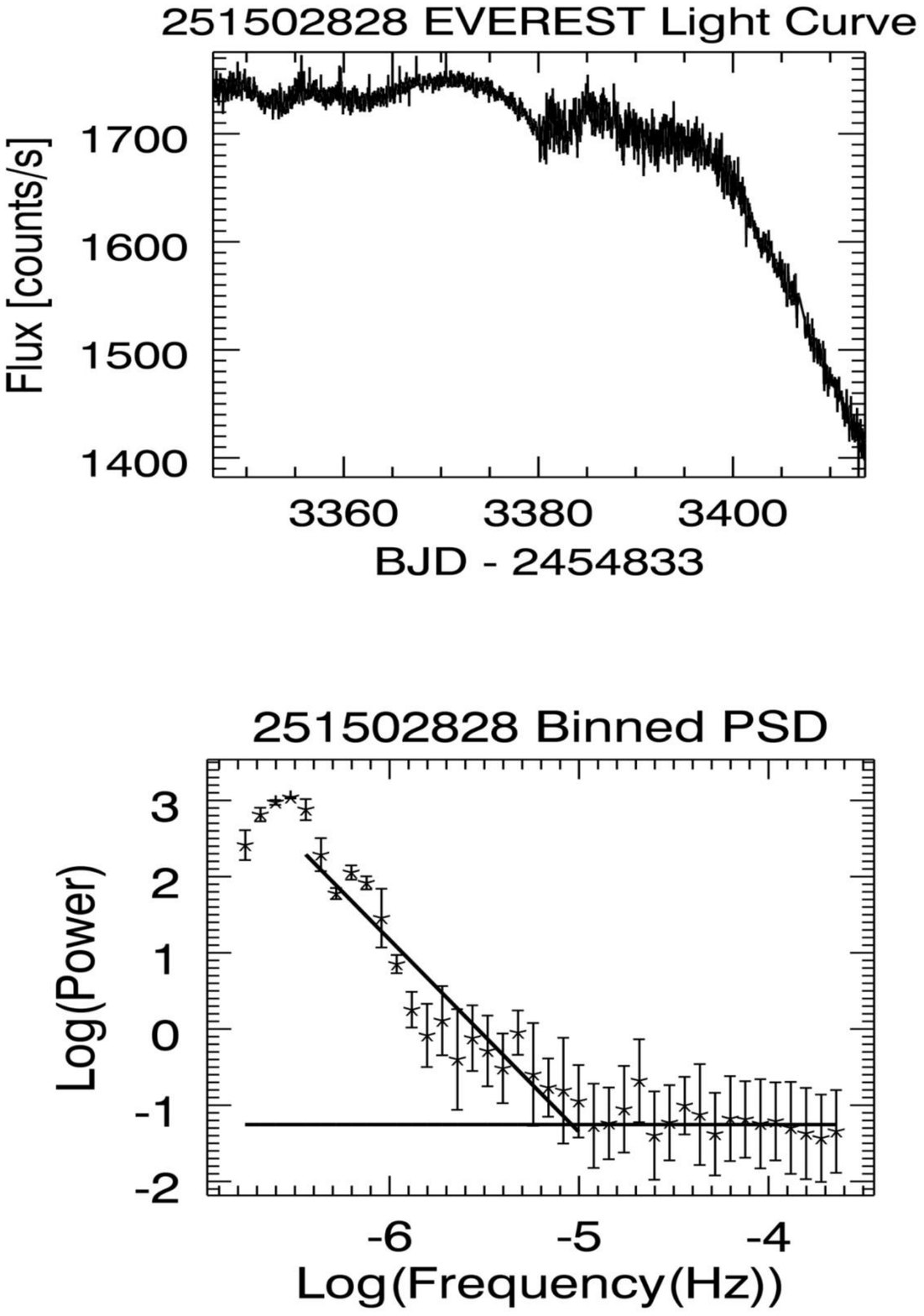}{0.47\textwidth}{(i)}
          }
          \caption{(i) EPIC 251502828 (PKS B1310$-$041)}.
\end{figure*}

\begin{figure*}
\figurenum{5}
\gridline{\fig{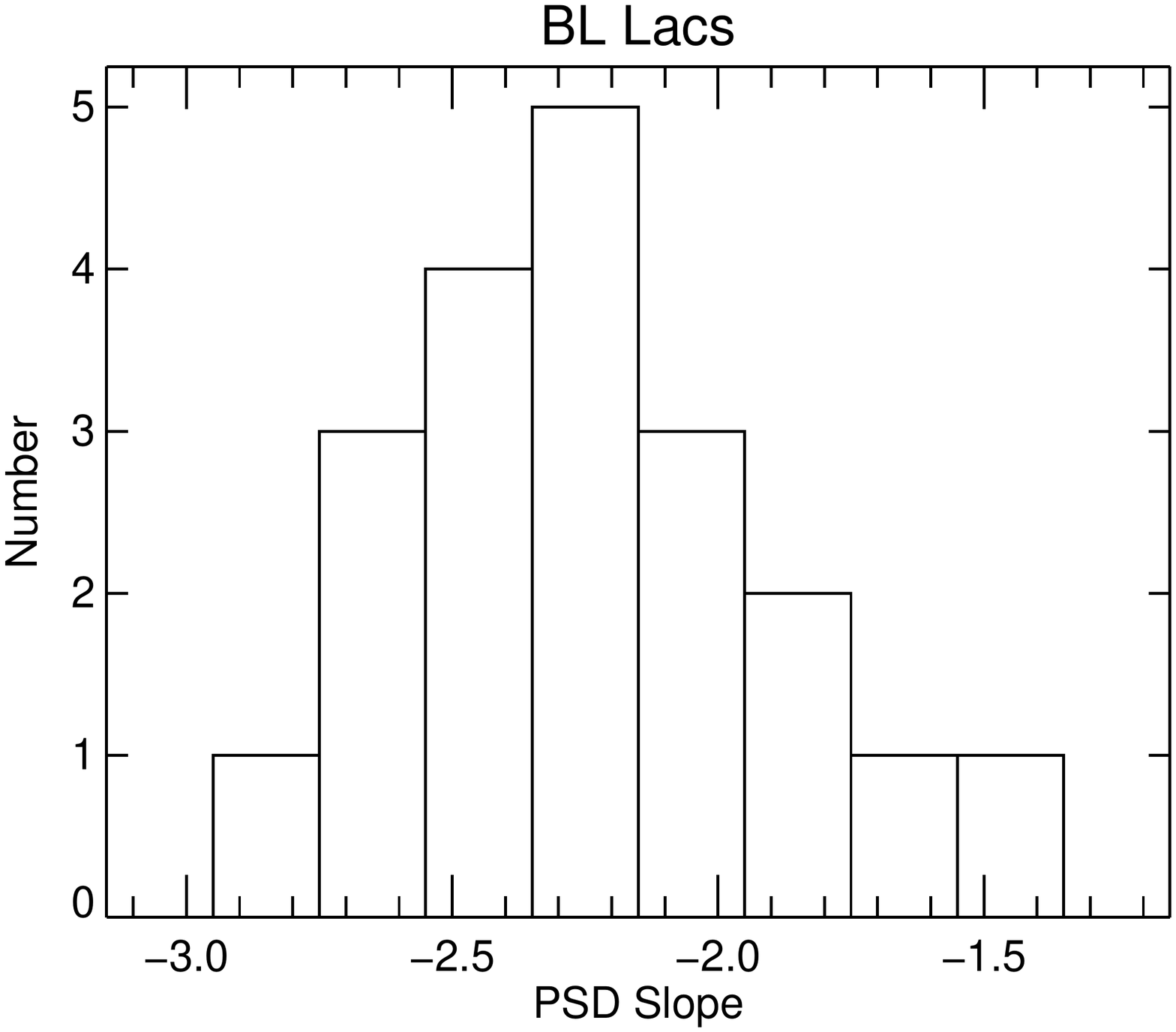}{0.45\textwidth}{(a)}
	\fig{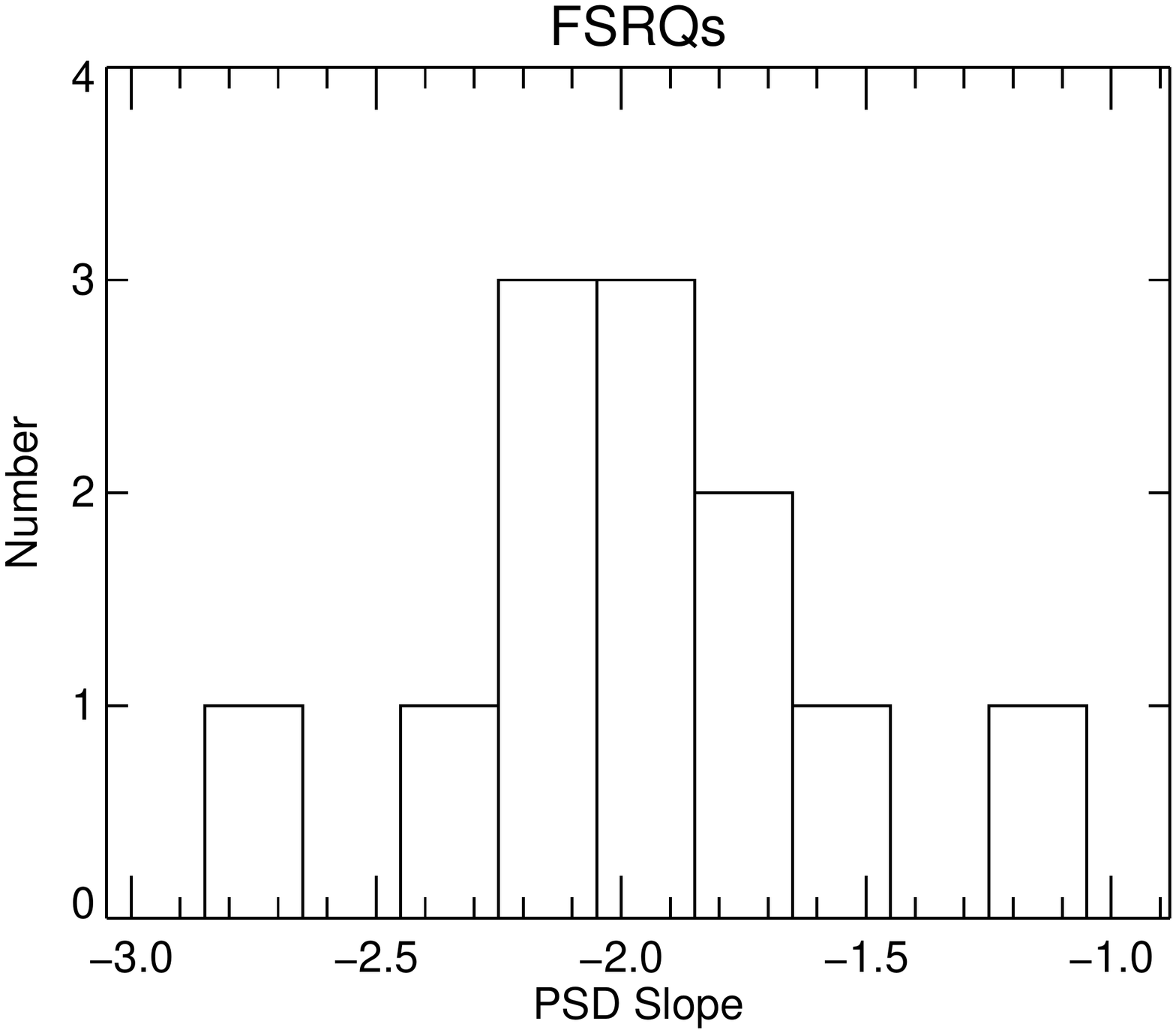}{0.45\textwidth}{(b)}
          }
          \gridline{\fig{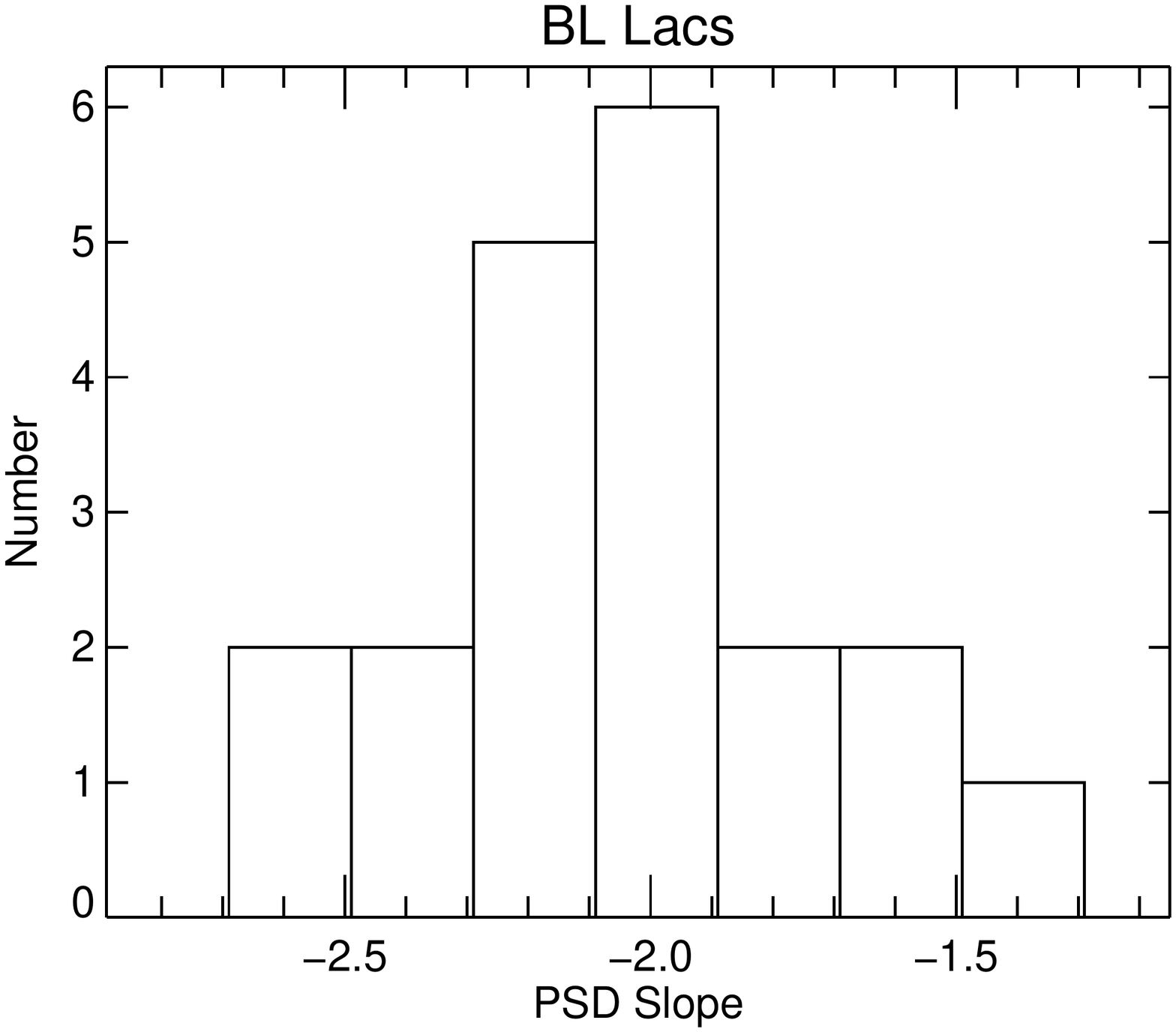}{0.45\textwidth}{(c)}
	\fig{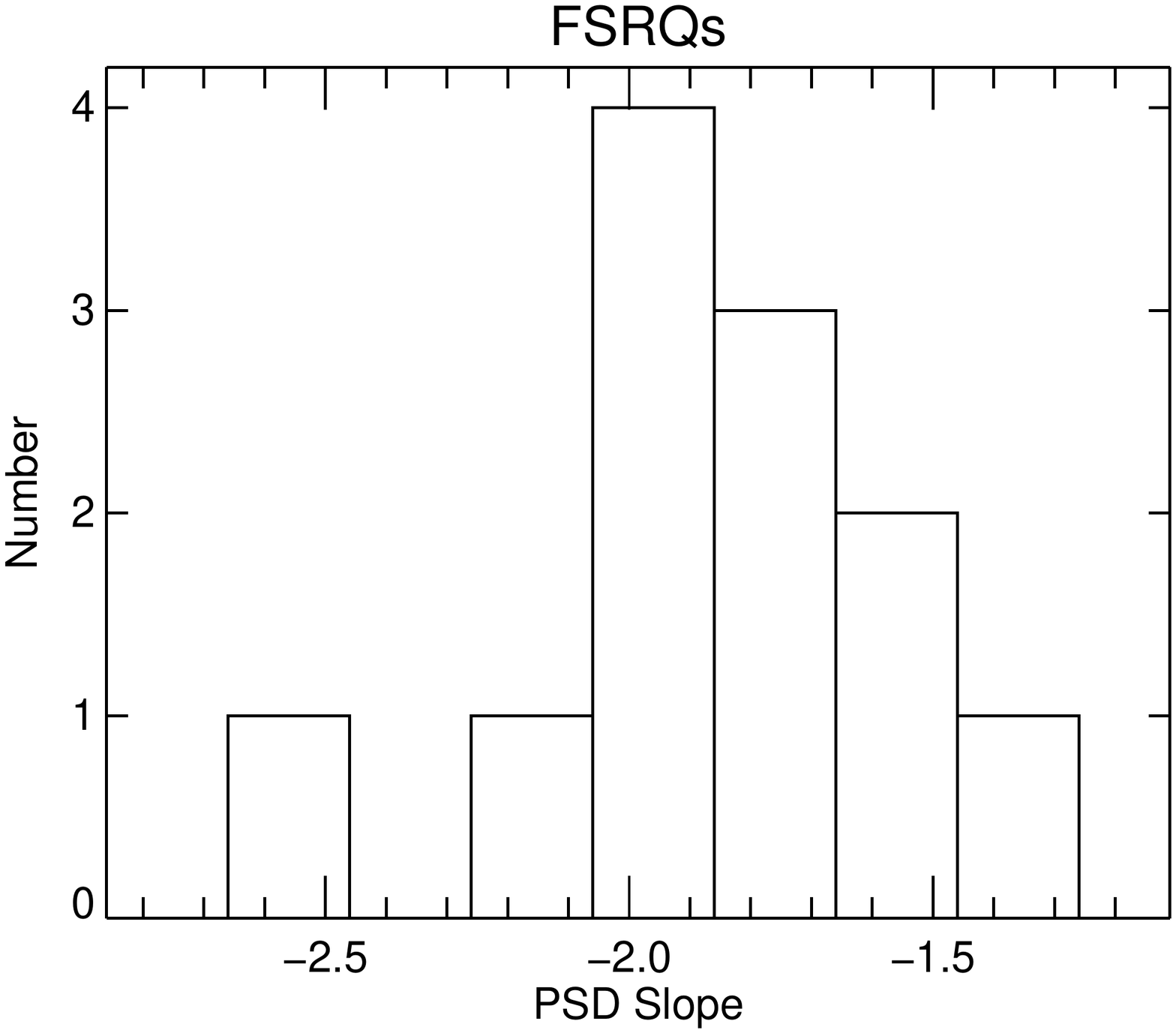}{0.45\textwidth}{(d)}
          }
\caption {Histograms of PSD slopes for all the long cadence light curves binned in 0.2 units. Panels(a) and (b) show the BL Lac and FSRQ PSD slopes determined by periodogram analysis. Panels (c) and (d) show the BL Lac and FSRQ PSD slopes determined by PSRESP analysis. See Section 4 for details.}
\end{figure*}

\section{Contemporaneous Ground-Based Observations}

Three of the sources in this paper --- OJ 287 (EPIC 211991001), NVSS J090226+205045 (EPIC 212035517) and PKS 1335$-$127 (EPIC 212489625)  --- have contemporaneous ground-based observations, allowing us to validate the EVEREST extraction method and place the observed flux level and variability in the context of the source's long term behavior. The ground-based observations were obtained from 5 different sources: the Robotically Controlled Telescope (RCT) blazar monitoring program led by  M.\ Carini \url {http://rct.wku.edu/wordpress/}, P. Smith's $\gamma$-ray blazar optical monitoring program at Steward Observatory (Smith et al.\ 2009), \url {http://james.as.arizona.edu/~psmith/Fermi/}, the SMARTS program (Bonning  et al.\ 2012), the  DEdicated MONitor of EXotransits and Transients telescope (DEMONEXT; Villanueva et al.\ 2018), and the Zwicky Transient Facility (ZTF; Masci et al.\ 2018; Bellm et al.\ 2019). For OJ~287 and NVSS J090226+205045, the offset between the times series from different telescopes was calculated by minimizing the KS statistic between overlapping segments of the time series. For PKS 1335$-$127, a different procedure was used (see below for details).

\subsection{OJ~287 (EPIC 211991001)}
Figure 6a shows the contemporaneous ground-based and K2 data for OJ 287 during Campaigns 5 and 18 and also plots observations between these campaigns and for a while before the first one. Here we utilized  ground-based R band data from  the RCT, from P.\ Smith (Steward Observatory) and the SMARTS program. Figure 6b displays the K2 observations of OJ 287 from Campaign 18, together with ground-based data from  the RCT and  from  Steward.   The agreement between these various ground-based measurements and those from K2, as processed by EVEREST, is very good.  During both K2 campaigns, OJ 287 was in a low-to-middle brightness range.  
 
\subsection{NVSS J090226+205045 (EPIC 212035517)}
Figure 6c displays the light curve composed of contemporaneous ground-based  and Campaign 18 K2 data for 4FGLJ0902.4+2051 (EPIC 212035517). The ground-based data are R-band observations from DEMONEXT. DEMONEXT is a 0.5 m PlaneWave CDK20 f/6.8 Corrected Dall-Kirkham Astrograph telescope at Winer Observatory in Sonoita, Arizona. It has a 2048 $\times$ 2048 pixel Finger Lakes Instruments Proline CCD3041 camera, with a $30.7\arcmin \times 30.7\arcmin$ field of view and a pixel scale of 0.90\arcsec pixel$^{-1}$. We binned the DEMONEXT data into 30-minute time bins to match the K2 long cadence length.  We find excellent agreement  of the overall shapes of the light curves between the K2 and DEMONEXT observations,  indicating that the EVEREST algorithm  again performed well in correcting the forward-facing and backward-facing K2 data. 
 
\subsection{PKS 1335$-$127 (EPIC 212489625)}
PKS 1335$-$127 (212489625) was observed by K2 in two different epochs: Campaign 6 in 2015 and Campaign 17 in 2018. The combined K2 and ground-based light curves are shown in Figure 6d. Our contemporaneous ground-based observations using the RCT and ZTF were only during Campaign 17.  As there were not enough overlapping observations between the K2 Campaign 17 and ZTF observations to perform a meaningful KS analysis, we determined the offset by eye.
In order to include the Campaign 6 observations in the same long-term light curve, we identified 5 objects  which were observed on the same K2 channel (module and output combination) as the blazar and had similar brightnesses as well as little or no long-term variability. We calculated an average offset between the Campaigns 6 and 17 K2 count rates for these 5 sources, and applied that to the blazar Campaign 6 observations.  The K2 observations in both campaigns were obtained when the source was in mid- to low-brightness states. The ground-based observations show it was substantially brighter in between the K2 measurements, and much more active in 2018 and 2019. It varied by a factor of about 12 between MJD 58300 and 58675.

\begin{figure*}
\figurenum{6}
\gridline{\fig{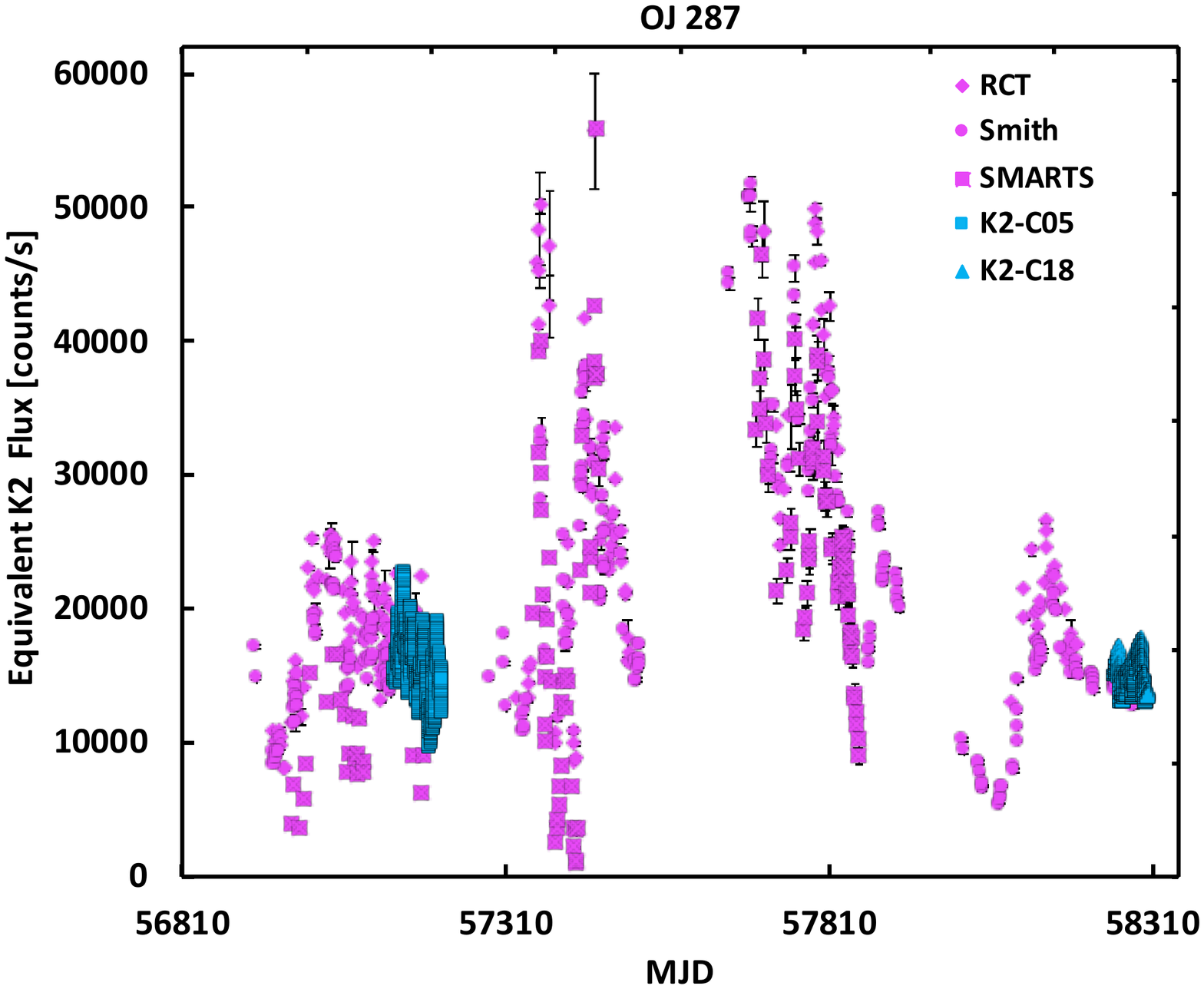}{0.5\textwidth}{(a)}
	\fig{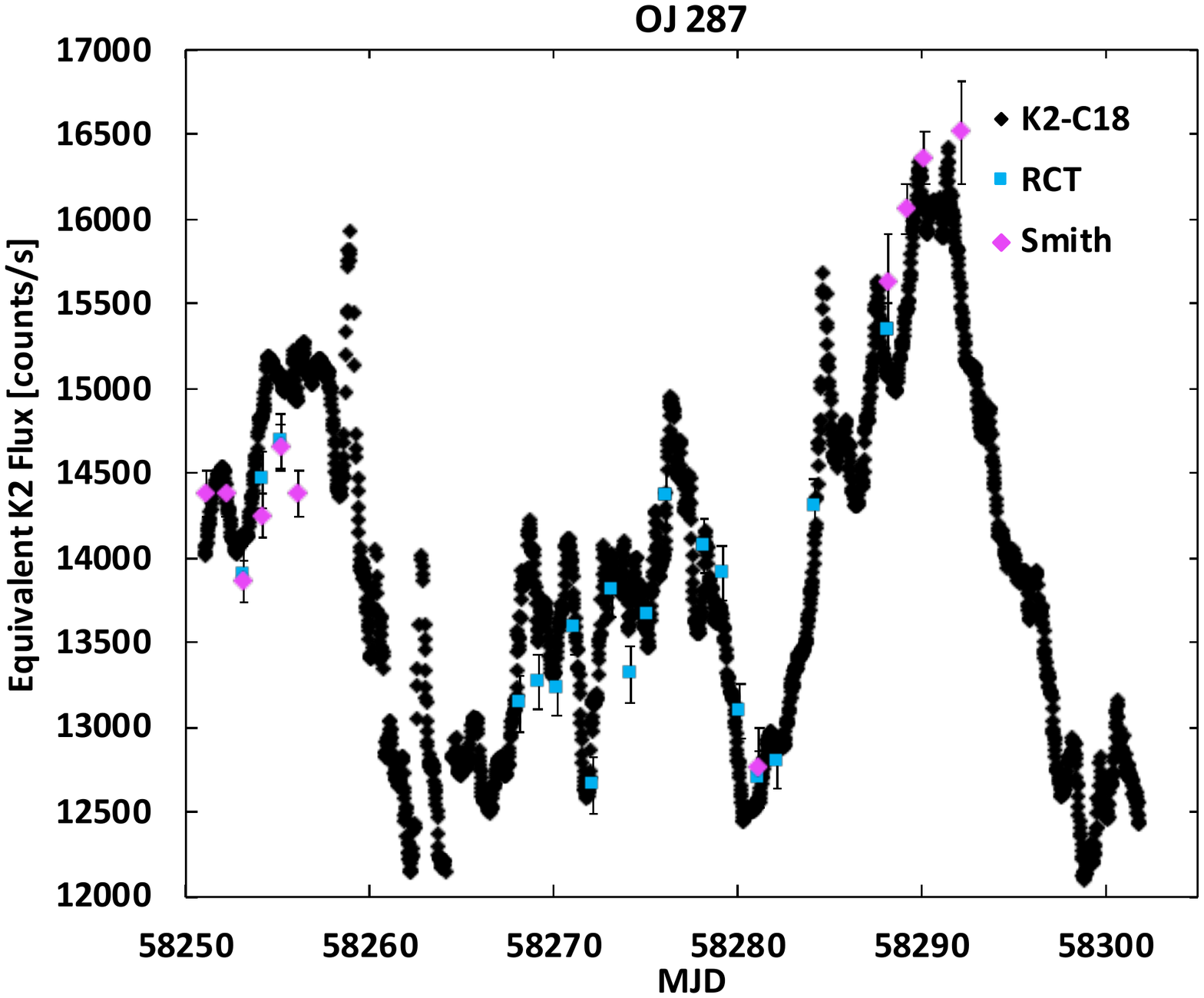}{0.5\textwidth}{(b)}
          }
   \gridline{\fig{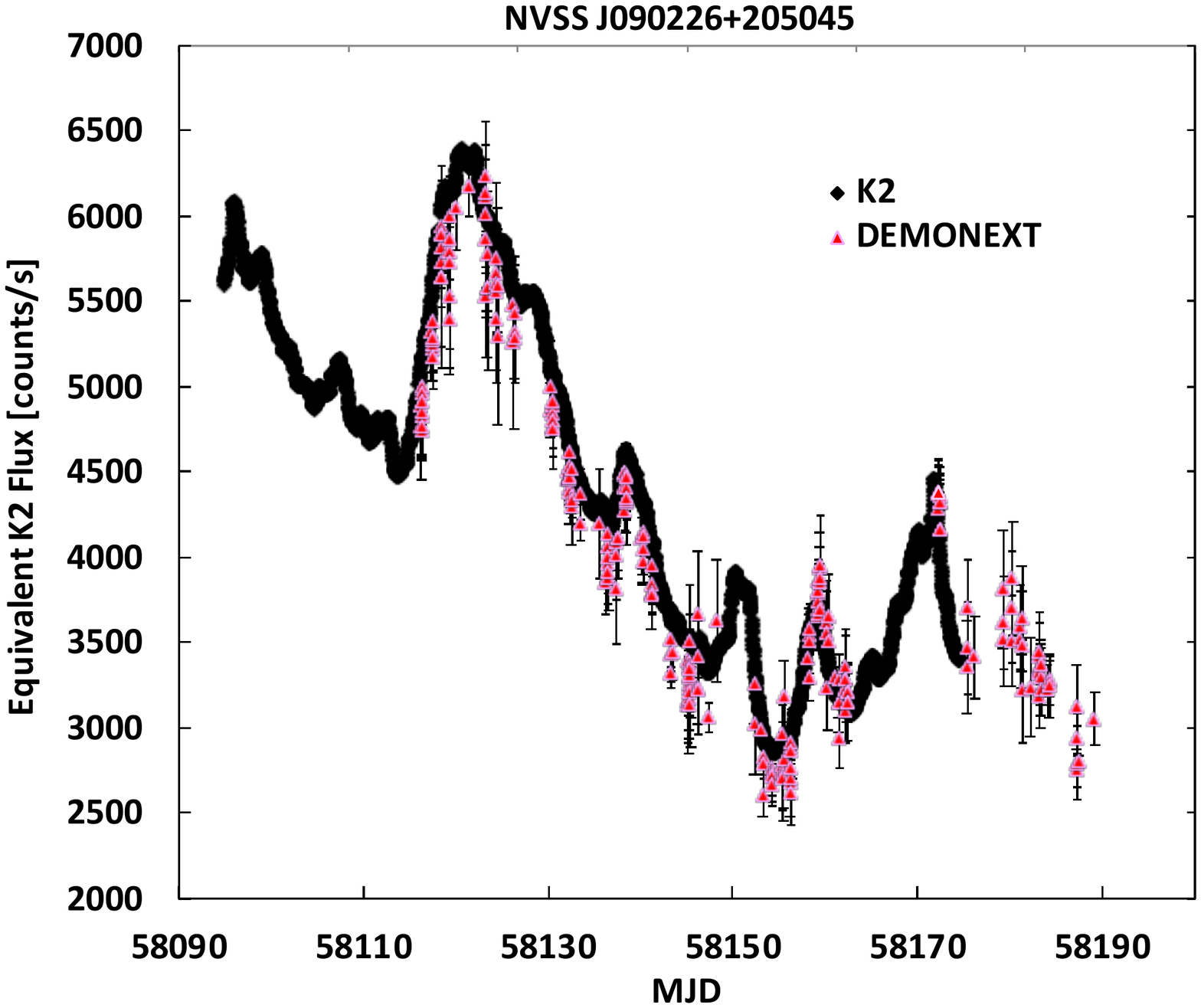}{0.5\textwidth}{(c)}
	\fig{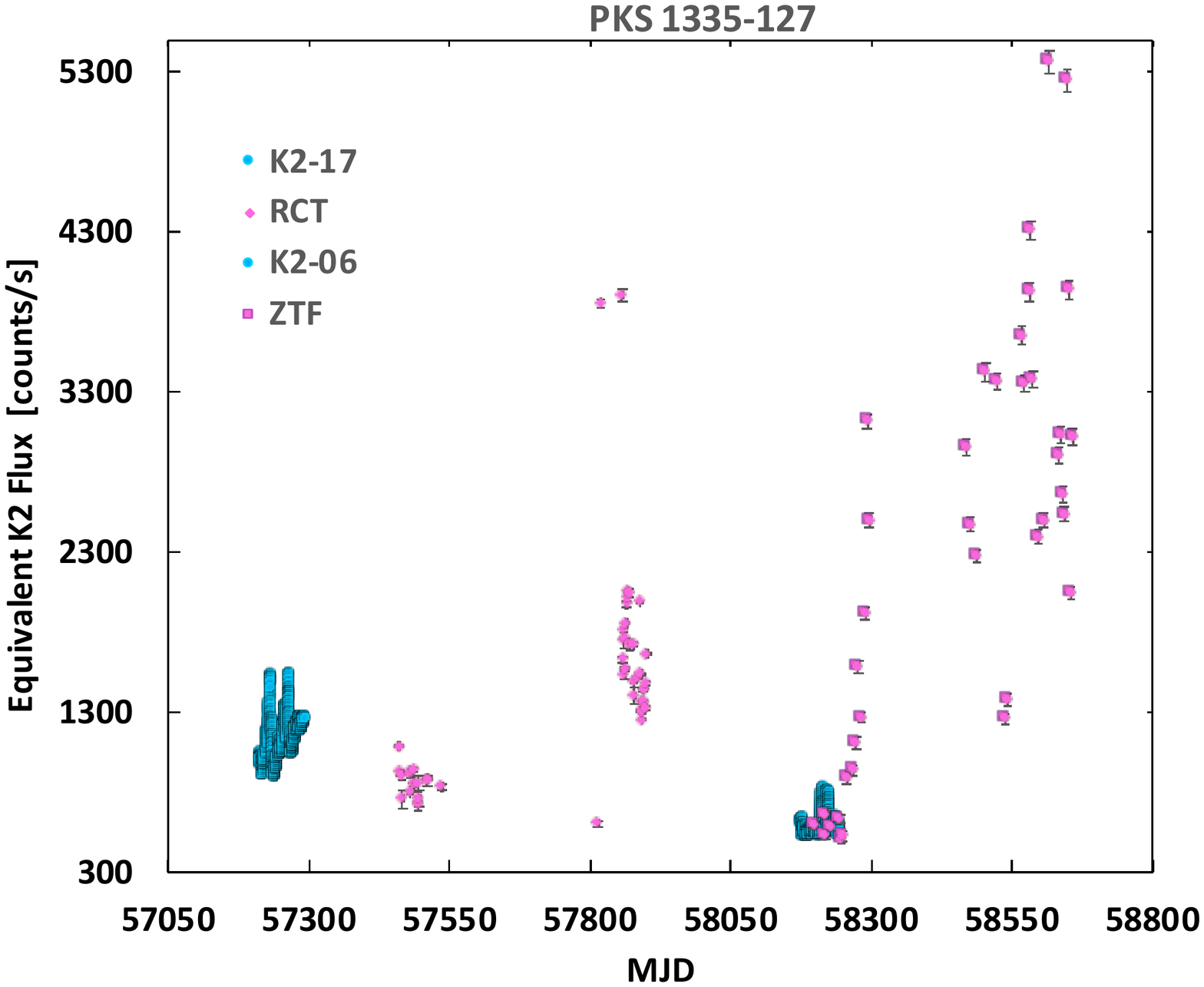}{0.5\textwidth}{(d)}
         }
       
\caption {Combined ground-based and K2 EVEREST light curves.  Ground-based photometry magnitudes were converted to flux via the standard magnitude-flux relationship and scaled to match the K2 data. (a) OJ 287 during Campaigns 5 and 18 in 2014 and 2018, respectively. Ground-based R-band photometry is from the RCT program led by  M.\ Carini, from the Steward Observatory program (Smith et al.\ 2009), and from SMARTS (Bonning  et al.\ 2012).  
(b) Combined ground-based and K2 EVEREST light curve of OJ 287 during  only Campaign 18 in 2018. 
 The R-band photometry has been scaled to match the first K2 observation. For reference, 2016 Jan 1 is MJD 57388.0 and 2018 Dec 31 is MJD  	58483.0.   
(c) NVSS J090226+205045 (EPIC 212035517) during Campaign 18. Ground-based R-band photometry is from  DEMONEXT (Villanueva et al.\ 2018).  
(d) PKS 1335$-$127 (EPIC 212489625) during Campaigns 6 and 17. Ground-based R-band photometry is from the  RCT program  and from the Zwicky Transient Factory  (ZTF; Masci et al.\ 2018; Bellm et al.\ 2019). The R-band photometry has been scaled to match the  second K2 observations on 2018 March 3 -- 2018 July 8 (MJD 58180--58307). 
}
\end{figure*}
\newpage
\section{Fermi-LAT Observations}

We reduced all-sky survey mode Fermi-LAT data using the method described in Papers 1 and 2. Pass 8 data (Atwood et al.\ 2013) were downloaded from the Fermi Science Support Center,  with radius 20$\degr$, for the 67-81 day timespans concurrent with the K2 campaigns.  The downloaded Fermi-LAT datasets were centered on three of the four  $10\degr$-square K2 campaign fields. We offset the fourth Fermi-LAT dataset's center by $5\degr$ from the  K2 Campaign 18 field's center which brought the Fermi-LAT field center closer to our prime target OJ 287.   We used the Fermi Science Tools for unbinned likelihood, instrument response function {P8R2\_SOURCE\_V6}, ``SOURCE class" events (parameters ``evclass = 128, evtype = 3") in the energy range 0.1--500 GeV and set the maximum zenith angle to $90\degr$.  
We used diffuse models {gll\_iem\_v07.fits} and {iso\_P8R3\_SOURCE\_V3\_v1.txt}. For the likelihood calculations, we used the spectral parameters of the 4FGL catalog (Ajello et al.\ 2020; Ballet et al.\ 2020) obtained by running the python script ``make4FGLxml.py"  with ``{gll\_psc\_v27.fit}'') 
as initial values  for the bright sources and power laws for the K2 targets with one exception, within a 30{$\degr$} radius. We allowed sources within a 20{$\degr$} radius and the diffuse emission models to vary.  
At the time we selected targets for observation with K2, the $\gamma$-ray source 2FGL J1040.7+0614 (later catalogued as 3FGL J1040.4+0615 but not included in the 4FGL catalog) had not been identified with an optical counterpart. We picked the nearby radio source 4C+06.41 (RA, DEC $= 10h40m31.630s, +6$\degr$17'21.70''$) as a likely counterpart and observed it with K2. Subsequently, Garrappa et al.\ (2019) identified two radio sources that could be associated with 3FGL J1040.4+0615 and discovered $\gamma$-ray emission from both radio sources by analyzing 9.8 years of Fermi LAT data.   The two objects, 4C+06.41 and GB6 J1040+0617, are only 22\arcsec apart. We used the spectral parameters derived by Garrappa et al.\ (2019) as initial values for the likelihood calculations for the two objects and used LAT data only when the Sun was farther than 15$\degr$ from them  (the  LAT data train was shortened by three days at the end), following recommendations by S. Garrappa (personal communication). We then applied the {\it fermiPy UpperLimits} scripts (Wood et al.\ 2017, \url {https://github.com/fermiPy/fermipy/tree/master/fermipy}) with the 4FGL catalog spectral shapes to calculate 95\% confidence level upper limits for undetected targets in the  0.1--500 GeV energy range.  

 We  present the power-law spectral indices, fluxes, and test statistic (TS) values and flux upper limits in  Table 4 for the targets observed by K2 in Campaigns 14--18. Of these 18 unique K2 targets, we detected 6 above TS = 16 ($\gtrsim 4\sigma$). Two twice-observed targets, PKS 1335-127 and OJ~287, were detected in both earlier and later campaigns. Three twice-observed targets were detected during earlier campaigns but not in later campaigns.  BZB J0816+2015 was observed in Campaign 5 as well  as 18 but was not included in Paper 1; only upper limits were found for both periods. 
Searching the online database for the Fermi All-sky Variability Analysis  (Abdollahi et al.\ 2017, \url {https://fermi.gsfc.nasa.gov/ssc/data/access/lat/fava_catalog/ }) showed no evidence of variability during the K2 observations for any of our targets.

\begin{deluxetable}{lrrrccrrr}
\tablecaption{Fermi-LAT Results}
\tablewidth{0pt}
\singlespace
\tablecolumns{9}
\tabletypesize{\footnotesize}
\tablehead{
             \colhead{Name}
              & \colhead{EPIC}
              & \colhead{Fermi}
             & \colhead{$\gamma$-ray}
             & \colhead{$\gamma$-ray}
             &\colhead{Test} 
             & \colhead{4FGL $\gamma$-ray}
             & \colhead{4FGL $\gamma$-ray}
              & \colhead{4FGL}
             \\
             \colhead{ }
               & \colhead{ID}
               & \colhead{Source}
 	    & \colhead{Flux\tablenotemark{a}}
             & \colhead{Flux Error\tablenotemark{a}}
             &\colhead{Stat-} 
            & \colhead{Flux\tablenotemark{b}}
             & \colhead{Flux Error\tablenotemark{b}}
              & \colhead{Detection}
            \\
             \colhead{ }
              & \colhead{ }
             & \colhead{Name}
             & \colhead{ph~s$^{-1}$~cm$^{-2}$}
             & \colhead{ph~s$^{-1}$~cm$^{-2}$}
              & \colhead{istic\tablenotemark{a}}
             & \colhead{ph~s$^{-1}$~cm$^{-2}$}
             & \colhead{ph~s$^{-1}$~cm$^{-2}$}
              & \colhead{Signif-}
             \\
              \colhead{ }
             & \colhead{ }
              & \colhead{ }
              & \colhead{ }
              & \colhead{ }
              & \colhead{ }
              & \colhead{ }
              & \colhead{ }
              & \colhead{icance $\sigma$ }
            \\
             }
\startdata
NVSS J110735+022225	 & 	201621388	 & 	4FGL~J1107.6+0222	 & 	$<$ 1.97E-9	 & 	\nodata	 & 	\nodata	 & 	5.34E-10	 & 	4.87E-11	 & 	17.159	 \\ 
4C +06.31	 & 	248611911	 & 	see note\tablenotemark{c}		 & 	5.76E-08	 & 	2.64E-10	 & 	42.1	 & 	\nodata	 & 	\nodata	 & 	\nodata	 \\ 
TXS 1013+054	 & 	251457104	 & 	4FGL~J1016.0+0512	 & 	4.41E-08	 & 	6.30E-10	 & 	31.0	 & 	1.20E-09	 & 	6.57E-11	 & 	30.8646	 \\ 
3C 207 (in C16)	 & 	211504760	 & 	4FGL~J0840.8+1317	 & 	$<$ 6.29E-9	 & 	\nodata	 & 	\nodata	 & 	1.95E-10	 & 	3.49E-11	 & 	7.3078	 \\ 
TXS 0836+182	 & 	211852059	 & 	4FGL~J0839.4+1803	 & 	$<$ 5.29E-9	 & 	\nodata	 & 	\nodata	 & 	3.08E-10	 & 	3.75E-11	 & 	11.5571	 \\ 
NVSS J090226+205045	 & 	212035517	 & 	4FGL~J0902.4+2051	 & 	1.05E-08	 & 	7.93E-11	 & 	26.1	 & 	1.34E-09	 & 	6.41E-11	 & 	37.7239	 \\ 
TXS 0853+211	 & 	212042111	 & 	4FGL~J0856.8+2056	 & 	$<$ 4.36E-09	 & 	\nodata	 & 	\nodata	 & 	4.68E-10	 & 	4.88E-11	 & 	14.5981	 \\ 
NVSS J090900+231112	 & 	251376444	 & 	4FGL~J0908.9+2311	 & 	1.80E-09	 & 	2.57E-11	 & 	16.6	 & 	4.83E-10	 & 	5.04E-11	 & 	18.2402	 \\ 
PKS 1335-127	 & 	212489625	 & 	4FGL~J1337.6-1257	 & 	1.12E-08	 & 	3.03E-09	 & 	18.0	 & 	1.72E-09	 & 	8.17E-11	 & 	37.3988	 \\ 
PMN J1318-1235	 & 	212507036	 & 	 4FGL~J1318.7-1234	 & 	$<$ 1.54E-8	 & 	\nodata	 & 	\nodata	 & 	5.52E-10	 & 	4.88E-11	 & 	16.794	 \\ 
PKS 1352-104	 & 	212595811	 & 	4FGL~J1354.8-1041	 & 	$<$ 5.54E-8	 & 	\nodata	 & 	\nodata	 & 	8.66E-10	 & 	6.20E-11	 & 	23.4487	 \\ 
RBS 1273	 & 	212800574	 & 	 4FGL~J1329.4-0530	 & 	$<$ 1.63E-8	 & 	\nodata	 & 	\nodata	 & 	4.55E-10	 & 	5.97E-11	 & 	10.8777	 \\ 
PKS B1329-049	 & 	229227170	 & 	4FGL~J1332.0-0509	 & 	$<$ 2.15E-9	 & 	\nodata	 & 	\nodata	 & 	2.65E-09	 & 	9.33E-11	 & 	54.4994	 \\ 
PKS B1310-041	 & 	251502828	 & 	 4FGL~J1312.8-0425	 & 	$<$ 8.13E-8	 & 	\nodata	 & 	\nodata	 & 	1.52E-09	 & 	7.41E-11	 & 	40.2091	 \\ 
RGB J0847+115	 & 	211394951	 & 	 4FGL~J0847.2+1134	 & 	$<$ 2.58E-9	 & 	\nodata	 & 	\nodata	 & 	4.20E-10	 & 	4.38E-11	 & 	19.037	 \\ 
3C 207 (in C18)	 & 	211504760	 & 	4FGL~J0840.8+1317	 & 	$<$ 7.13E-9	 & 	\nodata	 & 	\nodata	 & 	1.95E-10	 & 	3.49E-11	 & 	7.3078	 \\ 
WB J0905+1358	 & 	211559044	 & 	 4FGL~J0905.6+1358	 & 	$<$ 2.59E-8	 & 	\nodata	 & 	\nodata	 & 	1.03E-09	 & 	6.44E-11	 & 	31.9515	 \\ 
OJ 287	 & 	211991001	 & 	4FGL~J0854.8+2006	 & 	7.10E-08	 & 	2.09E-10	 & 	139.0	 & 	5.65E-09	 & 	1.23E-10	 & 	100.2436	 \\ 
BZB J0816+2051	 & 	212035840	 & 	4FGL~J0816.9+2050	 & 	$<$ 4.77E-8	 & 	\nodata	 & 	\nodata	 & 	3.48E-10	 & 	4.47E-11	 & 	13.0976	 \\ 
\enddata
\tablenotetext{a}{Fluxes, upper limits, and test statistics were in the energy range 0.1--500 GeV, measured during  K2 campaigns in  2017--2018.}
\tablenotetext{b}{Data obtained from Fermi-LAT Fourth Source Catalog - Data Release 2 (4FGL-DR2, Ajello et al.\ 2020 and Ballet et al.\ 2020), in the energy range 1--100 GeV, observed between 2008 August 4, to 2018 August 2, retrieved from https://fermi.gsfc.nasa.gov/ssc/data/ }
\tablenotetext{c}{4C +06.31 (also known as PKS 1038+0641) was tentatively associated with 2FGL J1040.7+0614, but it is not in the 4FGL.  Another $\gamma$-ray source, also not in the 4FGL, associated with quasar GB6 J1040+0617, has been found 22' away from PKS 1038+064 (Garrappa et al. 2019).}
\end{deluxetable}

\newpage 
\section{Discussion}

One of the key advantages of the K2 mission schedule for our investigation was the fact that 7 of these blazars could be observed during two different periods (Campaigns 5 and 18 or Campaigns 6 and 17) separated by just about 3.0 years and about 2.7 years, respectively. Additionally, one object (211504760) was observed in Campaign 16 as well as in 5 and 18, though it is classified as a lobe-dominated radio quasar   instead of a FSRQ or BL Lac. In this paper we have reduced the data and computed the PSDs for all these observations in a uniform fashion and thus can fairly investigate whether the nature of the fluctuations are stable or change over time scales of a few years. Table 3 gives the directly fitted periodogram PSD slopes and their nominal errors  in the ninth and tenth columns and it appears that  these values do not change much between campaigns. To quantify this impression, if we consider that these PSD slopes are not significant if the difference between those slopes is less than the sum of their errors, then 3 of the  9 PSDs  produced from the long cadence K2 data have changed significantly:  the BL Lac WB J0905+1358 (EPIC 211559047), but just barely; the FSRQ PKS 1335-127 (EPIC 212489625); and the X-ray QSO RBS 1273 (EPIC 212800574). If we instead compare the difference of the slopes to twice the sum of errors taken in quadrature, then only EPIC 212489625 and EPIC 212800574 are clearly substantially different.  The PSRESP slopes tend to have smaller nominal errors than the directly estimated slopes and so a larger number of differences between epochs are seen. Using either of the above criteria the BL Lacs OJ 287 (EPIC 211991001), RGB J0847+115 (EPIC 211394951)  and BZB J0816+2051 (EPIC 212035840) as well as the FSRQ PKS 1352-104 (EPIC 212595811) and the LDRQ 3C 207 (EPIC 211504760 - between the C16 and C18 observations) as well as EPIC 212800574 have different slopes. However, EPIC 212800574 is exceptional in the sense that while it among the brightest of our sample it varied by less than 1\% and 2\% during  these observations, making it more difficult to produce reliable PSDs. By inspecting the PSDs during these observations, we see that the slopes primarily changed because the transition from red to white noise was substantially different in the two campaigns (log $\nu = -4.6$ and $-5.5$, respectively, in Campaigns 6 and 17) corresponding to greater white noise domination in Campaign 17 than in Campaign 6.  As will be discussed in Section 7.4, this difference is probably not of astrophysical origin. 

New short cadence data were only collected for OJ~287 and those are discussed below.

In Papers 1 and 2 we reviewed previous studies of AGN using  Kepler and K2 and we refer the reader to them for key results from that earlier work, particularly that of Smith et al.\ (2018), who analyzed 21 {\it Kepler} light curves of Type 1 AGN and that of Aranzana et al.\ (2018) who analyzed 252 K2 AGN light curves,  most of which, however, were not blazars.  Here we note that in the few cases where members of our samples overlapped with theirs the PSD slopes we found were in good agreement with theirs.  

\subsection {Slopes of Optical Power Spectral Densities}

In keeping with the usual approach, we have computed the PSDs directly from the observations and we have made no attempt to correct the fluxes to jet rest-frame emission.  Doing so would require being able to calculate the opposing effects both of source redshifts and the jets' relativistic motions (e.g., Gopal-Krishna et al. 2003).  While redshifts are given in Table 1 for a substantial majority of the sources, they are not available for a few of them and are uncertain for a few others.   For those with solid $z$ values, the median redshift for FSRQs is substantially higher, at 1.072, while that for BL Lacs is  0.329.  For those sources at higher redshift, UV emission is observed in the K2 optical band and any big-blue-bump accretion disk contribution could dilute emitted flux variations. On the other hand, only for very few of these targets are even rough estimates of their jet Doppler factors available, so we cannot examine their effects on an object-by-object basis.  The long term study of radio knot velocities in the MOJAVE program includes 409 jets and a key conclusion of this work is that the highest apparent jet speeds are seen only in blazars with low peak synchrotron frequencies (Lister et al.\ 2019).  Specifically, FSRQs exhibit larger typical Doppler factors, while high-frequency BL Lacs have the smallest, with the low- and intermediate- frequency BL Lacs lying in between those extremes (Lister et al.\ 2019).  A larger Doppler factor means that rest-frame variability appears both stronger and to occur over a shorter time scale when observed.   We performed simulations to investigate if the Doppler factor and redshift differences between BL Lacs and FSRQs could explain the difference we saw and found no evidence that they could (see Appendix). 

As described  in Paper 1, we and others found slopes steeper than $-2$ for three FSRQs and a Seyfert 1.5 from {\it Kepler} data in its original observing mode (Wehrle et al.\ 2013; Revalski et al.\ 2014). Mushotzky et al.\ (2011), Kasliwal et al.\ (2015), and Smith et al.\ (2018) also found some significantly steeper slopes for various Type 1 AGN (including some blazars) observed for multiple quarters with {\it Kepler}.
 Recently, Goyal (2021) has determined PSDs from intranight optical light curves of 14 blazars taken from ground-based telescopes during the 29 monitoring sessions during which they showed significant variability.  These measurements probe sub-hour timescales and  she used a power spectral response method to estimate PSDs for the more variable 19 of those light curves.  Those PSD slopes showed a wide range between $-1.4$ and $-4.0$, but with an average of $-2.9\pm0.3$, somewhat steeper than the red-noise variability usually found on longer timescales (Goyal 2021)  as well as those found here.

 For power spectra that are very steep or show bends or breaks at lower frequencies it may be necessary to go beyond analyzing light curves in terms of single PSD slopes as we have done here and has usually been done in similar studies.  The most common way to do so is to consider continuous-time autoregressive moving average (CARMA) models (e.g., Kelly et al.\ 2009,  2014; Simm et al.\ 2016; Goyal et al.\ 2018; Moreno et al.\ 2019).  These characterize light curves with both a perturbation spectrum and an impulse-response function that allows for the interpretation of variability timescales.  The lowest order CARMA(1,0) model is isomorphic to the Ornstein-Uhlenbeck, or damped-random-walk, stochastic model (Kelly et al.\ 2009).   Higher order CARMA($p,q$) models, which  connect the light curve and its first $p$ time derivatives to the noise and its first $q$ time derivatives, can simulate light curves with multiple breaks in the power spectra slopes and also yield PSD slopes steeper than the $-2$ limit of a damped-random-walk model. 

In Paper 1 we gave an extensive discussion of the literature discussing the PSDs likely to arise from accretion disks and relativistic jet models; here we summarize a few of those points and also discuss some more recent papers.  Most physical models for light curve variability, particularly those relativistic jet-based ones most relevant to blazar fluctuations, have not been extended or analyzed to produce simulated PSDs.  Simulations of propagating jets, including relativistic turbulence,  were able to produce light curves with a range of PSD slopes between $\sim -1.5$ to $\sim -2.6$  (Pollack et al.\ 2016).  The turbulent extreme multi-zone model for a relativistic jet with a standing shock can produce multi-band light curves that look like observed variations (Marscher 2014) but there have been no published PSDs from this very promising approach.    Recently, Zhang \& Giannios (2021) have considered the light curves and PSDs  
arising from a ``striped blazar jet'', which is characterized by alternating magnetic polarities.  Such a jet, with a structure proposed by Giannios \& Uzdensky (2019), would be launched from the immediate vicinity of the SMBH, essentially via the Blandford \& Znajek (1977) mechanism, and be seeded with reversing magnetic polarities through the magneto-rotational instability in the accretion disk (e.g., Balbus \& Hawley 1991).  Their attractive model yields PSD slopes near $-2$ via magnetic reconnection for short-term variability ($\lesssim$ weeks) but a wider range of PSD slopes ($-3$ to $-1.3$) over longer terms, and can consistently explain jet acceleration,  jet emission on large scales, as well as typical emission signatures of blazars (Zhang \& Giannios 2021).

Fluctuations in brightness detected by K2 that directly arise from accretion disks may also be present in many blazars, particularly FSRQs, where disk radiation can make a significant contribution to the optical emission during low states.  A recent study of a sample of 67 AGN with measured SMBH masses finds typical optical PSD slopes around $-2.0$ on timescales of days to several weeks, but a flattening at lower frequencies on $\sim 100$d timescales that can be associated with the expected thermal timescale for the ultraviolet emitting radius from accretions disks (Burke et al.\ 2021). In general, when PSDs are calculated for simulated light curves from accretion disk fluctuations, both phenomenological models (e.g., Mangalam \& Wiita 1993) and sophisticated three-dimensional general relativistic simulations (Noble \& Krolik 2009; Cowperthwaite \& Reynolds 2014; O'Riordan et al.\ 2017) tend to produce PSD slopes in the range $-1.3$ to $-2.1$, at least for sources where the disk is viewed nearly face-on, as expected for blazars.  As shown in  Figure 5, while both BL Lacs and FSRQs exhibit a range of PSD slopes, those of BL Lacs are typically steeper, with a mean  of $-2.22$ while those of FSRQs have a mean of $-1.98$ using the  periodogram approach in the sweet spot, and respectively $-2.05$ and $-1.88$ using the PSRESP method. 

We can understand this difference theoretically, as the jet simulations do tend to produce steeper PSD slopes than do those involving accretion disks, although both can yield slopes around $-2$.  This type of difference between the blazar sub-types is aligned with the recent result of Xiao et al.\ (2022) who make a strong case that the jets of FSRQs are mostly powered by accretion disks while those of BL Lacs are powered by extracting black hole rotational energy.  Steeper PSD slopes either mean there is relatively  more power at long timescales or, equivalently, there is less power at short timescales.  We expect that BL Lacs, typically seen at lower redshifts, will have had more time for their SMBHs to grow and thus should have somewhat higher SMBH masses, as well as typically higher spins, than FSRQs.  All relevant timescales,  including those for variations increase with SMBH mass, either linearly for processes depending on the ergosphere (Blandford-Znajek) or inner accretion disk, or sub-linearly for thermal accretion disk processes (e.g.\ Burke et al.\ 2021).  Then the jets that are likely to be powered by rapidly spinning black holes with higher mass (i.e., BL Lacs) are expected to have relatively more power at longer timescales and relatively less power at short timescales than jets arising only from accretion disks around less massive (FSRQ) SMBHs. 

The difference in PSD slopes of the optical emission between BL Lacs and FSRQs we have observed also could be understood in a framework where the key physical distinction is related to the location from which those photons emerge within otherwise similar jets.  The optical photons from low synchrotron peaked blazars, such as FSRQs, are at frequencies above the synchrotron peak, whereas they are below that peak for the majority of BL Lacs, which are high synchrotron peaked.  In standard shock-in-jet models (e.g.\ Marscher \& Gear 1985) the radiating particles are accelerated to the highest energies they can achieve in a small region in the immediate vicinity of a shock front where both particle densities and magnetic field strengths are greatest.  Photons of lower frequencies emerge at greater distances behind the shock that involve larger volumes.  Hence, purely from a simple light crossing time argument, we expect faster variations to be more dominant from smaller regions and so from photons on the higher frequency side of the peak, i.e., for FSRQs as opposed to BL Lacs.  Such a preponderance of fluctuations on shorter timescales corresponds to relatively more spectral power at higher frequencies, hence, a typically  shallower  PSD slope for  FSRQs than for BL Lacs.    This distinction between the amount of variability in BL Lacs and FSRQs can also be explained in terms of the optical emission from the former coming from below the synchrotron peak and so arising from relativistic electrons that are cooling more slowly than those in FSRQs, where the equivalent electrons are nearer the peak and rapidly losing energy  (e.g.,\ Hovatta et al.\ 2014).

\subsection {$\gamma$-ray Activity Level and Blazar Classes}

During the Campaign 14--18 K2 observations, six of the 18 unique AGN were detected with Fermi-LAT in the energy band 0.1-500 GeV at significance levels $\gtrsim 4\sigma$ (TS $>16$).
The maximum/minimum optical variability amplitude during the K2 campaigns was lower for the AGN not detected with the Fermi-LAT than for the LAT-detected AGN. Specifically, the maximum to minimum ratios for $\gamma-$ray detected vs. non-detected AGN were 2.73  vs.\ 1.13 for the Campaign 14--18 sample. This could indicate that more active states occurred simultaneously at both bands. Hovatta et al.\ (2014)  also found that $\gamma-$ray detected FSRQs are more optically variable than the BL Lacs. We note that Bhatta (2021) has considered the optical variability of 12 $\gamma$-ray blazars over the course of a decade or so by combining data from the SMARTS and Steward Observatory monitoring programs, supplemented with AAVSO and Catalina Surveys data.  The only source in common with our  entire sample is 3C 273, for which Bhatta found no correlation between the optical and $\gamma$-ray fluxes.  In contrast, Bhatta (2021) found that for the other 11 sources in his sample, these two bands were highly correlated, implying a co-spatial origin of those emissions. 
     
The slopes of the optical power spectral densities  of the $\gamma-$ray blazars detected and not detected by the contemporaneous Fermi-LAT observations were very similar, averaging $-2.16$ and $-2.24$, respectively.  This result is in agreement with what we found for the original PSD slopes of targets in Papers 1 and 2 when refitted with ``sweet spot'' frequency ranges.  The ranges of maximum to minimum ratios are also similar, with the exception of the outlier X-ray QSO RBS~1273 with slope $-4.20$. The similarity of the optical variability characteristics argues in favor of the dominant optical emission during the K2 observations coming from the relativistic jet.

A study by Ryan et al.\ (2019) of 13 blazar $\gamma$-ray light curves argued that CARMA(1,0) models do not always adequately describe the variability properties but that the slightly more complex CARMA(2,1) models apparently may do so, as they can naturally produce the low-frequency PSD breaks that seemed to be present in 4 of them.
In contrast, a recent analysis of monthly binned $\gamma$-ray light curves of a much larger sample of 236 bright blazars found PSD slopes concentrated around $-1$ but ranging between 0 and $-2$ (Burd et al.\ 2021).  These Fermi-LAT observations were examined in terms of Ornstein-Uhlenbeck, or CARMA(1,0), models and it was found that the light curves could all be reasonably be fit by them.  A critical result was that these $\gamma$-ray flux variations thus could be described in terms of a stochastic model that required only three parameters --- mean flux, correlation length and random amplitude --- each of which was rather narrowly peaked around well-defined values (Burd et al.\ 2021).

The dissipation of magnetic energy in jets can provide the source of accelerated particles and radiation from them.  For quite some time, shear driven Kelvin-Helmholtz instabilities have been known to destabilize propagating relativistic jets (e.g., Hardee \& Clarke 1992).  Magnetohydrodynamic (MHD) current-driven kink instabilities arising in strongly magnetized jets are expected to produce significant fluctuations in the emission from those jets (e.g., Appl et al.\ 2000).  Recently, several papers have used simulations to discuss these MHD kink instabilities in connection with particle acceleration via magnetic reconnection (Acharya et al.\ 2021; Bodo et al.\ 2021; Kadowaki et al.\ 2021; Medina-Torrej{\'o}n et al.\ 2021).   For instance, Acharya et al.\ (2021) have performed a suite of special relativistic MHD simulations of a portion of a jet column that show the development of kink instabilities and their impact on the jet density, pressure, velocity, magnetic fields, and magnetic energy dissipation. The initial conditions considered various magnetic field energies, bulk Lorentz factors, and number of axial wavelengths fitting in the simulation box.  These simulated jet zones fairly quickly become very inhomogeneous, with filamentary structures dominating the plasma densities, magnetic fields, and estimated photon fluxes.  Once Doppler boosting is taken into account, substantial variations in the observed intensity are produced by these twisting filaments and give rise to a quasi-periodic structure in the light curve, with a strong peak in an otherwise quite flat PSD (Acharya et al.\ 2021).   

Another approach to understanding  the development of magnetic reconnection in relativistic jets has been taken by Medina-Torrej{\'o}n et al.\ (2021) and Kadowaki et al.\ (2021).
This group has used a series of three-dimensional special-relativistic MHD simulations of a portion of a relativistic jet to investigate how current-driven kink instabilities grow in these conditions.  Those instabilities appear to naturally lead to turbulence that yields zones where magnetic reconnection appears to arise. That magnetic reconnection can accelerate low-energy protons to relativistic energies in blazar jets and might well explain both $\gamma$-ray and neutrino emission (Medina-Torrej{\'o}n et al.\ 2021).  Developing this approach even further, they show how the turbulence in both magnetic and kinetic energies eventually approach 3-D Kolmogorov spectra (Kadowaki et al.\ 2021) and yield significant variations in emitted and observed fluxes.  They produce a simulated $\gamma$-ray light curve that looks generally similar to that of   the Fermi-LAT observations of the very bright  and clearly variable high-energy source, Mrk 421 (Kushwaha et al.\ 2017). They find a PSD slope of $\approx -2$ for that simulated $\gamma$-ray light curve while the observations had a modestly flatter ($\approx -1.5$) PSD slope.  An extension of this work to produce optical light curves and PSDs would be welcome.

However, it is important to note that all of the relativistic jet light curve simulations of which we are aware have only had the computing power needed to focus on the emissions from rather limited portions of the jet, whether from turbulence in the region of a shock (e.g., Marscher 2014; Pollack et al. 2016), or from some form of magnetic reconnection in a portion of the jet (Zhang \& Giannios 2021; Acharya et al.\ 2021; Kadowaki et al.\ 2021). So if a larger portion  of the jet can be included in future simulations we would expect that the fractional amplitudes of flux variations would be reduced as fluctuations  arising from multiple volumes partially cancel out, thereby reducing any signals of quasi-periodicities in the process and very possibly altering the overall PSD slopes.

\subsection{Results on OJ~287}

1.  While there are no substantial differences in the PSD slopes from the long cadence data during Campaign 5 and 18 observations, there are clearly different slopes when the short cadence data are considered.  The long cadence data of Campaign 5 and 18 respectively have slopes of  $-2.28 \pm 0.17$ and $-1.96 \pm 0.20$ in the  ``sweet spot'' range from  log $\nu = -5.0$ to $-6.4$, while the short cadence data have slopes of $-2.65 \pm 0.05$ and $-2.26 \pm 0.06$, respectively, when we fitted the full allowed frequency range from log $\nu = -3.4$ to $-6.4$.   The substantially larger frequency range spanned by the short-cadence data allow for the reduced errors.  We note that when the short cadence data PSDs are evaluated only over the same ``sweet spot'' used for the long cadence PSDs, the slopes are in agreement. Thus, there seems to have been a change in the  PSD slopes before and after the predicted passage of the secondary SMBH through the accretion disk of the primary SMBH (e.g.,\ Lehto \& Valtonen 1996; Valtonen et al.\ 2016). Of course, this modeled event may not be the cause of  the PSD slope change.

2. OJ 287 was about 13-14\% brighter in Campaign 5 than in Campaign 18 three years later. During the intervening three years, the blazar brightened as predicted by the $\sim 12$ year periodicity (see Figure 6), (e.g., Valtonen et al.\ 2016), then faded.  As mentioned earlier, the blazar was in the low-to-middle range of its historical brightness during both K2 campaigns.

3. There was no sign of a plateau at 5-day timescales in the Campaign 18 data as there was in Campaign 5 (see Paper 1 for details).  We note that Campaign 5 was 47\% longer than Campaign 18, so any long-lived plateau would have been easier to detect in Campaign 5. The causative agent for the 5-day timescale phenomenon may have dissipated sometime in the 3-year interval that included the predicted passage of the secondary SMBH through the primary SMBH's accretion disk.    A recent paper on the MOMO (Multiwavelength Observations and Modeling of OJ 287) project using Swift satellite data (Komossa et al.\ 2021) provides an up-to-date discussion of variability in optical and X-ray bands;  the introduction of that paper gives an excellent brief review of OJ 287's variability in many wavebands.  Among the key findings of  their work is that a structure function analysis of the optical--UV data indicates a characteristic timescale of $\sim 5$ days when the source was at low-levels but apparently somewhat longer when OJ 287 was undergoing outbursts.  Komossa et al.\ (2021) note that this $\sim 5$ d timescale agrees with our earlier (Paper 1) finding of a flattening of OJ 287's K2 PSD slope around 5.8 days during a relatively low flux period in 2015.

4. The minimum observed timescale of astrophysical variation, as seen by inspecting the short cadence light curve and using the transition from white noise to red noise at log $\nu = -3.4$ in the PSD, was about 17 minutes, similar to the Campaign 5 data. The observed brightness variations in Campaign 18, listed in Table 3 as the ratio of maximum to minimum brightness,  were 1.36 and 1.52 for long and short cadence, respectively.   Such differences in long cadence and short cadence range values are clearly due to 30-minute smoothing in the long cadence data over sharp peaks and valleys  sometimes visible in the short cadence  light curves.

5. As noted above, the maximum to minimum brightness ratios in Campaign 18 were 1.36 and 1.52 for long and short cadences, respectively. The corresponding values for Campaign 5 were  2.31 and 2.36. Hence, there was more variation in Campaign 5 than Campaign 18, as can be seen in Figure 6. The overall ``jagged", extremely variable, character of the light curves remained the same after three years.   This optically violent variable nature is typical of OJ~287, as ground-based observations have shown over many years (see Komassa et al.\ 2021 for a review).  The K2 observations showed that the source was {\it consistently} active at these fast timescales, as presaged by observations during dedicated ground-based campaigns. We note that variations on timescales of an hour are sometimes, but not always, seen in ground-based observations of OJ 287 (see, for example, Gupta et al.\ 2019 and references therein). 

6.   OJ~287 was strongly detected by Fermi-LAT during Campaign 18, with flux of $7.10 \times 10^{-8} \pm 2.09 \times 10^{-10}$ ph s$^{-1}$ cm$^{-2}$ in the energy band 0.1-500 GeV with a test statistic of 139.0.  During Campaign 18, it was brighter in the $\gamma$-ray band than during Campaign 5,  when the  flux was $5.29 \times 10^{-8} \pm 9.21 \times 10^{-9}$ ph s$^{-1}$ cm$^{-2}$ with test statistic of 146.9.

\subsection{Results on 8 Other Objects Observed 2 or 3 Times}

In the following descriptions, we use the Fermi-LAT data presented in Papers 1 and 2 for Campaigns 1--12 and in this paper (Table 4) for Campaigns 14--18. We use the K2 data presented in Table 3 for all campaigns because it has been processed in a uniform manner that benefitted from improved correction for instrumental effects. For historical background on each object, please see Papers 1 and 2.

3C~207  (EPIC 211504760): This lobe-dominated radio quasar at redshift 0.6808 was not detected by Fermi-LAT during any of our three K2 observations. The Campaigns 5, 16 and 18 upper limits were $<3.12 \times 10^{-8}, < 6.29 \times 10^{-9}$, and $< 7.13 \times 10^{-9}$ ph s$^{-1}$ cm$^{-2}$, respectively,  in the energy band 0.1-500 GeV. With K2, we observed substantial slow and ``smooth''  variations in the count rates with maximum/minimum values  of 1.32, 1.25, and 1.13, respectively.  The PSD slopes  in Campaigns 5, 16, and 18 were not substantially different from each other: $-1.89 \pm 0.15, -1.59 \pm 0.26, $ and $-1.57 \pm 0.22$, respectively.  Superluminal motion has been observed with VLBI (e.g., Lister et al.\ 2016), but its lack of core-dominance may indicate that its jet is tilted farther from our line of sight than those of the bona fide blazars in our sample. A larger tilt may give rise to a smoother light curve. Alternatively, the smoother light curve may arise from  a  jet with only one or two synchrotron-emitting blobs adding  to emission from the corona or disk, unlike the ``jagged'' light curves  that may come from core-jets with many independently-emitting blobs. 

PKS 1335-127 (EPIC 212489625):  One of the best-known and widely studied FSRQ blazars, it has a redshift of 0.539.  It was detected with Fermi-LAT in Campaign 17 with flux $1.12 \times 10^{-8} \pm 3.03 \times 10^{-9}$  ph s$^{-1}$ cm$^{-2}$ in the energy band 0.1-500 GeV with test statistic 18.0. In Campaign 6, it was detected with flux $4.25 \times 10^{-8} \pm 7.49 \times 10^{-9}$  ph s$^{-1}$ cm$^{-2}$ with test statistic 33.3.  With K2 during both campaigns, we detected   ``jagged'' light curves with fractal-like variability at all scales. The PSD slopes were $-2.35 \pm 0.21$ and $-1.74 \pm 0.17$ in Campaigns 6 and 17, respectively. The maximum/minimum ratios were 1.64 and 1.61, respectively. 

PKS 1352$-$104 (EPIC 212595811):
This strong, bright FSRQ at redshift $= 0.332$ flared dramatically during our first K2 observations in Campaign 6, and was by far the most variable object in our sample. The K2 light curve in Campaign 6, excluding the flare in the last 16.5 days of Campaign 6, had a PSD slope of $-1.58 \pm 0.24$. In Campaign 17, the maximum/minimum was 1.86, with PSD slope of $-1.18 \pm 0.38$. It is the only object for which a big flare caused a significant change in PSD slopes between campaigns, though the two light curves have a similar ``jagged'' appearance.  We used the slope ($-1.86 \pm 0.10$) from the full light curve in Campaign 6 in Figure 5. With Fermi-LAT,  the blazar was weakly detected with flux $2.88 \times 10^{-8} \pm 6.45 \times 10^{-9}$  ph s$^{-1}$ cm$^{-2}$ in Campaign 6 and not detected in Campaign 17 with upper limit of $5.54 \times 10^{-8}$  ph s$^{-1}$ cm$^{-2}$. 

RBS 1273 (EPIC 212800574):
This X-ray QSO, also known as 1RXS 132928.0-053132, has $K_p = 15.201$ and redshift 0.57587.   The light curves showed three or four big broad bumps in both Campaigns 6 and 17.   This X-ray QSO, consequently, has the steepest PSD slopes in the sample.  Comparing the two PSDs, we see that the white noise extends to much different frequencies in the two K2 observations. The white noise  characteristics are consistent with those of stars on the same channels and thus probably do not indicate astrophysical changes in the AGN. RBS 1273 was not detected during either campaign 5 or 17 by Fermi-LAT with upper limits of $9.88  \times 10^{-9}$  ph s$^{-1}$ cm$^{-2}$ and $1.63  \times 10^{-8}$  ph s$^{-1}$ cm$^{-2}$, respectively.

PKS B1329$-$049 (EPIC 229227170,  with alternate EPIC id 229228144):
 This high redshift FSRQ  ($z= 2.15$)  with $K_p =18.2$  was unexpectedly faint during Campaigns 6 and 17 with average count rates of 310 and 570 cts s$^{-1}$ while the maximum/minimum ratios were 1.14 and 1.29, respectively. There was no significant difference in the PSD slopes of $-2.01 \pm 0.46$ and $-1.93 \pm 0.23$.   It was not detected by Fermi-LAT in either campaign 6 or 17, with upper limits of $2.92  \times 10^{-8}$  ph s$^{-1}$ cm$^{-2}$ and $2.15  \times 10^{-9}$  ph s$^{-1}$ cm$^{-2}$, respectively.

 RGB J0847+115 (EPIC 211394951):
This strong X-ray BL Lac object, detected at TeV energies, is also known as RX J0847.1+1133. It is slightly extended (2.3\arcsec) on SDSS observations with  $z$ = 0.1982 based on weak lines.  It was highly variable in Campaign 5 and  moderately variable in Campaign 18 (maximum/minimum ratios of 1.47 and 1.17, respectively). The PSD slopes were  similar: $-2.20\pm 0.26$ and $-2.02\pm 0.17$. It was detected by Fermi-LAT in Campaign 5 with flux $9.38  \times 10^{-9}  \pm 7.06  \times 10^{-9}$ ph s$^{-1}$ cm$^{-2} $ but not in Campaign 18 with upper limit $2.58  \times 10^{-9}$  ph s$^{-1}$ cm$^{-2}$.

WB J0905+1358 (EPIC 211559044,  with alternate EPIC id 211559047):
Also known as MG1 J090534+1358, this BL Lac has slow, smooth light curves with maximum/minimum ratios of 1.44 and 1.28 in Campaigns 5 and 18.  Small variations on timescales of a few days are superimposed on slow large amplitude bumps, with resulting PSD slopes of $-1.76 \pm 0.13$ and $-2.04 \pm 0.13$, respectively. It was detected by Fermi LAT in Campaign 5 with flux $4.89  \times 10^{-9}  \pm 3.53  \times 10^{-9}$ ph s$^{-1}$ cm$^{-2}$ but not in Campaign 18 with upper limit $2.59  \times 10^{-8}$  ph s$^{-1}$ cm$^{-2}$.

BZB J0816+2051 (EPIC 212035840): 	
 This BL Lac, with $K_p = 17.486$ has no known redshift; the optical spectrum is featureless, see, for example, SDSS DR16 (\url http://skyserver.sdss.org/dr16).  It had jagged light curves in Campaigns 5 and 18, with maximum/minimum ratios of 1.57 and 1.34. Its  respective PSD slopes were steep $-2.69 \pm 0.26$ and $-2.57 \pm 0.17$. It was not detected by Fermi-LAT in Campaign 18 with an upper limit of $4.77 \times 10^{-8}$ {ph~s$^{-1}$~cm$^{-2}$}.   In Campaign 5 the upper limit was $3.67 \times 10^{-9}$ {ph~s$^{-1}$~cm$^{-2}$}.

\subsection{Results on Other Individual AGN with Single Observations}

We present our results on the other individual AGN with single observations as follows. Light curves and binned PSDs are shown in Figure 4 (a-i), ordered by EPIC number.

4C +06.41= PKS 1038+064 (EPIC 248611911): 
One of two possible counterparts for the IceCube neutrino event 141209A detected in 2014 (IceCube Collaboration 2017, 2018), this is an FSRQ with $K_p =16.888$ at redshift 1.27.   Garrappa et al.\ (2019) resolved the 2FGL and 3FGL $\gamma$-ray sources near the position of 4C +06.41 by using 9.8 years of Fermi-LAT data into two sources 22\arcsec apart which are coincident with  4C +06.41 at RA, DEC = 10h41m17.1625s, +06d10m16.924s, and GB6 J1040+0617 at RA, DEC = 10h40m31.630s, +06d17m21.70s. Over the course of the 9.8 years of Fermi LAT observations, the two $\gamma$-ray sources varied in brightness. The multi-wavelength light curves  in Garrappa et al.\ (2019) showed GB6 1040+0617 had an increase in brightness  in the $\gamma$-ray band and flared  in the optical band while  4C+06.41 was not detected with Fermi-LAT during the arrival of IceCube neutrino event 141209A.  Hence, Garrappa et al.\ (2019) found that GB6 1040+0617 is more likely than PKS 1038+064 to be the origin of IceCube neutrino event 141209A. Nonetheless, 4C +06.41 remains a plausible neutrino counterpart. During our K2 observations, 4C +06.41 was at least  7 times brighter than GB6 1040+0617, specifically, 4C+06.41 was strongly detected  with Fermi-LAT with flux of $5.76 \times 10^{-8} \pm 2.64 \times 10^{-10}$ {ph~s$^{-1}$~cm$^{-2}$}  with test statistic of 42.1 while GB6 J1040+0617 had  a Fermi-LAT upper limit of $8.42 \times 10^{-9}$ {ph~s$^{-1}$~cm$^{-2}$} . Hubble Faint Object Spectrograph observations showed an absorption line system at $z =1.224$, in addition to the previously identified system at $z = 0.441$  (Bechtold et al.\ 2002); we do not expect that the absorption line systems affected the K2 photometry observations.  During our observations in Campaign 14, the maximum/minimum ratio was 1.18 while the PSD slope was $-2.17 \pm 0.13$. 

NVSS J110735+022225 (EPIC 201621388): 
This is a BL Lac object with  $K_p = 18.584$ with a spectroscopic redshift lower limit of 1.0735 obtained from MgII absorption line detection by Paiano et al.\ (2017). The NED-preferred redshift of 1.08213 was obtained with SDSS DR7 photometry and is not reliable because there are no well-detected lines; subsequent SDSS data releases' redshifts have similar problems.  The object is in a crowded optical field with one object about 10\arcsec away.  During Campaign 14, the K2 light curve had a maximum/minimum ratio of 1.11 and PSD slope of $-2.95 \pm 0.35$. It was not detected by Fermi-LAT with an upper limit of $< 1.97\times 10^{-9}$  {ph~s$^{-1}$~cm$^{-2}$}  during Campaign 14.

TXS 1013+054 (EPIC 251457104):
 This is an FSRQ with $K_p = 18.75$ and a reliable SDSS redshift of 1.71272 determined from 6 emission lines. During Campaign 14, the   jagged K2 light curve showed very high variability, with maximum/minimum of 8.50,  and a PSD slope of $-1.88 \pm 0.17$. Simultaneously, it was detected with Fermi-LAT with flux of $4.41 \times 10^{-8} \pm  6.3 \times 10^{-10}$  {ph~s$^{-1}$~cm$^{-2}$}  and test statistic 31.0.

TXS 0836+182 (EPIC 211852059): This is a BL Lac with $K_p =17.046$ and unknown redshift. An estimated redshift of  0.28 was obtained by Abraham et al.\ (1991) based on galaxy morphological fitting. The other redshift given in NED is  $1.155507 \pm 0.264737 $ from an SDSS spectrum which has no reliably detected emission lines.  The K2 maximum/minimum ratio was 1.26, with PSD slope of $-2.38 \pm 0.17$. It was not detected with Fermi-LAT during the K2 observations with an upper limit of $< 5.29 \times 10^{-9}$  {ph~s$^{-1}$~cm$^{-2}$}.

NVSS J090226+205045 (EPIC 212035517): 
This object has $K_p = 15.741$.  The redshift and class are not known. The two redshifts in NED were obtained by different generations of SDSS analysis; the most recent was $1.52204 \pm 1.41678$.  It was detected by Fermi-LAT with flux of $1.05 \times 10^{-8} \pm 7.93 \times 10^{-11}$ ph~s$^{-1}$~cm$^{-2}$ with test statistic 26.1. The K2 variability was strong in the jagged light curve, with maximum/minimum ratio 2.48 and PSD slope of $-2.67 \pm 0.14$.

TXS 0853+211 (EPIC 212042111):
This BL Lac has $K_p = 18.345$ and unknown redshift. The spectrum is nearly featureless: various SDSS data releases listed redshifts of $0.2762 \pm 4.00000$, $1.58702 \pm 1.00000$, or classified the object as an AO star. The K2 light curve is very jagged, with maximum/minimum ratio of 1.43 and PSD slope of $-1.76 \pm 0.18$. It was not detected by Fermi-LAT during the K2 observations with an upper limit of $<  4.36 \times 10^{-9}$  {ph~s$^{-1}$~cm$^{-2}$}. 

PMN J1318-1235 (EPIC 212507036): 
This object of unknown class has $K_p =18.027$	 and no known redshift. The jagged K2 light curve has maximum/minimum ratio of 1.71 and a  relatively steep PSD slope of $-2.55 \pm 0.21$.  It was not detected by Fermi-LAT during the K2 observations with an upper limit of $< 1.54\times 10^{-8}$  {ph~s$^{-1}$~cm$^{-2}$}.

NVSS J090900+231112 (EPIC 251376444):  As mentioned earlier, this BL Lac has a fainter, non-AGN companion galaxy 3.8\arcsec away (Rosa-Gonz{\'a}lez et al.\ 2017) with the same absorption  line redshift. The K2 light curve is jagged, with a maximum/minimum ratio of 1.24 and steep PSD slope of $-2.57 \pm 0.32$. It was detected  by Fermi-LAT with flux of $1.80 \times 10^{-9} \pm 2.57 \times 10^{-11}$ ph~s$^{-1}$~cm$^{-2}$ with test statistic of 16.6.

PKS B1310-041 (EPIC 251502828):
 This blazar has $K_p = 18.3$. Sowards-Emmerd et al.\ (2004) show an optical spectrum obtained with the HET/LRS with derived redshift of 0.8249 and classed the object as an FSRQ.  The K2 light curve is smooth, dropping precipitously toward the end of the K2 observations, with maximum/minimum ratio of 1.29 and PSD slope of $-1.73 \pm 0.10$. It was not detected by Fermi-LAT during the K2 observations with an upper limit of $< 8.13\times 10^{-8}$  {ph~s$^{-1}$~cm$^{-2}$}.

\section {Summary}

Our main results are as follows: 

1. Using a KS test  with a $p$ value of 0.039, we find a statistically significant difference between the periodogram PSD slopes of the BL Lacs and FSRQs in our  combined sample of Fermi-LAT detected blazars observed by K2,  with the BL Lac slopes steeper; however, this result is not as strongly supported by PSRESP analysis probably because it includes data at low frequencies below the well-sampled ``sweet spot'' which tend to flatten the slope. We find  that neither redshift nor doppler boosting effects can account for this difference.   Differences in the origin of the jets from the ergosphere or accretion disk in these two classes could produce such a disparity, as could the different sizes and locations of emission regions within the jets.

 2. We see small differences in PSD slopes of some AGN for which we had long cadence K2 observations at two or three epochs. The significance of the differences depends on whether the periodogram or PSRESP slopes are used.

3.  For OJ~287, although the two long cadence periodogram PSD slopes are consistent, the short cadence PSD slopes show a substantial difference over frequencies where the astrophysical red-noise  dominated over instrumental white noise, specifically, on timescales shorter than a few hours.  This change may be related to the passage of the secondary SMBH through the accretion disk of the primary SMBH which the binary black hole model predicted (Valtonen et al. 2016) would have occurred between those K2 observations.

4. The K2 light curves appeared either fast and  jagged or slow and smooth.  Three-quarters of the blazars had jagged light curves. No class differences were observed: specifically, 12 BL Lacs and 7 FSRQs had jagged light curves; 4 BL Lacs and 2 FSRQs had smooth light curves.   
 
5.  Sources with Fermi-LAT detections during their K2 observations usually showed greater optical variability.
\\
\\
Our other results include:

6. We found no correlation between redshift and PSD slope.

7. Contemporaneous ground-based observations for a subset of the sources confirm that the EVEREST Version 2 processing properly reproduces the astrophysical variability in blazars.

8. Simulations show that for an input red noise process with a specific slope, a Fourier-based time series analysis produces a spread of slopes around the input PSD slope  (see Appendix). Hence, PSD slopes for a single observation of a single source should be cautiously interpreted.
 
In conclusion, the K2 mission provided an unique window onto the optical variability properties of blazars, thanks to the  sub-hour cadence and essentially uniform temporal sampling of these observations made over quite extended durations.  Such measurements really cannot be obtained from the ground.  Currently, the Transiting Exoplanet Survey Satellite (TESS) is observing a larger number of blazars with measurements over shorter total durations (28 days) than K2 with the first of these results now available (Weaver et al.\ 2020; Raiteri et al.\ 2021). The TESS short cadence observations can probe the  higher frequency regions of the the PSDs of the very brightest blazars.  As these short timescales are where we noted a significant difference between the two observations of OJ~287, multiple observations of other very bright blazars by TESS should be illuminating.  In the future, the multi-day cadence monitoring of very large samples of blazars and other AGN will come from the LSST program on the Vera Rubin Telescope.  The LSST observations will provide superb long duration data sets  and produce critical complementary information on the optical variability discussed here. 

\begin{acknowledgements}

We thank the anonymous reviewer for their comments. We thank Josefa Becerra Gonzalez, Deirdre Horan, and Arti Goyal for helpful discussions. We are  grateful to Rodrigo Luger (University of Washington and Flatiron Institute) for advice on processing short cadence data through EVEREST software. We thank Paul Smith (Steward Observatory) for helpful discussions and for making his data publicly available. We thank Simone Garrappa for advice and discussion about the 4C+06.41 field.

We acknowledge support from NASA K2 Guest Investigator Program grant NNX16AI61G. This paper includes data collected by the K2 mission. Funding for the K2 mission is provided by the NASA Science Mission Directorate.  

B.S.G. was supported by Thomas Jefferson Chair for Discovery and Space Exploration at the Ohio State University.
 
The 0.5-m DEdicated MONitor of EXotransits and Transients telescope (DEMONEXT) is located at and operated by the Winer Observatory in Arizona and funded by the Massachusetts Institute of Technology, The Ohio State University, and Vanderbilt University.

The 1.3-m Robotically Controlled Telescope (RCT) located at Kitt Peak National Observatory is operated by the RCT Consortium.

Data from the Steward Observatory spectropolarimetric monitoring project were used. This program is supported by Fermi Guest Investigator grants NNX08AW56G, NNX09AU10G, NNX12AO93G, and NNX15AU81G. 

The Fermi LAT Collaboration acknowledges generous on-going support from a number of agencies and institutes that
have  supported  both  the  development and  the  operation  of
the LAT as well as scientific data analysis. These include the
National Aeronautics and Space Administration and the Department of Energy in the United States, the Commissariat a
l'Energie Atomique and the Centre National de la Recherche
Scientifique /  Institut National de  Physique Nucl\'eaire et  de
Physique des Particules in France, the Agenzia Spaziale Italiana and the Istituto Nazionale di Fisica Nucleare in Italy, the
Ministry of Education, Culture, Sports, Science and Technology (MEXT), High Energy Accelerator Research Organization (KEK) and Japan Aerospace Exploration Agency (JAXA)
in Japan, and the K.\ A.\ Wallenberg Foundation, the Swedish
Research Council and the Swedish National Space Board in
Sweden.  Additional support for science analysis during the
operations  phase  is  gratefully  acknowledged  from  the  Istituto Nazionale di Astrofisica in Italy and the Centre National  d'  \'Etudes Spatiales in France.

This research has made use of the NASA/IPAC Extragalactic Database (NED) which is operated by the Jet Propulsion Laboratory, California Institute of Technology, under contract with the National Aeronautics and Space Administration.  
Some of the data presented in this paper were obtained from the Mikulski Archive for Space Telescopes (MAST). STScI is operated by the Association of Universities for Research in Astronomy, Inc., under NASA contract NAS5-26555. Support for MAST for non-HST data is provided by the NASA Office of Space Science via grant NNX09AF08G and by other grants and contracts. The 
specific observations analyzed can be accessed via \dataset[https://dx.doi.org/10.17909/ajrj-2a29]{https://dx.doi.org/10.17909/ajrj-2a29}.

Some  observations were obtained with the Samuel Oschin 48-inch Telescope at the Palomar Observatory as part of the Zwicky Transient Facility project. ZTF is supported by the National Science Foundation under Grant No. AST-1440341 and a collaboration including Caltech, IPAC, the Weizmann Institute for Science, the Oskar Klein Center at Stockholm University, the University of Maryland, the University of Washington, Deutsches Elektronen-Synchrotron and Humboldt University, Los Alamos National Laboratories, the TANGO Consortium of Taiwan, the University of Wisconsin at Milwaukee, and Lawrence Berkeley National Laboratories. Operations are conducted by COO, IPAC, and UW. 

\end{acknowledgements}

\facility{Kepler, K2, Fermi, RCT, SMARTS, ZTF}.

\software{EVEREST (Luger, R., Agol, R., Kruse, E., Barnes, R., Becker, A., Foreman-Mackey, D., \& Deming, D. 2016, AJ, 152,100), K2SFF (Vanderburg, A. \& Johnson, J.~A. 2014, PASP, 126, 948)}

\section {Appendix: Simulations}

\subsection{Comparison of PSD Analysis Approaches}

To better understand the potential limitations of the PSD analysis, we undertook a series of simulations comparing calculating the PSD from a Fast Fourier Transform (FFT,) from a  straightforward implementation of the discrete Fourier transform (DFT),  and from a periodogram (Scargle 1982) analysis.  Light curves were initially simulated following the method of Timmer \& Koenig (1995), and then PSDs were calculated.

In one of the most frequently cited publications on PSDs of  light curves, Fougere (1985) simulated 100 red noise light curves with a 5 minute duration and 2.8 second cadence, matching that of ionospheric scintillation data from the MARIST satellite, then calculated PSDs and determined their slope. He showed that for end-matched and windowed data with an input PSD slope = $-2.00$, an average PSD slope of $1.9841 \pm 0.0688$ was found when the PSD was determined via an FFT.  No binning of the PSD was done prior to his fitting process. As described in Section 4, we used a periodogram rather than an FFT (since the K2 data are not precisely evenly spaced) to determine the PSD, and we binned the PSD by 0.08 in log $\nu$ space prior to fitting. In order to test the effects of this binning, we  produced 100 red noise light curves and PSDs with the same duration and cadence as in Fougere (1985), determined the PSD via  both a FFT and a periodogram analysis, and then binned  and fit the PSD slope in the same manner as in this paper. For the PSD produced by the FFT, we found an average slope of $-1.92 \pm 0.04$ and for the PSD produced by the periodogram analysis, we found an average slope = $-2.00 \pm 0.04$. It does not appear that the use of binning or the use of periodogram rather than the FFT  to produce the PSD has any significant effect on the slope calculation  though the periodogram may be preferable.  The resulting histograms of PSD slopes are shown in Figure 7a,b.

\begin{figure}[ht]
\figurenum{7}
\vspace{0mm}.
\begin{center}
\gridline{\fig{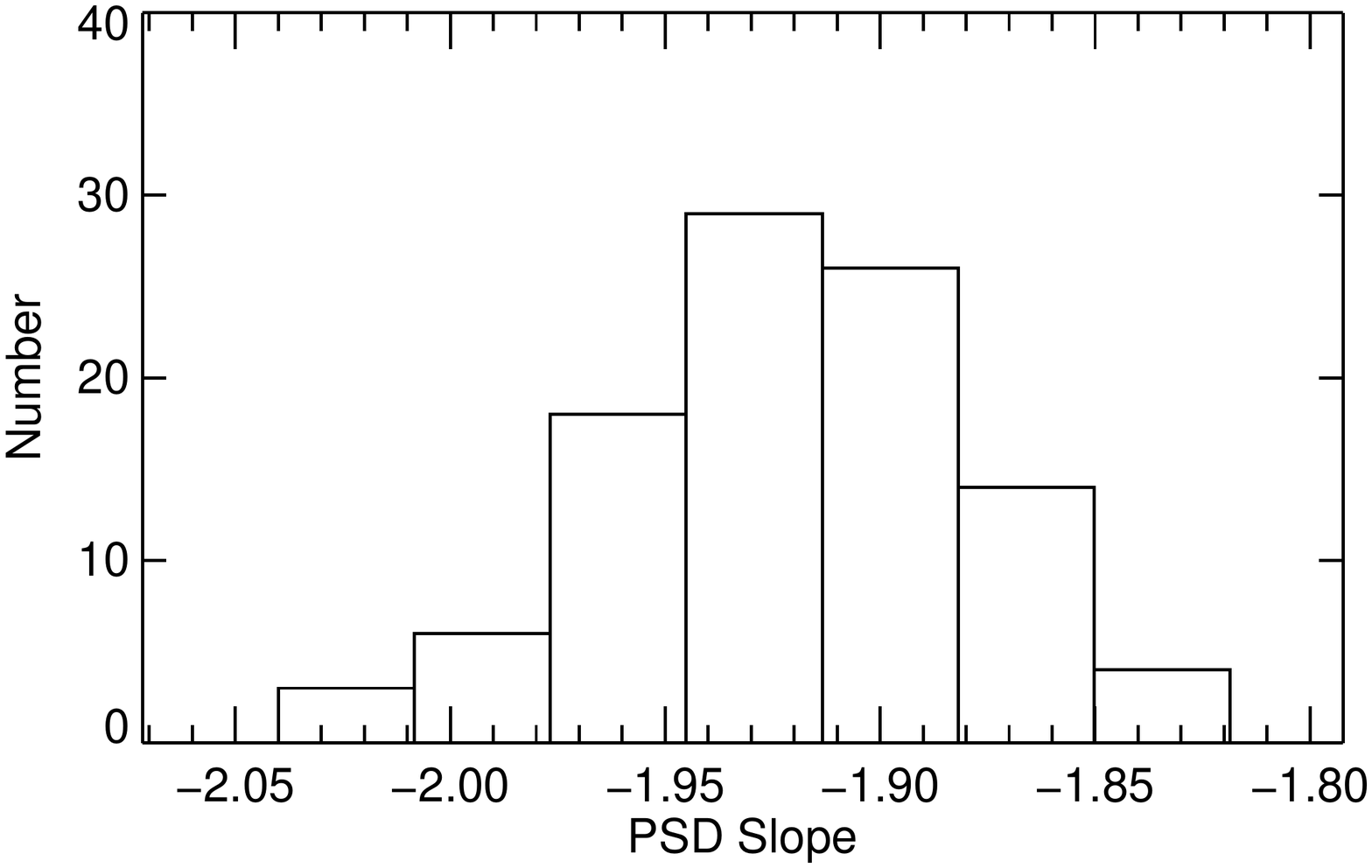}{0.45\textwidth}{(a)}
          \fig{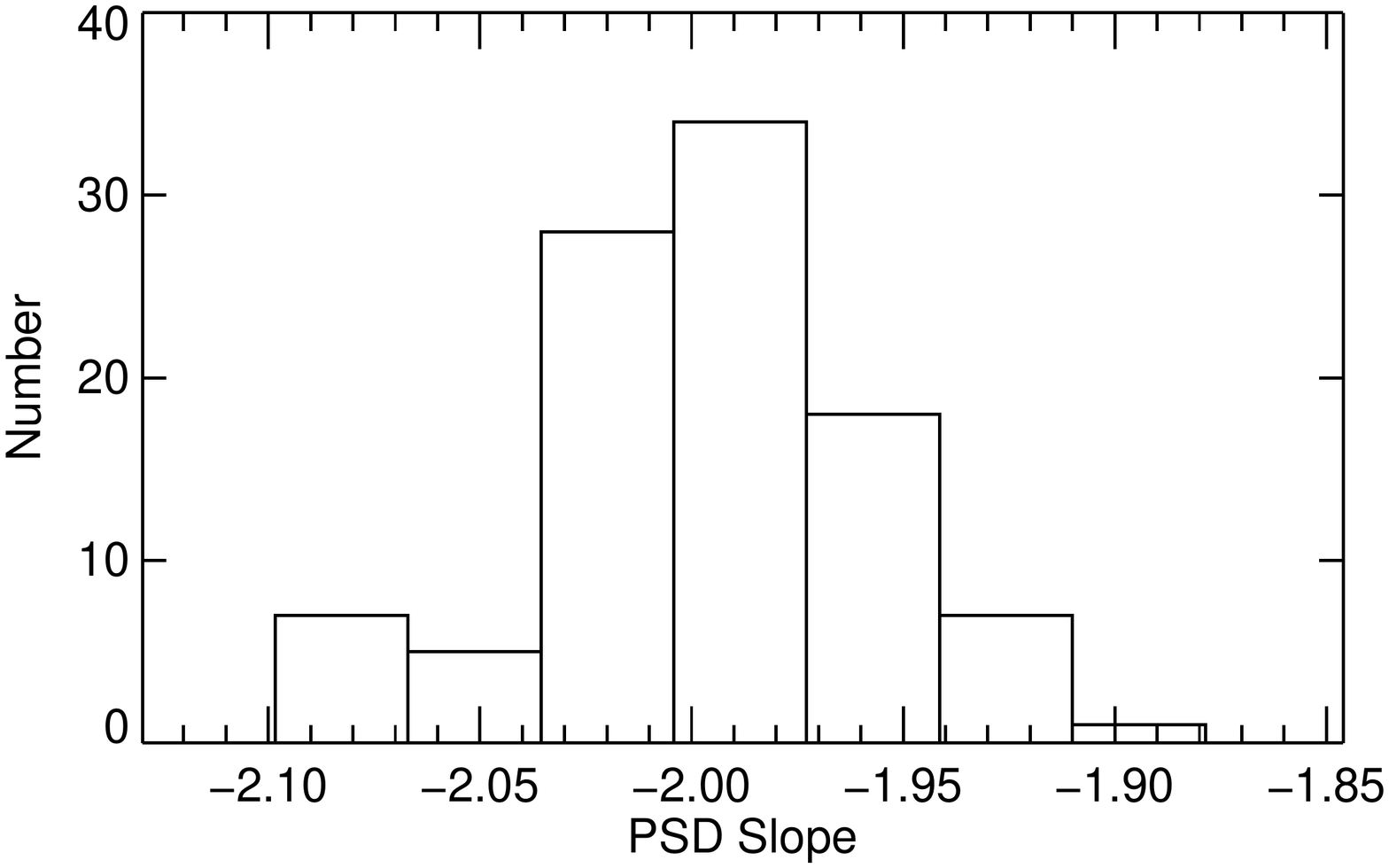}{0.45\textwidth}{(b)}
          }
\gridline{\fig{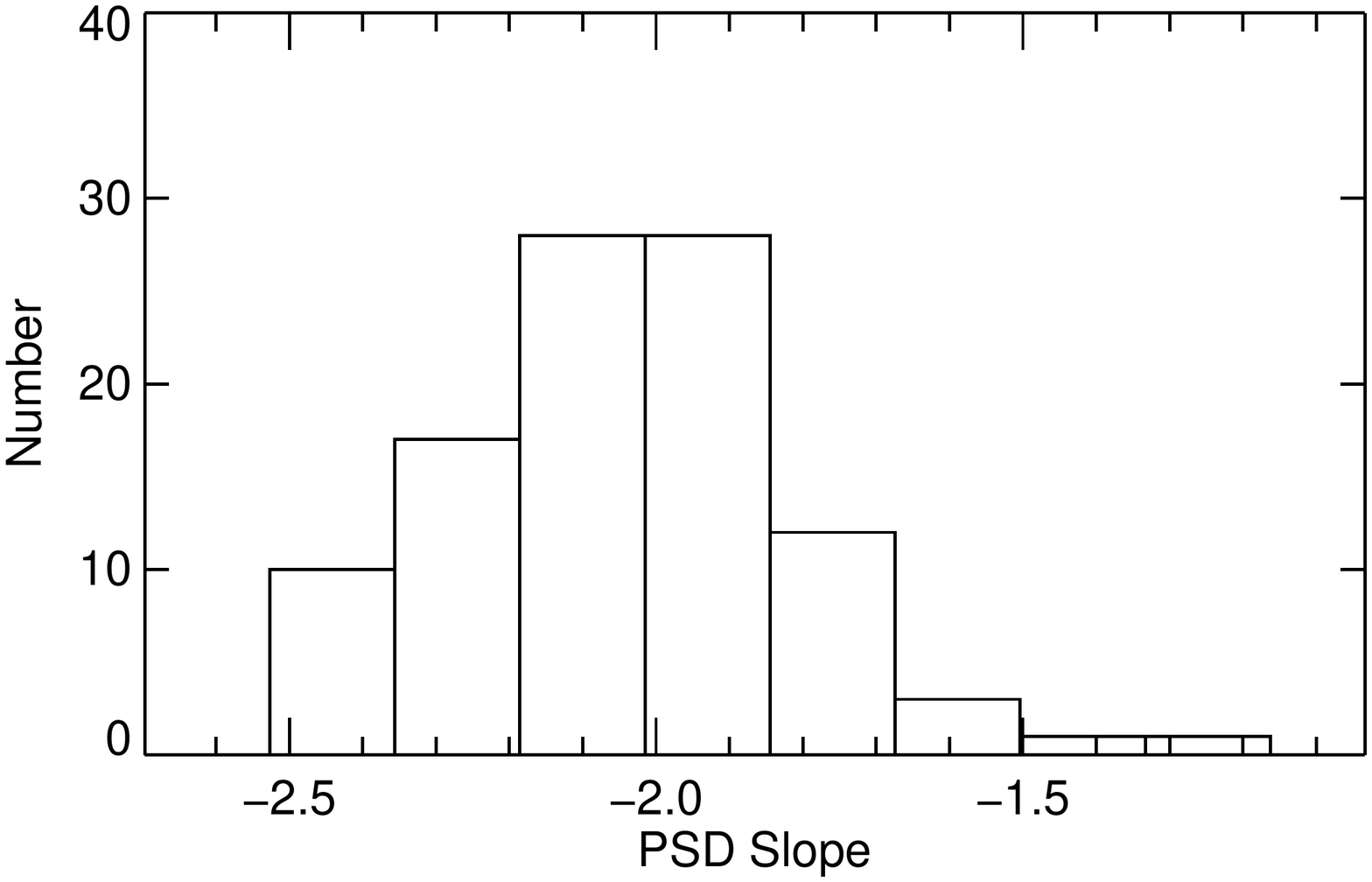}{0.45\textwidth}{(c)}
          \fig{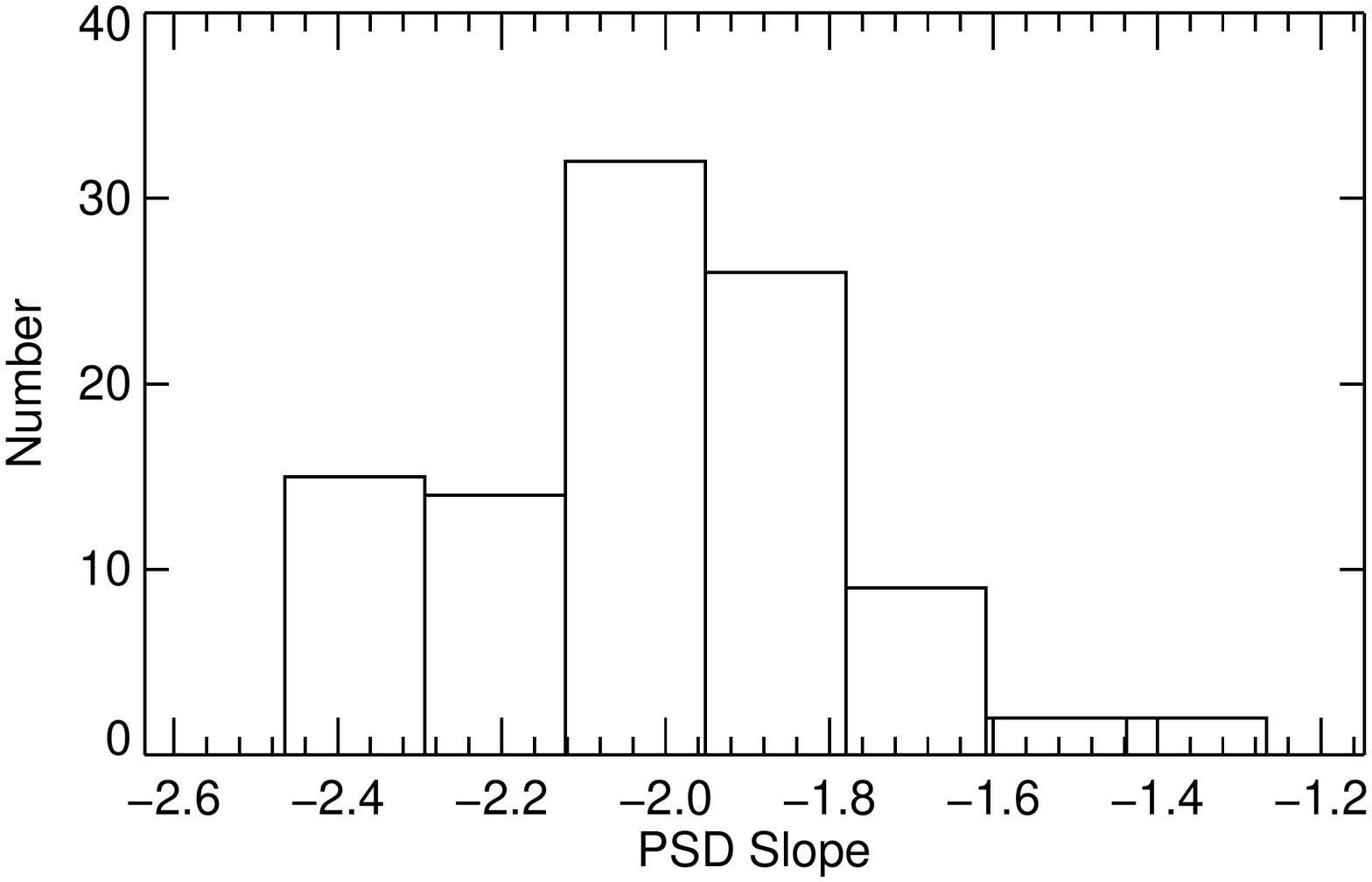}{0.45\textwidth}{(d)}}
\gridline{\fig {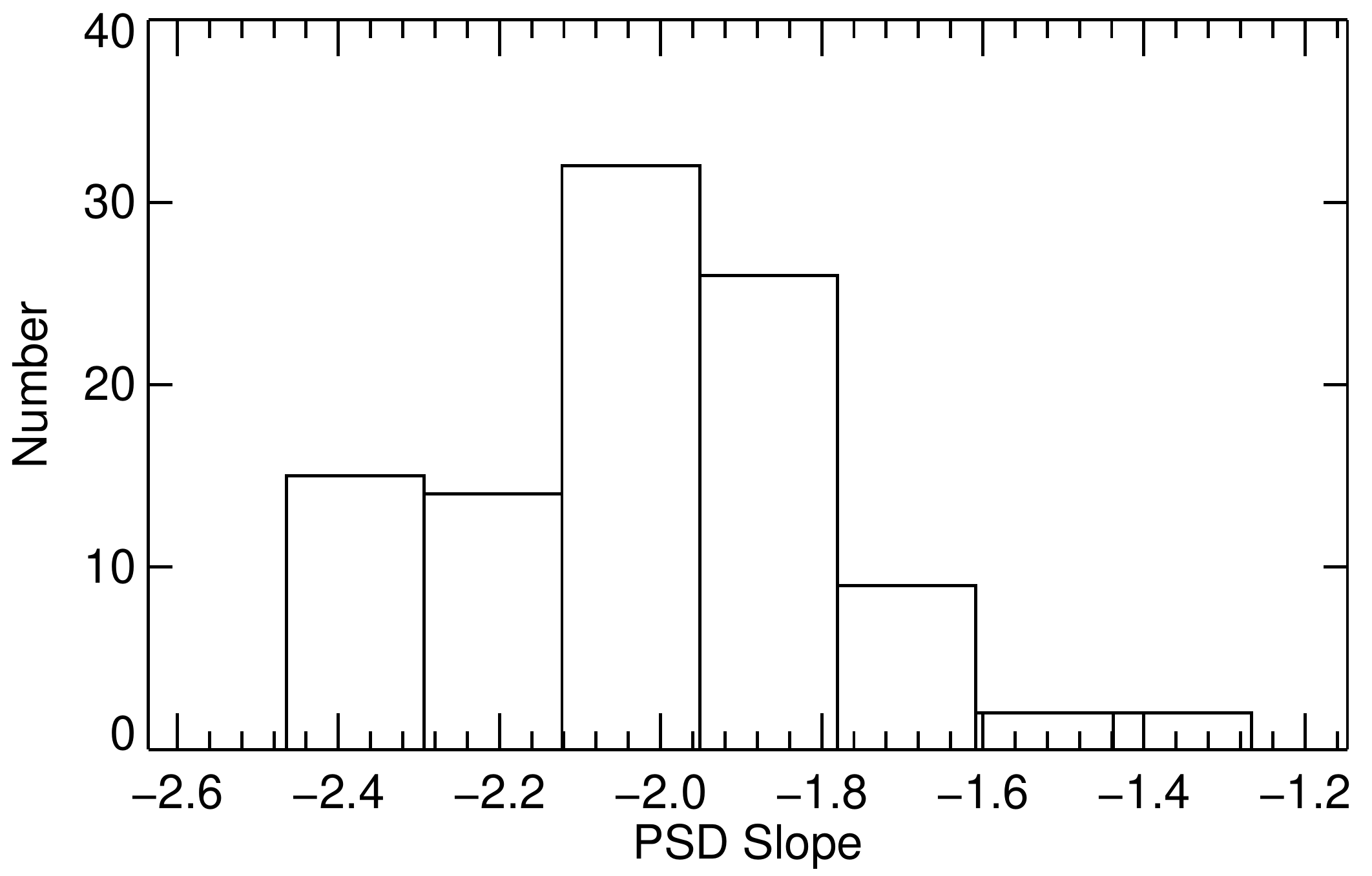}{0.35\textwidth}{(e)}
            \fig{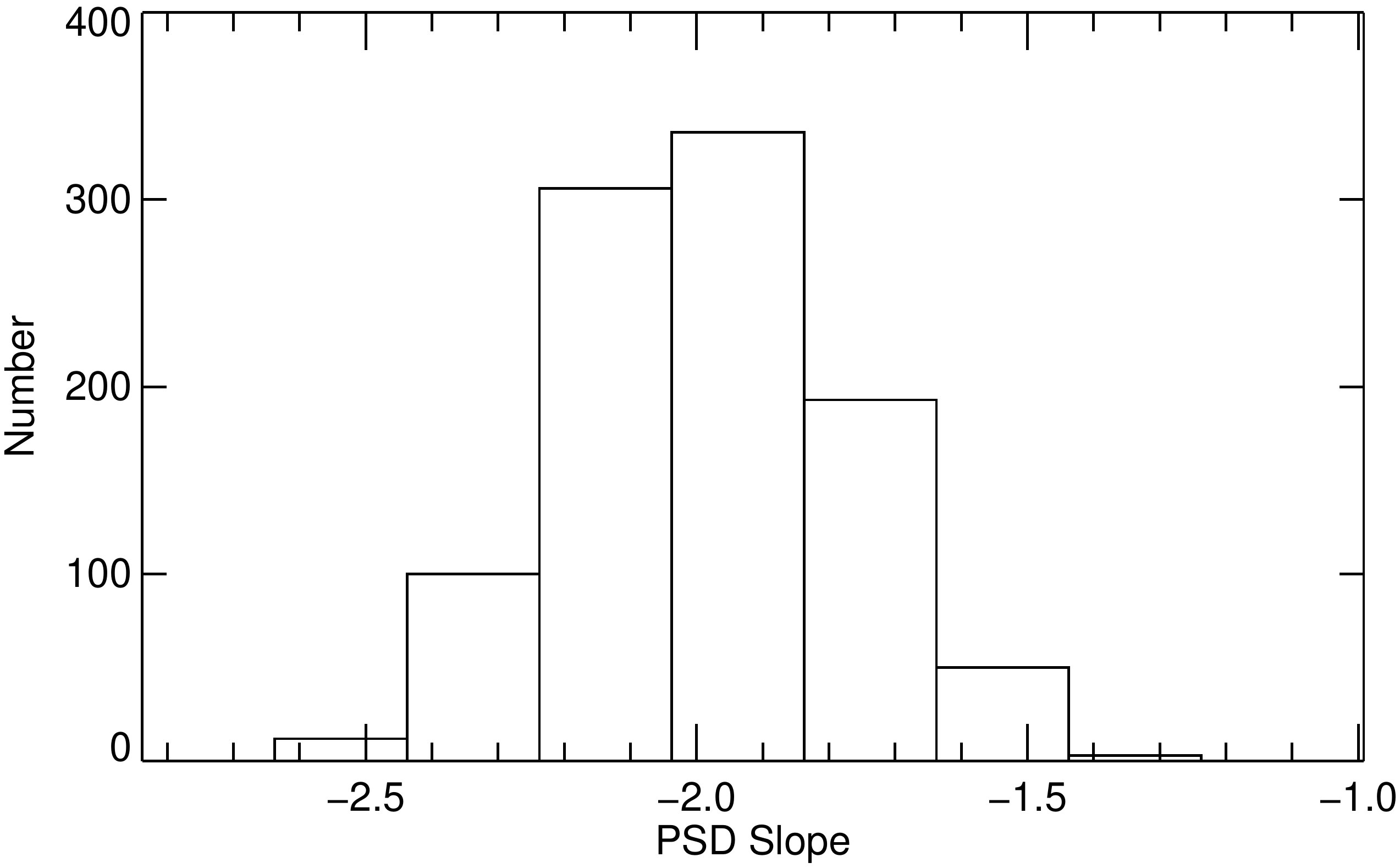}{0.35\textwidth}{(f)}}
\caption {Histograms of simulated PSD slopes calculated via (a) an FFT  and (b) a periodogram  with the same duration and cadence as Fougere (1985).  Use of a periodogram to determine the PSD and binning of the PSD prior to fitting do not affect the determination of the PSD slope.  Histograms of simulated PSD slopes calculated via  (c) an FFT,  (d) a DFT and (e) a periodogram for 100 simulated light curves with the K2 observation cadence and duration, generated with an input PSD slope of $-2.0$.  On average, the periodogram returns the closest match to the input PSD slope. (f) Histogram of PSD slopes determined from our periodogram analysis of 1000 light curves simulated with an input PSD slope of $-2.0$, with each 800-day light curve broken into 10 segments, and with a K2 template sampling pattern applied.}
\end{center}
\end{figure}

We noticed that in the simulations described above, an input slope of $-2.0$ returned a range of measured output PSD slopes. In order to understand the implications of this for the use of the PSD to determine the underlying process,  we simulated 100 light curves with an input slope $= -2.0$, a turnover to white noise at log($\nu$) = $-5.0$, 30-minute sampling and a duration of 80 days, thus matching the K2 light curve characteristics of the objects in our sample.  We calculated the PSDs using three techniques:  FFT, discrete Fourier transform and a periodogram analysis.  We performed the fits to the slopes in exactly the same manner as for the K2 light curves in this paper. The resulting histograms of the distribution of fitted PSD slopes are shown in Figure 7c,d,e. For the FFT case, the average slope = $-2.05 \pm 0.23$, for the DFT case the average slope = $-2.22 \pm 0.23$ and, for the periodogram case, the average slope = $-2.02 \pm 0.23$. The periodogram  approach returns the average output slope closest to the input slope of  $-2.0$; however, in all three cases,  a  very similar range of slopes is returned by the analysis. These simulations show that while a slope returned via a PSD analysis does not always precisely reflect the input PSD, the computed slopes will usually be quite close to the input power-spectrum. 
We then examined what, if any, effect the slightly uneven sampling of the K2 observations would have on the determination of the PSD slope via our periodogram methodology. Examining our K2 light curves, we found that after excluding data points with poor data quality (determined using the project-provided quality flags), between 5 and 10\% of cadences were removed from the light curves. Using the observation of OJ 287 as a typical (6.8\% cadence loss) sampling template, we simulated 1000 light curves with a duration 10 times that of the K2 observations (800 days) with an input PSD slope of $-2.0$. Each 800 day light curve was broken into ten 80-day segments (to mimic the usual K2 observation campaign length) and the typical K2 sampling pattern described above was applied. The resulting distribution of output slopes after application of our periodogram method is shown in Figure 7(f). The result is that for an input slope of $-2.0$, the mean returned slope was  $-1.99 \pm 0.21$.

We also investigated the effect that our choice of bin size for the PSD has on the resulting PSD slope determination.
We simulated an 80-day light curve with 30-minute sampling, a slope of $-2.0$ and a turnover to white noise at log($\nu$)= $-5.0$. We fit the PSD in the same manner as the PSDs in this paper.   We find that  using  bin sizes between 0.05 and 0.20 in log space does not significantly affect the slope determination. 

\subsection{Effects of Doppler Boosting and Redshifts}

 To test if the effects of different Doppler boosting or redshifts could be the cause of the differences in PSD slopes between the BL Lac objects and FSRQs in our sample, we simulated  1000 rest frame light curves and applied Doppler boosting corrections to simulate the rest frame light curve observed as a BL Lac and as an FSRQ. The rest frame light curves were simulated via the process described in  Timmer \& Koenig (1995). We assumed an input PSD slope of $-2.0$ and a turn over to white noise at log ($\nu$)$ =-5.0$, which is consistent with the typical turnover to white noise seen in our sample objects.  The Doppler boosting corrections applied were those discussed in Gopal-Krishna et al.\ (2003). We used the mean redshifts of BL Lacs (0.329) and FSRQs (1.05) and we assumed a spectral index of $-1.0$. We assumed average Doppler boosting factors of 9.2 for BL Lacs and 15.2 for FSRQs, as given in Liodakis et al.\ (2017).  We computed the  PSDs and their slopes using the same method as for our sample objects, described in Section  4 and Papers 1 and 2.

We found no evidence that Doppler boosting  (or redshift differences, which partially cancel each other) would create the difference in PSD slopes seen in our sample.  The mean PSD slope of the rest frame light curves boosted to  the above BL Lac parameters was  $-2.02 \pm 0.25$  and for rest frame light curves boosted to FSRQ parameters also was $-2.04 \pm 0.25$, essentially identical to the input slope in both cases. Figure 8  shows histograms of the distribution of PSD slopes. The distributions look almost identical; a two sided KS test  returns $p =0.60$ for the null-hypothesis that these two distributions arise from the same parent population. 

\begin{figure}[ht]
\figurenum{8}
\vspace{0mm}
\begin{center}
\gridline{\fig{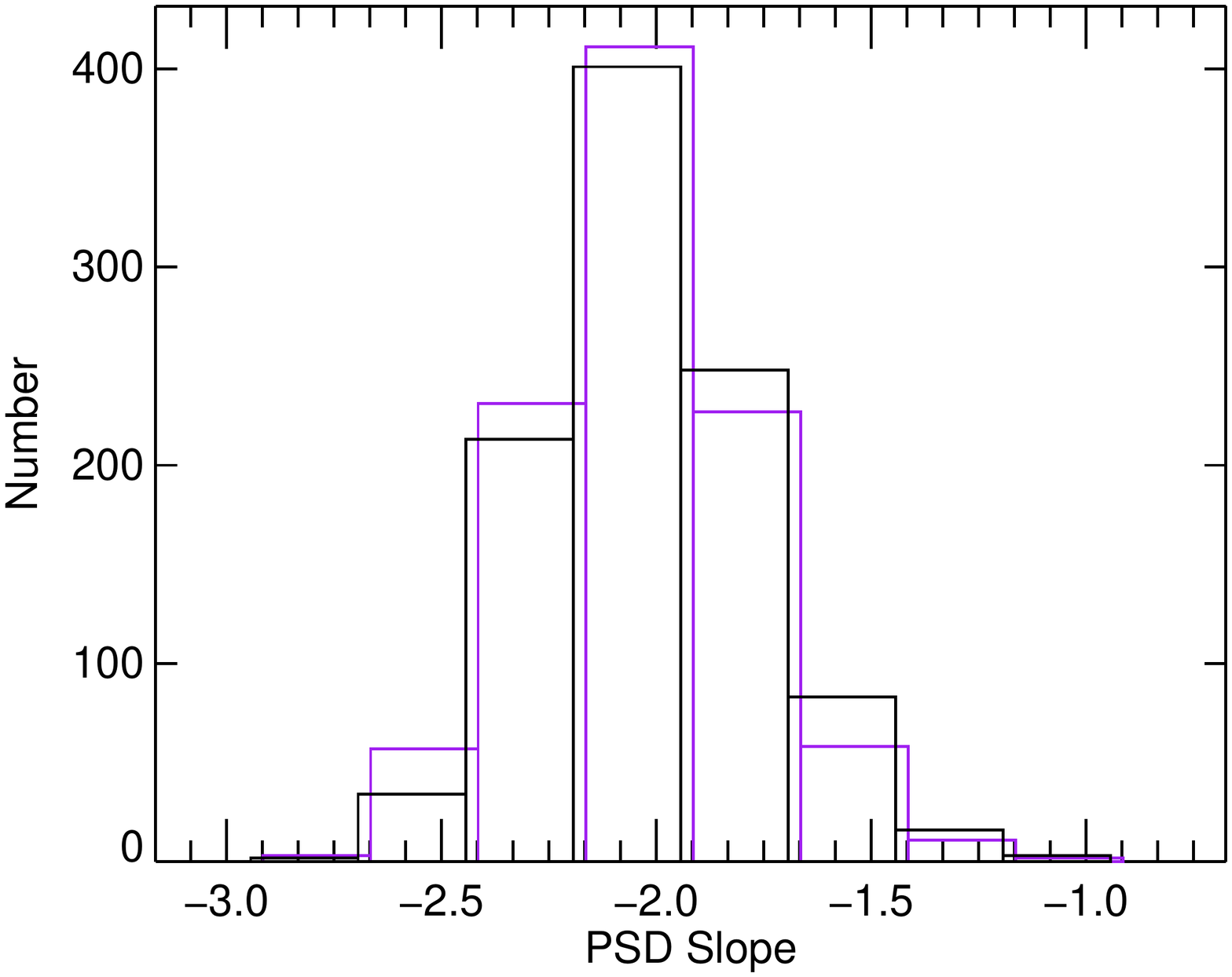}{0.50\textwidth}{}
          }
\caption{Histograms of the PSD slopes of 1000 simulated  red noise AGN light curves with an input PSD slope of $-2.0$, boosted by Doppler factors of  9.2 for BL Lacs (black) and 15.2 for FSRQs (purple).  The average slope and  range of slopes are not affected by Doppler boosting or redshift differences.}
\end{center}
\end{figure}
 
\newpage

\end{document}